\pdfoutput=1
\documentclass[universe,article,accept,moreauthors,pdftex,10pt,a4paper]{mdpi}

\firstpage{1}
\articlenumber{23}
\doinum{10.3390/universe2040023}
\pubvolume{2}
\pubyear{2016}
\copyrightyear{2016}
\externaleditor{Academic Editors: Lorenzo Iorio and Elias C. Vagenas}
\history{Received: 21 August 2016; Accepted: 14 September 2016; Published: 28 September 2016}
\newcommand\myurl[1]{\changeurlcolor{black}\url{#1}\changeurlcolor{blue}}
\makeatletter
\g@addto@macro{\UrlBreaks}{\UrlOrds}
\makeatother

\usepackage{amssymb}
\usepackage{definitions}
\usepackage{pdflscape,multirow}
\usepackage{upgreek,soul}
\usepackage{microtype}
\newcommand*{\dd}{\, \mathrm{d}}
\newcommand{\DE}{\mathrm{DE}}
\newcommand{\vect}[1]{\textbf{\textrm{}{#1}}}
\newcommand{\OM}{\Omega_\mathrm{m}}
\newcommand{\OD}{\Omega_\mathrm{DE}}
\newcommand{\OR}{\Omega_{\mathrm{rad}}}
\newcommand{\OC}{\Omega_{\mathrm{c}}}
\newcommand{\OL}{\Omega_{\Lambda }}
\newcommand{\Onu}{\Omega_\nu}

\newcommand{\OB}{\Omega_b}
\newcommand{\hpM}{\,h\,\mathrm{Mpc}^{-1}}
\newcommand{\e}{\mathrm{e}}
\newcommand*\Laplace{\mathop{}\!\mathbin\bigtriangleup}
\newcommand*\DAlambert{\mathop{}\!\mathbin\Box}
\newcommand{\bq}{\begin{equation}}
\newcommand{\eq}{\end{equation}}
\newcommand{\LCDM}{\Lambda\mathrm{CDM}}
\def\expec#1{\langle#1\rangle}

\def\brn#1{\left(#1\right)}   %% replaces \fett(   and \right)
\let\mr=\mathrm
\let\mc=\mathcal
\let\ol=\overline

\newcommand{\bs}{\boldsymbol{}}

 \theoremstyle{mdpi}
 \newcounter{thm}
 \newcounter{ex}
 \newcounter{re}
 \theoremstyle{mdpidefinition}

\Title{General Relativity and Cosmology: Unsolved Questions and Future Directions}

\Author{Ivan Debono $^{1,*,\dagger}$ and George F. Smoot $^{1,2,3,}$$^{\dagger}$}
\AuthorNames{Ivan Debono, George F. Smoot}
\address{%
$^{1}$ \quad Paris Centre for Cosmological Physics, APC, AstroParticule et Cosmologie, Universit\'e Paris Diderot, CNRS/IN2P3, CEA/lrfu, Observatoire de Paris, Sorbonne Paris Cit\'e, 10, rue Alice Domon et L\'eonie Duquet, 75205 Paris CEDEX 13, France; gfsmoot@lbl.gov\\
$^{2}$ \quad Physics Department and Lawrence Berkeley National Laboratory, University of California, Berkeley, \mbox{94720 CA,} USA\\
$^{3}$ \quad Helmut and Anna Pao Sohmen Professor-at-Large, Hong Kong University of Science and Technology, Clear~Water Bay, Kowloon,  999077 Hong Kong, China}

\corres{Correspondence: ivan.debono@apc.univ-paris7.fr; Tel.: +33-1-57276991}

\firstnote{These authors contributed equally to this work.}
\abstract{For the last 100 years, General Relativity (GR) has taken over the gravitational theory mantle held by Newtonian Gravity for the previous 200 years. This article reviews the status of GR in terms of its self-consistency, completeness, and the evidence provided by observations, which have allowed GR to remain the champion of gravitational theories against several other classes of competing theories.  We pay particular attention to the role of GR and gravity in cosmology, one of the areas in which one gravity dominates and new phenomena and effects challenge the orthodoxy.  We also review other areas where there are likely conflicts pointing to the need to replace or revise GR to represent correctly  observations and consistent theoretical framework. Observations have long been key both to the theoretical liveliness and viability of GR. We conclude with a discussion of the likely developments over the next 100 years.}

\keyword{General~Relativity; gravitation; cosmology; Concordance~Model; dark energy; dark matter; inflation; large-scale~structure}

\begin{document}

%%%%%%%%%%%%%%%%%%%%%%%%%%%%%%%%%%%%%%%%%%
%% Sections that are not mandatory are listed as such. The section titles given are for Articles. Review papers and other article types have a more flexible structure.

%% Only for the journal Gels: Please place the Experimental Section after the Conclusions

%%%%%%%%%%%%%%%%%%%%%%%%%%%%%%%%%%%%%%%%%%

\section{Perspective}

Scientists have been fascinated by General Relativity ever since it was developed.  It has been described as poetic, beautiful, elegant, and, at times, as impossible to understand.

General Relativity is often described as a simple theory. It is hard to define simplicity in science. One can always construct an entire theory encapsulated in one equation.
Richard Feynman famously demonstrated this in a thought experiment where he rewrote all the laws of physics as $\vec{U}=0$, where each element of $\vec{U}$ contained the hidden structure \cite{Feynman1963}.
His point was that simplicity does not automatically bring truth.

An examination of the mathematical structure of General Relativity gives us a more sober definition of ``simplicity''. Under certain assumptions about the structure of physical theories, and of the properties of the gravitational field, General Relativity is the only theory that describes gravity. Alternative theories introduce additional interactions and fields.

General Relativity is also unique among theories of fundamental interactions in the Standard Model. Like electromagnetism, but unlike the strong and weak interactions, its domain of validity covers the entire range of length scales from zero to infinity. However, unlike the other forces, gravity as described by General Relativity acts on all particles. This implies that the theory does not fail below the Planck scale. All gravitational phenomena, from infinitesimal scales to distances beyond the observable universe, may be modelled by General Relativity. We may therefore formulate a mathematically rigorous description of General Relativity: it is the most complete theory of gravity ever developed.

 All gravitational phenomena that have ever been observed can be modelled by General Relativity.
It describes everything from falling apples, to the orbit of planets, the bending of light, the dynamics of galaxy clusters, and even black holes and gravitational waves.
The domain of validity of the theory covers a wide range of energy levels and scales.
That is why it has survived so long, and that is why it survives today, one hundred years after it was formulated, in an age in which the amount of data and knowledge increases by orders of magnitude every few years.

\begin{figure}[H]
\centering
\includegraphics[width=0.7\textwidth]{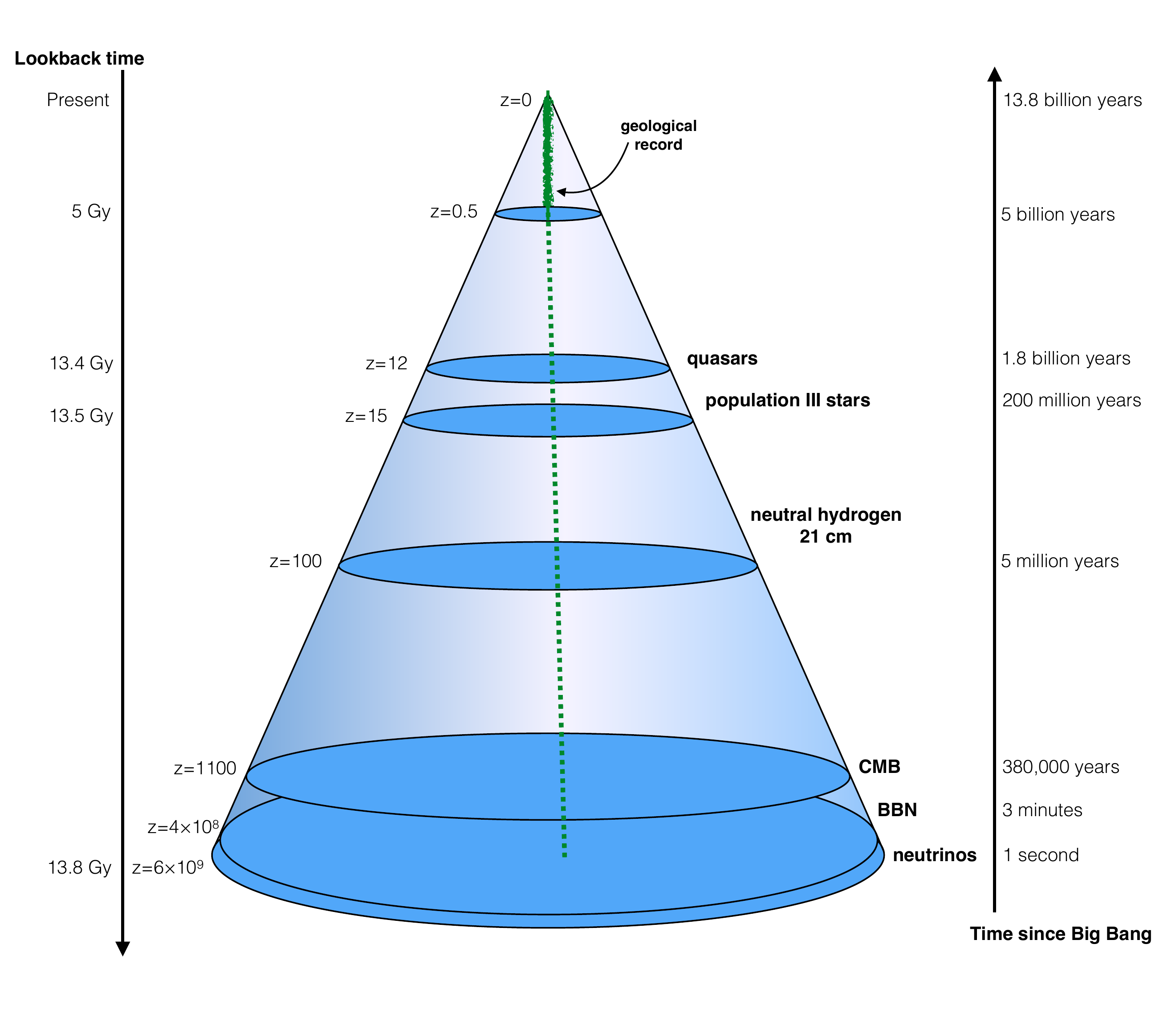}
\caption{How we observe the universe. The lookback time is the difference between the age of the universe now, and the age of the universe when photons from an object were emitted. The more distant an object, the farther in its past we are observing its light. This distance in both space and time is expressed by the cosmological redshift $z$. We obtain most of our astrophysical information from the surface of our past light cone, because it is carried by photons. The only information from within the cone come from local experiments and observations, such as geological records. The~green dotted line is the world-line of the atoms and nuclei providing the material for our geological data. Local~experiments are carried out along this bundle of world-lines. They provide a useful test of physical constants. One~example is the observation of the Oklo phenomenon \cite{Gauthier-Lafaye1996}. The earliest information we have collected so far comes from the cosmic microwave background (CMB). Earlier than the CMB time-like slice is the cosmic neutrino background. We observe Big Bang nucleosynthesis (BBN) indirectly, through observations of the abundances of chemical elements.}
\label{GR_cone}
\end{figure}

Why, then, are we still testing General Relativity? Why do we still develop, discuss, test and fine-tune alternative theories?
Because there are some very fundamental open questions in physics, particularly in cosmology.
Moreover, the big questions in cosmology happen to be the ones that are not answered by General Relativity: the accelerated expansion of the universe, the presence of a mysterious form of matter which cannot be observed directly, and the initial conditions in the \mbox{early universe. }

The theoretical completeness described above is both a necessary and aesthetic feature of a fundamental theory. However, it creates experimental difficulties, for it compels us to test the theory at extreme scales, where experimental errors may be large enough to allow several alternative~ theories.

At extremely small scales below the Planck length, classical mechanics should break down. \mbox{This compels} us to question whether General Relativity is still accurate at these scales, whether it needs to be modified, and whether a quantum description of gravity can be formulated. At the other end of the scale, at cosmological distances, we may question whether General Relativity is valid, given that the universe cannot be modelled sufficiently accurately by General Relativity without invoking either a cosmological constant, or some additional, unknown component of the universe. Finally, we may question the accuracy of our solutions to the equations of General Relativity, which depend on some approximation scheme. These approximations provided analytical solutions which enabled most of the early progress in General Relativistic cosmology and astrophysics. However, one century after the formulation of the theory, we now have a flood of data from increasingly accurate observations (as shown in Figure \ref{GR_cone}), coupled with computing power which was hitherto unheard of. Tests of the higher-order effects predicted by General Relativity and some of its competitors are now within reach.

The purpose of this review is to examine the motivation for the development of alternative theories throughout the history of GR,
to give an overview of the state of the art in General Relativistic cosmology, and to look ahead. In the next few decades, some of the open questions in cosmology may well be answered by a new generation of experiments, and GR may be challenged by \mbox{alternative theories.}

\section{A Brief History}

Let us start this review by breaking our own rule about unscientific adjectives.
General Relativity is a beautiful theory of gravity. It has not only thrilled us, but has survived 100 years of challenges, both by experimental tests and by alternative theories.
The beauty of the theory was clear at the beginning, but the initial focus was on whether it was right.
When General Relativity provided an explanation for the 43 seconds of arc per century discrepancy in the advance of the perihelion of Mercury \cite{Ein15}, it got the attention of the scientific community. However, it was the prediction and the observation of the bending of light by the Sun \cite{Dyson1920} that confirmed GR's place as the new reigning theory of gravity \cite{Will2015}.

The setting at the Royal Society under the portrait of Newton for the report of the eclipse light bending observations led by Arthur Eddington, and reported by the great writer Aldous Huxley, was perfect to describe to the world the ascendancy of a new theory replacing Newton's gravity \mbox{(see, e.g., \cite[][]{Coles2001}).}
From this point onwards, the scientific community started to take General Relativity seriously, and theorists worked hard to understand this new theory, beguilingly simple but hard to apply, and to advance its predictions.

Shortly after its publication, GR quickly became the framework for astrophysics, and for the Standard or Concordance Model of cosmology. However, it was still challenged by alternative theories. Initially, the alternatives were motivated by theoretical considerations.
This early period led to a fuller understanding of GR and its predictions. Some of the predictions, such as black holes and gravitational waves, divided the scientific community. Did they exist as physical objects, or just as mathematical artifacts of the theory?

By the time GR turned 50, the model of cosmology had been established, GR had been tested, and things had started to stagnate. However, advances in observations led to new discoveries, which in turn led to renewed challenges.

First came the missing mass in the universe. Could GR be modified to account for it? Then came the theories about the very early universe, and the behaviour of the quantum-scale, tiny initial universe. Finally, twenty years ago, came the confirmation of cosmic acceleration. This had a twofold effect. On one hand, it spurred the development of a whole range of alternative theories of gravity. On~the other hand, it confirmed GR like never before, for General Relativity, with a cosmological constant, can~account for the observations perfectly.

In 2015, on the 100th birthday of General Relativity, gravitational waves were observed for the first time. This had been the last major untested prediction of General Relativity. It was a remarkable achievement, and in many ways it heralds a new age of astrophysical observations. The experimental capabilities and the computing power have finally caught up with the theory. Cosmology and astrophysics have now entered the era of Big Data, and much of the theoretical effort is now driven by data. However, the foundation for almost the entire scientific endeavour is still this theory of chronogeometrodynamics, developed 100 years ago when today's instruments and computers were still a distant dream.

\vspace{6pt}
\noindent\emph{From Aristotle to Einstein}
\vspace{6pt}

General Relativity is the basis for the Standard Model of physical cosmology, and here we shall discuss the development of General Relativity (GR). The history of cosmology and GR are intertwined.  We shall discuss why the theory has been so successful, and the criteria that must be satisfied by any alternative theory of physics, and by cosmological models.

Cosmology, in its broadest definition, is the study of the cosmos. It aims to provide an accurate description of the universe. Throughout much of the history of science, the development of cosmology was hampered by the lack of a universal physical theory. Observational tools were extremely limited, and there was no mathematical formulation for physical laws. The cosmos was described in metaphysical, rather than physical terms.

Discussions on the history of physics often refer to Karl Popper's concept of `Falsifizierbarkeit' (falsifiability) \cite{Popper}. In this formalism, scientific discovery proceeds by successive falsifications of theories. A falsifiable theory that covers observations, and that has not yet been proven false can be regarded as provisionally acceptable. Yet we know that in reality it is not quite as straightforward. \mbox{A theory} that is considered to be correct acquires this status by accumulation of evidence rather than by a single falsification of a previous theory \cite{Lahav2014}. This is especially true in cosmology, where the selection of theoretical models often depends on the outcome of statistical calculations.

The scientific revolution which brought about the development of  a precise mathematical language for physical theories heralded the scientific age of cosmology. Physical laws, tested here on Earth and later in the Solar System, could be applied to the `entire universe', and could thus provide a precise physical description of the cosmos. Modern cosmology is based upon this epistemological framework. Cosmology depends upon a fundamental premise. As a science, it must deal strictly with what can be observed, but the observable universe forms only a fraction of the whole cosmos. One is forced to make the fundamental but unverifiable assumption that the portion of the universe which can be observed is representative of the whole, and that the laws of physics are the same throughout the whole universe \cite{Ellis:1975aa}. Once we make this assumption, we can construct a model of the universe based on a description of its observable part.

Any cosmological model which assumes the universality of physical laws must be based upon some physical theory. Since cosmology aims to describe the universe on the largest possible scales, \mbox{it must} be based upon an application long-range physical interactions. Since the theory of gravitation is the physical theory at the basis of standard cosmology, and is also at the centre of the big questions facing modern cosmology, we shall give an overview of the development of theories of gravitation.

The development of physical theories of gravity was far from smooth, nor did it always conform to Popper's scheme. Before the logical tools (mathematics) for the phenomenological description (physics) were invented, progress was rather haphazard.

According to Popper's scheme, this development should be driven by the search for ever more general principles. Yet Aristotelian theory, to take one example, considered itself to be general enough---its claimed region of validity was the entire universe, except that rising smoke, floating feathers, falling apples and orbiting celestial spheres each had their own rules.

The real revolution came when it was realised that the behaviour of all bodies could be described by a single rule---a universal theory of gravitation.  This theory is a description of the long range forces that electrically neutral bodies exert on one another because of their matter content.

Whether they choose to or not, scientists \scalebox{.95}[1.0]{will always stand on the shoulders of giants. No theory} is invented in a scientific vacuum.
This goes all the way back to the cosmology of the Euro-Mediterranean Ancient World, codified in the Aristotelian teachings of the 4th century B.C.
\mbox{This Hellenic} ``natural philosophy'' provided qualitative rather than
quantitative descriptions for what we would call today the free parameters of the theory \cite{Crombie1961}.
It stands to reason---the instruments had not yet been invented that could test the theory of gravity to within numerical accuracy.
Without accurate timekeeping instruments, processes could at best be described as ``slower than'' or ``faster than''.
However, instruments to measure the movement of the celestial bodies, such as sundials, quadrants and astrolabes, were invented and improved upon, and measurements were carried \mbox{out \cite{Bennett2011}.} Astronomy~flourished.

There is a certain logic to
the development of physical theories from the Ancient World, to the Middle
Ages, and right up to the Renaissance \cite{Frank76,Sorab88,Cushing1998}. The basic tenet of the physics of Aristotle is that actions follow logically from causes. He distinguished between natural and violent motion. Natural motion implies falling at a speed proportional to the weight of the
object and inversely proportional to the density of the medium. Violent motion happens whenever there is a force acting on an object, and the speed of the object is
proportional to this force.
Strato of Lampsacus replaced Aristotle's explanation of `unnatural' motion with one that is very close to the modern notion of inertia. \mbox{He identified} natural motion as a form of acceleration, and demonstrated experimentally that falling bodies accelerate. In the 14th century, Jean Buridan came up with the notion of impetus, where the initial force imparts motion to the object, which gradually diminishes as gravity and air resistance act against this initial force. Concurrently, Nicole Oresme was using a crude early form of graph to describe motion, and unwittingly showing the complicated notions of differentiation and integration in pictorial form\cite{Grant1978,Babb2005}.

The cosmological observations, limited to the innermost five planets of the Solar System (Mercury, Venus, Mars, Jupiter, and Saturn) and the sphere of stars, seemed to confirm the Aristotelian-Ptolemaic theory. Celestial bodies moved in regular patterns made up of repeating circles. Small discrepancies were explained by circles within circles.

The fact that the theories were based on these regular patterns is no accident. Patterns are the keyword in all of physics. Human beings are wired to recognise patterns. We can only build theories because we recognise patterns in the universe. This characteristic of valid theories has been called sloppiness. The patterns fall within some hyper-ribbon of stability in the theory \cite{Transtrum2015}.

The revolution in physics came with the development of mathematical, quantitative,
models to describe physical reality. Starting in the 1580 Galileo carried out a series of observations in which he subjected kinematics to rigorous experiment, and showed that naturally-falling objects really do accelerate. Crucially, he showed that the composition of the body has no effect whatsoever on this acceleration. He also realised that for violent motion, the speed is constant in the absence of friction. Galileo also took rigorous observations of astronomical objects. In 1610 he made the first observation of Jupiter's satellites, and the first observation of the phases of Venus, which is impossible according to the Ptolemaic geocentric model. His observations were important in putting to rest the Aristotelian theory of perfect and unchanging heavens.

By the time Newton came along, telescopes had been invented. Galileo had observed moons orbiting the Solar System planets, and hundreds of stars invisible to the naked eye. His 1610 treatise, aptly called \textit{Siderus Nuncius} (``Starry Message'', or ``Astronomical Report'' in modern language) \cite{Galileo}, was the first scientific work based on observations through a telescope. Mechanical clocks had been invented. The sphere of observed data had expanded \cite{Perryman2012}. Calculus provided the tool to make sense of this new flood of data.  Thus, physicists of Newton's
generation found a very different scientific environment than the
one in which Galileo had started off.

In 1687, Isaac Newton published in his ``Mathematical Principles of Natural Philosophy'', known by its abbreviated Latin title as
\textit{Principia} \cite{Newt1687}. This was a significant milestone in physics. Newton's model of gravitation was, in his own words, a ``universal'' law. It applied to all bodies in the universe, whether it was cannonballs on Earth, or planets orbiting the Sun.
For more than two centuries, Newton's theory, was the
standard physical description of gravity.
There was no other attempt to find a different theory for the gravitational force,
although the intervening years between Newtonian gravity and Relativity produced some important physical concepts such as de Maupertuis's ``Principle
of Least Action'' \cite{dMaup1746}, further developed by Euler \cite{Euler1750}, Lagrange \cite{Lagrange1788} and Hamilton \cite{Ham1834,Ham1835}. \mbox{The path} of each particle is assigned a number called an action, which is the integral of the Lagrangian.
\mbox{In classical mechanics,} the action principle is equivalent to Newton's Laws. Lagrangian field theory is an important cornerstone of modern physics. The Lagrangian of any physical interaction, when subjected to an action principle, give us field equations and conservation laws for the theory. It is an expression of the symmetries in physical laws.

Newtonian gravity was the great success story of nineteenth century physics, the golden age of mathematical astronomy.
It allowed astronomers to calculate the position of planets and asteroids with ever greater precision, and to confirm their calculations by observation. Thus the size of the known universe grew. Evidence started to accumulate suggesting that there might be other galaxies in the universe besides our own.
In 1845, the planet Neptune was discovered, after Urbain le Verrier suggested pointing telescopes in a region of the Solar System which he predicted by Newtonian calculations \cite{Galle1846,Danjon1946}. The search was motivated in the first place by an anomaly in the orbit of Uranus which could not be otherwise explained
using Newtonian theory \cite{Uranus}.
The discovery of Neptune showed that Newtonian theory was valid even in the very farthest limits of the Solar System.

There was another anomaly which could not be explained---the excessive perihelion precession of Mercury by $43$ arcseconds per century, confirmed by le Verrier himself.
Urbain le Verrier thus holds the distinction of being one of the few experimentalists to have confirmed Newton's theory and then disproved it.
Astronomers attempted to explain this perihelion anomaly using Newtonian mechanics, which led them to speculate on the existence of Vulcan, a hypothetical planet whose orbit was even closer to the Sun \cite{Sheehan2016}.

The first doubts on Newtonian theory began to take shape just at the time when theorists were examining the full implications of the theory for complex, multi-body dynamical systems such as the Solar System. In 1890, Henri Poincar\'e published his magnum opus on the three-body problem \cite{Poincare1890}, a~masterpiece of celestial mechanics. At the time, Poincar\'e was working on another open question in physics: the aether. This led him to formulate a theory which was very close to Special Relativity \cite{Poincare89}, but which did not quite fit with Maxwell's electromagnetism \cite{Max1865}, and was ultimately flawed.

By the end of the 19th century, the necessary mathematical tools were in place which would enable the development of Special and then General Relativity. There is an intimate connection between physics and the language of mathematics which is often overlooked. The former, especially in modern times, depends on the latter. Could Aristotle have developed General Relativity? No. Because he had not the mathematical language. Equations and mathematical formulations are relatively recent in the history of physics. Even Newton, for all his fame as a mathematical genius, never wrote the equation $F=-GMm/r^{2}$. He wrote a series of statements implying this law in (Latin) words: ``\textit{Gravitatem, qu{\ae} Planetam unumquemque respicit, ese
reciproc{\ae} ut quadratum distanti{\ae} locorum ab ipsius centro''}, and so on. It is hard to imagine how human beings could manipulate tensors and solve the field equations of Relativity in anything but numbers and symbols. Theories and physics do not happen in a cultural and scientific vacuum. They are human creations, and they depend intimately on tools for the transmission and communication of human knowledge.

The physical theory of gravity---the laws that govern gravitational interactions---remained unchanged until Einstein's time. In 1905, Einstein published his Theory
of Special Relativity \mbox{(SR) \cite{Ein05}}. Soon after, he turned to the problem of including gravitation within four-dimensional \mbox{spacetime \cite{Ein11,Ein12,Ein12b,Ein13}.}

Newton's formulation of the gravitational laws is expressed by the equations:
\begin{align}
\frac{\dd^2x^i}{\dd t^2}&=-\frac{\partial\varPhi}{\partial
x^i}\label{Newt1},\\ \Laplace\varPhi&=4\pi G\rho
\label{Newt2}\,,
\end{align}
where $\varPhi$ is the gravitational
potential, $G$ is the universal gravitational constant, $\rho$
is the mass density, and $\Laplace = \nabla^{2}$ is the Laplace operator. These equations cannot be incorporated into
Special Relativity as they stand. The equation of motion
(\ref{Newt1}) for a particle is in three-dimensional form, so it
must be modified into a four-dimensional vector equation for
$\dd ^2x^\mu / \dd \tau^2$. Similarly, the field Equation
(\ref{Newt2}) is not Lorentz-invariant, since the three-dimensional
Laplacian operator instead of the four-dimensional d'Alembertian ${\DAlambert = \partial^{\mu}\partial_{\mu}}$
means that the gravitational potential $\varPhi$ responds
instantaneously to changes in the density $\rho$ at arbitrarily
large distances. The conclusion is that Newtonian gravitational fields propagate with infinite velocity. In other words, instantaneous action in Newtonian theory implies action at a distance when reconsidered in the light of Special Relativity. This violates one of the postulates of SR. How do we reconcile gravity and Special Relativity?

\section{The Development of General Relativity}\unskip

\subsection{From Special to General Relativity}

The simplest relativistic generalisation of Newtonian gravity is obtained by representing the
gravitational field by a scalar $\varPhi$.
Since matter is described in Relativity by the stress-energy tensor $T_{\mu\nu}$, the only
scalar with dimensions of mass density (which corresponds to $\rho$) is $T^\mu_\mu$.
A consistent scalar relativistic theory of gravity
would thus have the field equation
\bq
\DAlambert\varPhi=4\pi
GT^\mu_\mu\, .
\eq

However, when the equation of motion from this theory are applied to a static, spherically symmetric field $\varPhi$, such as
that of the sun, acting on an orbiting planet, they would result in a negative precession, or retardation of the perihelion.
Experimental evidence since the time of \mbox{Le Verrier} and his observation of the orbit of Mercury \cite{LeVerrier1859} clearly shows that planets experience a prograde precession of the perihelion.
Moreover, in the limit of a zero rest-mass particle, such as a
photon, the equations of motion show that the particle experiences
no geodesic deviation. The existence of an energy-momentum tensor
due to an electromagnetic field would also be impossible, since
$(T_{\text{electromagnetic}})^\mu_\mu=0$. The theory therefore
allows neither gravitational redshift, nor deviation of light by
matter, both of which are clearly observable phenomena \cite{Will01}.
Another route to generalisation could be to represent the gravitational field by a vector field $\varPhi_\mu$, analogous to
electromagnetism. Following through with this strategy, the ``Coulomb'' law in this theory gives a repulsion between two massive particles,
which clearly contradicts observations. The theory also predicts that gravitational waves should carry negative energy, and, like the
scalar theory, predicts no deviation of light. Like the scalar theory, then, the vector theory must be discarded.

What about a flat-space tensor theory? The gravitational field in this theory is described by a symmetric tensor
$h_{\mu\nu}=h_{\nu\mu}$. The choice of the Lagrangian in this theory
is dictated by the requirement that $h_{\mu\nu}$ be a
Lorentz-covariant, massless, spin-two field.

In the 1930s, Wolfgang Pauli and Markus Fierz \cite{FierzPauli39} were the first to write down this Lagrangian and investigate the
resulting theory. The predictions of the theory for deviation of light agree with those of General Relativity, and are consistent
with observations. Since the field equations and gauge properties
are identical to those of Einsein's linearised theory, the
predictions for the properties of gravitational waves, including
positive energy, agree with those obtained using the linearised
theory in General Relativity.
However, the theory differs from
General Relativity in its predicted value for the perihelion
precession, which is $\frac{4}{3}$ of that given by GR. This disagrees with the value obtained from observations of Mercury's
orbit.

The theory has an even worse
deficiency. If two gravitating bodies (that is, not test particles)
are considered, and the field equations are applied to them, then
the theory predicts that gravitating bodies cannot be affected by
gravity, since they all move along straight lines in a global
Lorentz reference frame. This holds for bodies made of arbitrary
stress-energy, and since all bodies gravitate, then one must
conclude that no body can be accelerated by gravity, which is a
obvious self-inconsistency in the~theory.

The only way in which a consistent theory of gravity can be constructed within Special
\mbox{Relativity is to} consider the geometry of spacetime as the
gravitational field itself. In other words, \emph{all matter}
moves in an effective Riemann space of metric $g^{\mu\nu}\equiv
\eta^{\mu\nu}+h^{\mu\nu}$, where $\eta^{\mu\nu}$ is the Minkowski
metric.
The requirement of consistency leads us to universal coupling, which implies the \mbox{Equivalence
Principle.}

The existence of curved spacetime can be deduced from purely physical arguments.
In 1911, before he had fully developed General Relativity, Einstein \cite{Ein11} showed that a photon must be
affected by a gravitational field, using conservation of energy
applied to Newtonian gravitation theory. \mbox{Schild~\cite{Schild60, Schild62,
Schild67}} showed by a simple thought experiment, formulated within
Special Relativity, \mbox{that a consistent} theory of gravity cannot be
constructed within this framework. His argument is based upon a
gravitational redshift experiment carried out in the field of the
Earth, using a global Lorentz frame tied to the Earth's centre.
Successive pulses of light rising to the same height should experience a redshift, and therefore the pulse rate at the top
should be slower than that at the bottom.
But light rays are drawn at 45 degrees in Minkowski spacetime diagrams, so that top and
bottom time intervals are equal, which is impossible if redshift
occurs. Hence the spacetime must be curved.
One therefore concludes that in the presence of gravity, Special Relativity cannot be valid over any sufficiently extended region.

General Relativity may be understood as a generalisation of Special Relativity over extended regions. Since Special Relativity can
comfortably be described using tensor calculus, it was only natural
to extend the flat Minkowski spacetime of Special Relativity to the
curved spacetime of General Relativity.
This was a physical application of Riemannian
geometry \cite{Riemann_collected,Riemann_posthumous1868}, which had been developed in the
second half of the 19th century. The idea of tensor calculus on
curved manifolds was already mathematically well-established.
Einstein's innovation lay in identifying the Einstein tensor, itself related to the Riemann curvature tensor, as the
``gravitational field'' in the theory.

Einstein had been working on the problem for some years, starting in 1907. He arrived at the final, correct form in 1915 \cite{Ein15a,
Ein15b}. He was well-aware of the significance of his publication, \mbox{and he} gave it the succinct title of ``The Field Equations of Gravitation'' (\textit{Feldgleichungen der Gravitation}). \mbox{The correct} field equations for the theory contained in this publication served as the starting point or subsequent derivations.

\subsection{The Formalism of General Relativity}

General Relativity is based on two independent but mutually supporting postulates.

The first postulate is sometimes referred to collectively as the \textit{Einstein Equivalence Principle}:
\begin{itemize}[leftmargin=*,labelsep=4mm]
\item \emph{The Strong Equivalence Principle:} The laws of physics take the same form in a freely-falling reference frame as in Special Relativity
\item \emph{The Weak Equivalence Principle:} An observer in freefall should experience no gravitational field. That is to say, an observer cannot determine from a local experiment whether the his laboratory is being accelerated by a rocket of static at the surface of a gravitating body. Gravity is erased up to tidal forces, which are determined by the size of the laboratory and its distance to the centre of the gravitational attraction.
\end{itemize}

The Equivalence Principle allows us to construct the metric and the equation of motion by transforming from a freely-falling to an accelerating frame. It can be mathematically expressed by the assuming that all matter fields are minimally coupled to a single metric tensor $g_{\mu\nu}$. The distance between two points in 4-dimensional spacetime, called events, is:
\bq \label{metric1}
\dd s^2=c^2\dd\tau^2=g_{\mu\nu}\dd x^\mu\dd x^\nu \, .
\eq
Throughout the text, we follow the Einstein summation convention for repeated indices, so that $c_{i}x^{i}=\sum\limits_{i=1}^{n}c_{i}x^{i}$ for $i=1,\dots , n$. Greek indices are used for space and time components, while Latin indices are spatial ones only. We use the following metric signature: $(-+++)$.

The metric defines lengths and times measured by laboratory rods and clocks. This metric implies that the action for any matter field $\psi$ is of the form
\bq
S_{\mr{matter}}[\psi,g_{\mu\nu}] \, \label{metriccoupling},
\eq
which gives us three important results. First, it implies the universality of freefall. Second, it implies that all non-gravitational constants are spacetime independent. Third, it implies that the laws of physics are isotropic. This equation defines how matter behaves in a given curved geometry, how light rays propagate, how stars, planets and galaxies move, and gives us verifiable \mbox{observational consequences.}

The second postulate is related to the dynamics of the gravitational interaction. This is assumed to be governed by the Einstein-Hilbert action:
\bq
S_{\mr{gravity}}=\frac{c^{3}}{16\pi G}\int \dd^{4}x\sqrt{-g_{*}R_{*}} \, \label{EHaction}
\eq
where $g^{*}_{\mu\nu}$ is a massless spin-2 field called the Einstein metric. General Relativity identifies the Einstein metric with the physical metric, that is: {$g_{\mu\nu}=g^{*}_{\mu\nu}$}. This implements the Strong \mbox{Equivalence Principle. }

The Einstein-Hilbert action defines the dynamics of gravity itself. Relativity is thus a geometrical approach to fundamental interactions.
These are realised though continuous classical fields which are inseparably connected to the geometrical structures of spacetime, such as the metric, affine
connection, and curvature.

The General Relativistic equation of motion is simply parallel transport on curved spacetime. \mbox{It is given} by
\bq
\frac{\dd^2x^\mu}{\dd\tau^2}+\Gamma^\mu_{\alpha\beta}\frac{\dd x^\alpha}{\dd\tau}\frac{\dd x^\beta}{\dd\tau}=0 \, ,\label{1.2}
 \eq
  where $x^\mu$ is some set of coordinates for a point in spacetime.
  $\Gamma^\mu_{\alpha\beta}$ are the components of the affine connection (or metric connection).
 The fundamental theorem of Riemannian geometry states that the affine connection can be expressed entirely in terms of the metric:
 \bq
 \Gamma^\alpha_{\lambda\nu} =\frac{1}{2}g^{\alpha\nu}(g_{\mu\nu,\lambda}+g_{\lambda\nu,\mu}-g_{\mu\lambda,\nu}) \, ,
 \eq
  where the comma denotes a derivative, i.e., $g_{\mu\nu,\lambda}=\tfrac{\partial g_{\mu\nu}}{\partial x^\lambda}$.

We need to construct invariant quantities in GR (quantities that are the same for all observers).
To~achieve this, we need to contract the covariant $A_\mu$ and contravariant $A^\mu$ components of a vector or tensor $A$ by using the metric to raise or lower indices: $A_\mu=g_{\mu\nu}A^\mu$.
Thus the equation of motion (\ref{1.2}) can be made covariant by recasting it as the covariant derivative of the 4-velocity $U^\mu=\gamma(c,\mathbf{v})$:
\bq
\frac{D_\mu U^\mu}{\dd \tau}=0 \, ,
\eq
where the covariant derivative is defined as
\bq
D_\mu A^\mu=\dd A^\mu + \Gamma^\mu _{\alpha\beta}A^\alpha \dd x^\beta \, .
\eq
The quantity $\gamma$ is the Lorentz factor:
\bq
\gamma = \frac{1}{\sqrt{1-v^{2}/c^{2}}} \, .
\eq

The transformation from SR to GR is then carried out by mapping the Minkowski metric to a general metric: $\eta \rightarrow g$ and by mapping $\partial \rightarrow D$.

In GR, freely-falling bodies travel along a geodesic. Geometrically, this is the shortest distance between two points in spacetime. The path length along a geodesic is given by
 \bq \label{geodesic}
 S=\int(g_{\mu\nu}\dd x^\mu \dd x^\nu)^{1/2}\, .
 \eq

In cosmology, it is essential for us to be able to describe spacetime which is not ``empty''. In the presence of a perfect fluid (an inviscid fluid with density $\rho$ and isotropic pressure $p$), the energy and momentum of spacetime is described by the energy-momentum tensor (or stress-energy tensor)
\bq
T^{\mu\nu}=\left(\rho+\frac{p}{c^2}\right)U^\mu U^\nu - pg^{\mu\nu}\, .
\eq
\scalebox{.95}[1.0]{Classical energy and momentum conservation are generalized in GR as the four conservation laws}
\bq D_\mu T^{\mu\nu}=0\,.\eq
In other words, the stress-energy tensor has a vanishing covariant divergence. In the absence of a component possessing pressure or density, or both, the energy-momentum tensor is zero.

The central notion in General Relativity is that gravitation can be described by a metric.
The~ Einstein equations give us the relation between the metric and the matter and energy in the~ universe:
\bq
G^{\mu\nu}=-\frac{8\pi G}{c^4} T^{\mu\nu} \label{GR_field_eqn}.
\eq
\scalebox{.95}[1.0]{The left-hand side of this equation is a function of the metric: $G^{\mu\nu}$ is the Einstein tensor, defined as: }
\bq
G^{\mu\nu}=R^{\mu\nu}-\frac{1}{2}g^{\mu\nu}R \, ,
\eq
where $R^{\mu\nu}$ is the Ricci tensor, which depends on the metric and its derivatives, and the Ricci scalar $R$ is the contraction of the Ricci tensor ($R=g_{\mu\nu}R^{\mu\nu}$).
The right-hand side of Equation (\ref{GR_field_eqn}) is a function of the energy: $G$ is Newton's constant, and $T^{\mu\nu}$ is the energy-momentum tensor.

Einstein's Relativity has three main distinguishing characteristics:
\begin{itemize}[leftmargin=2em,labelsep=4mm]
\item	it agrees with experiment
\item	it describes gravity entirely in terms of
geometry
\item	 it is free of any ``prior geometry''
\end{itemize}

These characteristics are lacking in most of the other theories \cite{Ni:1972, ThNiWi71}.
Apart from the issue of agreement with experiment,
Einstein's theory is unique in its physical simplicity.

Every other theory introduces auxiliary gravitational fields, or involves prior geometry.
Prior~geometry is any aspect of the geometry of spacetime which is
fixed immutably, that is, it~cannot be changed by changing the
distribution of gravitating sources.

A rigorous mathematical definition of the unique simplicity of General Relativity is given by Lovelock's theorem
\cite{Lovelock1969,Lovelock1971,Lovelock1972}.
This is a generalisation of an earlier theorem by \'Elie Cartan \cite{Cartan1922a}, and may be formulated as follows:
\begin{quote}\emph{In 4 spacetime dimensions, the only divergence-free symmetric rank-2
tensor constructed solely from the metric $g$ and its derivatives up to second
differential order, and preserving diffeomorphism invariance, is the Einstein
tensor plus a cosmological term.}\end{quote}

In simple terms, the theorem states that GR emerges as the unique theory of gravity if the conditions of the theorem are followed.
In fact, Lovelock's theorem provides a useful scheme for classifying alternatives to General Relativity.

Einstein described both the demand for ``no prior geometry'' and for a ``geometric, coordinate-independent formulation of physics'' by the
single phrase ``general covariance'', but the two concepts are not quite the same.

While many physical theories can be formulated in a generally covariant way, General Relativity is actually based on the ``no prior geometry''
demand.
This distinction was not always made, especially in the
first decades after Einstein's publications \cite{Anders67,Norton2001}.
Erich Kretschmann's famous objection \mbox{in 1917 \cite{Kretsch17}} concerned this point, since he regarded general covariance merely as formal feature that any theory could have, not as a special feature belonging to GR.

\subsection{Newtonian Nostalgia: The First Wave of Alternative Theories}

Newton's theories had predicted observations of Solar System objects, comets and asteroids, with astounding precision.
Why should they be tampered with? The first wave of alternative theories were driven more by theoretical considerations
than by observations.
\mbox{Equations (\ref{Newt1}) and (\ref{Newt2})} can be generalized so
that they are consistent with the postulates of Special and General Relativity.
Several~ generalisations of this kind were attempted in the first few decades following the development of GR,
motivated by lingering resistance to any deviation from Newtonian gravity.

One early theory, involving prior geometry, was formulated by Nordstr{\o}m in 1913 \cite{Nord13}.
In this theory, the physical metric of spacetime ${g}$ is generated by a background
flat spacetime metric ${\eta}$, and by a scalar gravitational field $\phi$.
Stress-energy generates $\phi$:
\bq
\eta^{\alpha\beta}\phi_{,\alpha\beta}=-4\pi\phi\eta^{\alpha\beta}T_{\alpha\beta}
\eq
and ${g}$ is constructed from $\phi$ and ${\eta}$:
\bq g_{\alpha\beta}=\phi^2\eta_{\alpha\beta}.\eq

Prior geometry cannot
be removed by rewriting Nordstr{\o}m's equations in a form devoid of
${\eta}$ and $\phi$ \cite{EinFok14}.
Mass only influences one
degree of freedom in the spacetime geometry, while the other degrees
of freedom are fixed \textit{a priori}. This prior geometry, if it
existed, could be detected by \mbox{physical experiments.}

In the 1920s, Alfred North Whitehead \cite{White22}
formulated a two-tensor theory of gravity in which the prior
geometry is quite different from later theories such as Ni's \cite{Ni:1972}. Whitehead's
theory is remarkable in that it agrees with Einstein's in its
predictions for the four standard tests (bending of light,
gravitational redshift, perihelion shift, and time delay). It was
accepted as a viable alternative for Einstein's theory until Clifford Martin Will
\cite{Will71} showed that it predicts velocity-independent
anisotropies in the Cavendish constant (the gravitational constant
$G$ in Newtonian theory). This~would produce time-dependent Earth
tides which are clearly contradicted by everyday observations.
Any~valid theory of gravity must not only agree with relativistic
experiments, but also with past experiments in the Newtonian
regime.

One theory which disagrees violently with non-relativistic
experiments is due to George David Birkhoff \cite{Birk43}. It was developed in the 1940s, and it predicts the same redshift,
perihelion shift, deflection and time-delay as General Relativity, but it requires that the pressure inside gravitating bodies should
be equal to the total density of mass-energy ($p=\rho$). This means that
sound waves travel with the speed of light. This clearly contradicts
everyday experiments.

Most of the early alternative theories were abandoned either because they were contradicted by observations, or because of internal inconsistencies in the theories themselves.
One notable exception is Dicke-Brans-Jordan theory, sometimes called Brans-Dicke, or
Jordan-Fierz-Brans-Dicke \mbox{theory \cite{Jord59,BranDick61},} developed in the 1960s by Robert H. Dicke and Carl H. Brans following earlier work by Pascual Jordan and Markus Fierz.
The different names arise from the fact that the theory is a special
case of Jordan's, with $\eta=-1$.
An alternative mathematical representation of the theory is given \mbox{by \cite{Dicke62}. }

This theory introduced auxiliary gravitational fields. Brans and Dicke took the equivalence principle as the starting point of their
theory, and thus they describe gravity in terms of spacetime
curvature, but their gravitational field, unlike Einstein's, is a
scalar-tensor combination. In this way it overcomes the difficulties
associated with tensor or scalar-only theories mentioned earlier.
The trace of the energy-momentum tensor $(T_\mr{M})_{\mu\nu}$
(representing matter) and a coupling constant $\lambda$ generate the
long-range scalar field $\phi$ via the equation
\bq
 \Box^2\phi=4\pi\lambda(T_\mr{M})^\mu_\mu.\eq

 The scalar field
$\phi$ fixes the value of $G$, which is therefore not a constant,
but simply the coupling strength of matter to gravity. The
gravitational field equations relate the curvature to the
energy-momentum tensors of the scalar field and matter:
\bq R_{\mu\nu}-\tfrac{1}{2}g_{\mu\nu}R=-\frac{8\pi}{c^4\phi}\left[(T_\mr{M})_{\mu\nu}+(T_\Phi)_{\mu\nu}\right],\eq
where $(T_\mr{M})_{\mu\nu}$ is the energy-momentum tensor of matter
and $(T_\Phi)_{\mu\nu}$ is the energy-momentum tensor of the scalar
field $\phi$. For historical reasons, it is usual to write the
coupling constant as
 \bq \lambda=\frac{2}{3+2\omega} \, ,
 \eq
 where $\omega$ is the dimensionless `Dicke coupling constant'. In the
limit $\omega\rightarrow\infty$, we have $\lambda\rightarrow 0$, so
$\phi$ is not affected by the matter distribution, and can be set to
a constant $\phi=1/G$. Hence Dicke-Brans-Jordan theory reduces to
Einstein's theory in the limit~$\omega\rightarrow\infty$.

The equivalence principle is satisfied in this theory since the
special-relativistic laws are valid in the local Lorentz frames of
the metric $g$ of spacetime. The scalar field does not exert any
direct influence on matter. It only enters the field equations that
determine the geometry of spacetime. On a conceptual level,
Brans-Dicke theory can be seen as more fully Machian than Einstein's
theory.
Einstein himself attempted to incorporate Mach's Principle into his theory, but in Einstein's General Relativity, the inertial
mass of an object will always be independent of the mass distribution in the universe.
In~Brans-Dicke theory, the long-range scalar field is an indirectly coupling field, so it does not
directly influence matter, but the Einstein tensor is determined partly by the energy-momentum tensor, and partly by the long-range
scalar field.

Dicke-Brans-Jordan theory is self-consistent and complete, but
experimental evidence based on Solar System tests, shows that $\omega\geq 600$ \cite{Will93}, as a
conservative estimate. Some calculations raise this limit even higher, with $\omega\gtrsim 10^4$
\cite{Psaltis05}. The Cassini mission set a comparable limit of $\omega > 40,000$ \cite{Bertotti2003}. Recent cosmological data from the \textit{Planck} probe show that $\omega\geq 890$ \cite{Avilez2014,Ooba2016}. This is consistent with the Solar System bounds. Future cosmological experiments and data from large-scale structure could provide even better constraints \cite{Bull2016}.

Brans-Dicke theory is a special case of general scalar-tensor theories with $\omega(\phi)=\text{constant}$, where $\phi$ is a
value depending on the cosmological epoch.
In these theories, the function $\omega(\phi)$ could be such that the theory is very
different from GR in the early universe or in future epochs, but
very close to GR in the present.
In fact, it has been shown that GR is a natural attractor for such scalar-tensor theories, since cosmological evolution naturally drives the fields towards large values of $\omega$ \cite{DamNord93a, DamNord93b}.

\subsection{Self-Consistency, Completeness, and Agreement with Experiment}

Any viable theory must satisfy three fundamental criteria: self-consistency, completeness,
and~agreement with past experiment.

To be self-consistent, a theory must not contain any internal contradictions.
The spin-two field theory of gravity \cite{FierzPauli39} is equivalent to linearised
General Relativity but it is internally inconsistent since it predicts that gravitating bodies should have their motion unaffected
by gravity.
When one tries to remedy this inconsistency, the resulting theory is nothing but General Relativity.
Another self-inconsistent theory is due to Paul Kustaainheimo \cite{Kust66,kust1967}.
It predicts zero gravitational redshift when the wave version of light (Maxwell theory) is used, and nonzero redshift when the particle version
(photon) is used.

To be complete, a theory must be able to analyse the outcome of any experiment.
This means that it must be compatible with other physical theories which describe any other forces that are present in experiments.
This can only be achieved if the theory is derived from first principles, since the theoretical postulates must be as general as possible if the theory is to cover the widest range \mbox{of phenomena.}

A viable theory must agree with past experiment, which includes experiments in the Newtonian regime, and the standard tests of
General Relativity. Its results must agree with those obtained from Newtonian theory in the weak field limit, and
with GR in relativistic situations. It also means that the theory must agree with cosmological observations.

The experimental criterion also works the other way. Any alternative to General Relativity that claims to have a smaller set of limiting
cases must be experimentally distinguishable, perhaps by future experiment.
At some point, the divergence between GR and other theories must manifest itself physically, in the form of predictions which can be verified by experiment.
This is perhaps the greatest challenge of current alternatives to GR.

\subsection{Metric Theories and Quantum Gravity}

Most theories of gravity incorporate two principles:
\emph{spacetime
possesses a metric; and that metric satisfies the equivalence
principle}.
Such theories are called metric theories. There are some exceptions.

Soon after the publication of the theory of General Relativity, it
became apparent that its formulation is incompatible with a
Quantum Mechanical description of the gravitational field. It~was
Einstein himself who pointed out that quantum effects must lead to a
modification of General Relativity~\cite{Ein16}. Back then, the
first successful applications of Quantum Mechanics to
electromagnetism were starting to give useful results.
These developments led
to the question of whether General Relativity can be quantized.

This is a difficult question. First, Einstein's field equations are
much more complicated than Maxwell's equations, and in fact are
nonlinear. The physical reason for this is that the gravitational
field is coupled to itself---the stress-energy tensor acts as the source
for spacetime curvature, which in general contributes to the stress-energy tensor.
This means that the equations seem to violate the superposition
principle, which requires the existence of a linear vector space (see, e.g., ~\cite[][]{Kumar2000,Deser2010}).
This~ is the mathematical expression of wave-particle duality---a
central tenet of Quantum Theory.

Second, to
quantize the gravitational field we would have to quantize spacetime
itself. \mbox{The physical} meaning of this is not completely clear.

Finally, there are
experimental problems. Maxwell's equations predict electromagnetic
radiation, which was first observed by Hertz \cite{Hertz}. Quantization of the
field results in being able to observe individual photons, and these
were first seen in the photoelectric effect predicted by
Einstein \cite{Ein1905b}.
Similarly, Einstein's equations for the
gravitational field predict gravitational radiation \cite{Ein16}, so there should
be, in principle, the possibility of observing individual gravitons,
which are the quanta of the field.
The direct observation of gravitational waves was finally achieved in September 2015 by the LIGO instrument \cite{GW2015}.
The detection of individual gravitons is more
difficult and is beyond the capability of current experiments.

To develop a quantum theory of General Relativity, the fundamental
interactions in GR would have to follow quantum rules. In Quantum
Theory, particle interactions are described by gauge theories, so GR
would have to follow the gauge principle. Although the gauge
principle was first recognized in electromagnetism, modern gauge
theory, formulated initially by Chen Ning Yang and Robert Mills~\cite{Y&M54b,Y&M54a}, emerged
entirely within the framework of the quantum field programme. \mbox{As more} particles were discovered after the 1940s, various
possible couplings between those elementary particles were being
proposed. It was therefore necessary to have some principle to
choose a unique form out of the many possibilities suggested. The
principle suggested by Yang and Mills in 1954 is based on the concept of
gauge invariance, and is hence called the gauge principle.

\subsection{The Gauge Approach and Non-Metric Theories}

The idea of gauge invariance, and the term itself, originated
earlier, from the following consideration due to Hermann Weyl in 1918 \cite{Wey18a,
Wey18b}. In addition to the requirement of General Relativity that
coordinate systems have to be defined only locally, so likewise the
standard of length, or scale, should only be defined locally. It is
therefore necessary to set up a separate unit of length at every
spacetime point. Weyl called such a system of unit-standards a gauge
system (analogous to the standard width, or ``gauge'', of a railway track).

The gauge
principle therefore may be formulated as follows: {If a
physical system is invariant with respect to some global (spacetime
independent) group of continuous transformations $G$, then it
remains invariant when that group is considered locally (spacetime
dependent), that is $G\mapsto G(x)$. Partial derivatives are
replaced by covariant ones, \mbox{which depend} on some new vector
field.}

In Weyl's view, a gauge system is as necessary
for describing physical events as a coordinate system. Since
physical events are independent of our choice  of descriptive
framework, \mbox{Weyl maintained} that gauge invariance, just like general
covariance, must be satisfied by any \mbox{physical theory.}

In Euclidean geometry, we know that translation
of a vector preserves its length and direction. In Riemannian
geometry, the
Christoffel connection \cite{Christoffel1869} (or affine connection) guarantees length preservation, but a
vector's orientation is path dependent. However, the angle between
two vectors, following the same path, is preserved under
translation. Weyl wondered why the remnant of planar geometry,
length preservation, persisted in Riemannian geometry. After all,
our measuring standards (rigid rods and clocks), are known only at
one point in spacetime. To measure lengths at another point, we must
carry our measuring tools along with us. Weyl maintained that only
the \emph{relative} lengths of any two vectors at the same point,
and the angle between them, are preserved under parallel transport.
The length of any single vector is arbitrary. To encode this
mathematically, Weyl made the following~substitution:
\bq
{g_{\mu\nu}(x)\mapsto\lambda(x)g_{\mu\nu}(x)\label{gaugetrans},}
\eq
where the conformal factor $\lambda(x)$ is an arbitrary, positive,
smooth function of position. Weyl required that in addition to GR's
coordinate invariance, formulae must remain invariant under the
substitution (Equation \eqref{gaugetrans}). He called this a \emph{gauge
transformation}. The scale therefore becomes a local property of the
metric.

Weyl's theory enabled him to unify gravity and electromagnetism, the
only two forces known at the time. However, Weyl's original scale
invariance was abandoned soon after it was proposed, since its
physical implications seemed to contradict experiments. In
particular, if two identical clocks $C_1$ and $C_2$ are transported
on two different paths, which both end at the same point $Q$, the
time-like vectors $l_1$ and $l_2$ given by $C_1$ and $C_2$ at $Q$
would be different in the presence of an electromagnetic field.
Therefore the two clock rates would differ. As Einstein (probably
the only expert who could keep an eye on Weyl's theory at the time)
pointed out, this concept meant that spectral lines with definite
frequencies could not exist, since the frequency of the spectral
lines of atomic clocks would depend on the atom's location, both
past and present. However, we know the atomic spectral lines to be
definite, and independent of spacetime position \cite{Goenner2004,Smiciklas2011,Peck2012,Hohensee2013}.

Despite its initial failures, Weyl's idea of a local gauge symmetry
survived, and acquired new meaning with the development of Quantum
Mechanics. According to Quantum Mechanics, interactions are realized
through quantum (that is, non-continuous) fields which underlie the
local coupling and propagation of field quanta, but which have
nothing to do with the geometry of spacetime. The question is
whether General Relativity can be formulated as a gauge theory. \mbox{This
question} has been discussed by ever since it was first posed
 \cite{Ut56,Kib61,Carmeli1977,IvSar83,SardZakh92,Jackson2001,Sardan02}.

If features of General Relativity could be recovered from a
gauging argument, then that would show that the two formulations are
not inconsistent. The first to succeed in this was Kibble \cite{Kib61},
who elaborated on an earlier, unsuccessful attempt by Utiyama \cite{Ut56}.
Kibble arrived at a set of gravitational field equations, although
not the Einstein equations, constructing a slightly more general
theory, known as ``spin-torsion'' theory. The inclusion of torsion in
Einstein's General Relativity had long been theorized. In fact the
necessary modifications to General Relativity were first suggested
by \'Elie Cartan in the 1920s \cite{Car22,Cart23,Cart24,Cart25}, who identified torsion as a possible physical
field.

The connection between torsion and
quantum spin was only made \scalebox{.95}[1.0]{later \cite{Kib61,Wey50,Sci64}, once it}
became clear that the stress-energy \scalebox{.95}[1.0]{tensor for a massive fermion
field must be asymmetric \cite{Weys47,CdB63}}. The~ Einstein-Cartan
(1920s) and {the Kibble-Sciama (late 1950s) developments occurred
independently.} For historical reasons, spin-torsion theories are
sometimes referred to as Einstein-Cartan-Kibble-Sciama (ECKS)
theories, but Einstein-Cartan Theory (ECT) is the term more commonly
employed.

The Einstein-Cartan Theory of gravity is
a modification of GR allowing spacetime to have torsion in addition
to curvature, and, more importantly, relating torsion to the density
of intrinsic angular~momentum. This modification was put forward by
Cartan before the discovery of quantum spin, \mbox{so the} physical
motivation was anything but quantum theoretic. Cartan was influenced
by the works of the Cosserat brothers \cite{Cosserat09} who
considered a rotation stress tensor in a generalized continuous medium
besides a force tensor.

Cartan assumed the linear connection to be metric and derived, from
a variational principle, \mbox{a set of gravitational} field equations.
However, Cartan required, without justification, that the covariant
divergence of the energy-momentum tensor be zero, which led to
algebraic constraint equations, thus severely restricting the
geometry. This probably discouraged Cartan from pursuing his theory.
It is now known that the conservation laws in relativistic theories
of gravitation follow from the Bianchi identities and in the
presence of torsion, the divergence of the energy-momentum tensor
need not vanish.

In simple mathematical terms, a non-zero torsion tensor means that
\bq
{T^\mu_{\nu\sigma}=\Gamma^\mu_{\nu\sigma}-\Gamma^\mu_{\sigma\nu}\neq
0\,.}
\eq

Geometrically, it means that an infinitesimal geodesic
parallelogram forms a non-closed loop. Torsion is therefore a local
property of the metric.
The Lagrangian action of Einstein-Cartan theory takes the usual Einstein-Hilbert form:
\bq
S =\int \dd^{4} x \sqrt{-g}\left(-\frac{g^{\mu\nu}R_{\mu\nu}(\Gamma)}{16\pi G} + \mathcal{L}_{m}\right)\, ,
\eq
where $\Gamma$ is a general affine connection and $\mathcal{L}_{m}$ is the matter Lagrangian.
The theory differs from GR in the structure of $\Gamma$, leading to a field theory with additional interactions.

Torsion vanishes in the absence of spin and
the Einstein-Cartan field equation is then the classical Einstein field
equation. In particular, there is no difference between the Einstein
and Einstein-Cartan theories in empty space. Since practically all
tests of relativistic theory are based on free space experiments,
the two theories are, to all effects, indistinguishable via the
standard tests of GR. \mbox{The inclusion} of torsion only results in a
slight change in the energy-momentum tensor. Cartan's theory holds
the distinction of being complete, self-consistent and in agreement
with experiment, \mbox{but of being} a non-metric theory of gravitation.
The link between torsion and quantum spin means that it could be
possible to study the divergence between the GR and ECKS theories at
the quantum level.
Such experiments have recently been proposed \cite{Puetzfeld2014}.

Kibble's theory contains some features which were criticized \cite{HvdHKN76}. It is
now accepted that torsion is an inevitable feature of a gauge theory
based upon the Poincar\'{e} group. Classical GR must be modified by the introduction of a
spin-torsion interaction if it is to be viewed as a gauge theory.
\mbox{The gauge} principle alone fails to provide a conceptual framework
for GR as a theory of gravity.

In the 1990s, \scalebox{.95}[1.0]{Anthony Lasenby, Chris Doran and Stephen Gull proposed an alternative formulation} of General Relativity which is derived from gauge principles alone
\cite{LDG93,LDG93a,LDG93b,LDG93c,LDG95,LDG98}. Their
treatment is very different from earlier ones where only infinitesimal translations are considered \cite{Kib61,HvdHKN76}.
There are a few other theories similar in their approach to that of Lasenby,
Doran and Gull (e.g., \cite[][]{Car90,Mukunda}).

\section{Why Consider Alternative Theories?}

The motivation for considering alternatives to GR comes
mainly from theoretical arguments, like scale invariance
of the gravitational theory, additional scalar fields that
emerge from string theories, Dark Matter, dark energy or inflation, or additional degrees of freedom
that arise in the framework of brane-world theories.

In Table \ref{Table_Alternatives}, we draw up a list of some of the more well-known alternatives to General Relativity. This list is far from exhaustive, but it serves to highlight the major elements which differentiate these theories. There are several works containing a more detailed listing and discussion of the various alternative theories  (e.g., \cite[][]{MTW:73,Will01,Berti2015}).

\begin{table}[H]
 \caption{A ``comparative morphology'' of some of the major alternatives to General Relativity, 
 in~approximate chronological order. We have only listed the theories of particular historical significance. The current landscape, in which cosmologists seek to explain Dark Matter, dark energy, and inflation, offers far more theories. It~is generally easier to incorporate the non-gravitational laws of physics within metric theories, since other theories would result in greater complexity, rendering calculations difficult. The only way in which metric theories significantly differ from each other is in their laws for the generation of the metric. Abbreviations: Tensor (T), V (Vector), S (Scalar), P (Potential), Dy~(Dynamic), Einstein Equivalence Principle (EEP), i.e., uniqueness of freefall, Local Lorentz Invariance (LLI), Local Position Invariance (LPI), param (Parameter), ftn (Function).}
 \label{Table_Alternatives}
 \centering
 \scalebox{0.69}[0.69]{
 \begin{tabular}{p{4.5cm}p{2cm}p{2cm}p{2.3cm}p{9.3cm}}
 \toprule
 \small % Font size can be changed to match table content. Recommend 10 pt.

\textbf{Theory} & \textbf{Metric} & \textbf{Other} \textbf{Fields}	 & \textbf{Free Elements} & \textbf{Status} \\
%&			&					&			& 		\\
\midrule
Newton 1687 \cite{Newt1687} & Nonmetric & P& None &  Nonrelativistic, implicit action at a distance \\ \midrule

Poincar\'e 1890s--1900s \cite{Poincare89,Poincare1904} & & & & Fails; does not mesh with electromagnetism\\ \midrule
Nordstr{\o}m 1913 \cite{Nord13}& Minkowski & S & None &  Fails to predict observed light detection \\  \midrule
General Relativity 1915 \cite{Ein15a} & Dy & None & None & Viable \\ \midrule
Whitehead 1922 \cite{White22} &  &  &  & Violates LLI; contradiction by everyday observation of tides\\  \midrule
Cartan 1922{--1925} \cite{Cart23}&ST& & &Still viable; introduces matter spin\\  \midrule
Kaluza-Klein 1920s \cite{Kaluza1921,Klein1926}&T &S &Extra dimensions &Violates Equivalence Principle\\   \midrule
Birkhoff 1943 \cite{Birk43}&T&&  &Fails Newtonian test; demands speed of  \\
&                     &&&sound equal to speed of light\\  \midrule
Milne 1948 \cite{Milne}&Machian  && & Incomplete; no gravitational redshift prediction; \\
&background &&& contradicts cosmological observations.\\  \midrule
Thiry 1948 \cite{Thiry1948}&ST&& &Unlikely; extremely constrained by results on $\gamma^{\mr{PPN}}$ \\   \midrule
Belifante-Swihart 1957 \cite{Belifante:1957} & Nonmetric & T & K param & Violates EEP; contradicted by Dicke--Braginsky experiments\\   \midrule
Brans--Dicke 1961 \cite{BranDick61} & Generic S & Dy & S &  Viable for $\omega > 500$\\ \midrule
Ni 1972 \cite{Ni:1972}&Minkowski&T, V, S &1 param, 3 ftns &Violates LPI; predicts preferred-frame effects\\   \midrule
Will-Nordtvedt 1972 \cite{Will:1972} & Dy T & V & & Viable but can only be significant at high energy regimes \\  \midrule
Barker 1978 \cite{Barker1978}& ST&&& Unlikely; severely constrained.\\  \midrule
Rosen 1973 \cite{Rosen:1973, Rosen:1973b}& Fixed & T & None & Contradicted by binary pulsar data \\  \midrule
Rastall 1976  \cite{Rastall1976}& Minkowski & S, V & None & Contradicted by gravitational wave data \\  \midrule
$f(R)$ models 1970s \cite{Buchdahl1970,Starobinsky1980}&$n+1$ST&S &Free ftn & Consistent with Solar System tests; \\
&&&& viable but severely constrained\\   \midrule
MOND 1983 \cite{Milgrom:1983,Milgrom:1983a,Milgrom:1983b} & Nonmetric & P & Free ftn & Nonrelativistic theory \\  \midrule
DGP 2000 \cite{DGP}&ST/Quantum&&&Appears to be contradicted by BAOs, CMB \\
&&&&and Supernovae Ia unless DE added\\  \midrule
TeVeS 2004 \cite{Bekenstein2004} &T,V,S & Dy S & Free ftn & Highly unstable \cite{Seifert2007}; ruled out by SDSS data \cite{Reyes2010}\\
\bottomrule
  \end{tabular}}
\end{table}
%%%%%%%%%%%%%%%%%%%%%%%%%%%%%%%%%%%%%%%%%%%%%%%

\section{From General Relativity to Standard Cosmology}
\label{Cosmology_section}

When Einstein published his seminal GR papers it became almost immediately apparent that the theory could be applied to the whole universe, under certain assumptions, to obtain a relativistic cosmological description. If the content of the universe is known, then the energy-momentum tensor can be constructed, and the metric derived using Einstein's equations. Einstein himself was the first to apply GR to cosmology in 1917 \cite{Einstein:1917}. The first expanding-universe solutions to the relativistic field equations, describing a universe with positive, zero and negative curvature, were discovered by Alexander Friedmann \cite{Fried1,Fried2}. This occurred before Edwin Hubble's observations and the empirical confirmation, in 1929, that the redshift of a galaxy is proportional to its distance. Hubble~ formulated the law which bears his name: $v=H_0r$, where $H_0$ is the constant of proportionality \cite{Hubble29}.
\mbox{The problem} of an expanding universe was independently followed up during the 1930s by Georges
Lema\^{i}tre \cite{Lemaitre}, and by Howard P. Robertson \cite{Robertson1, Robertson2, Robertson3} and Arthur Geoffrey Walker \cite{Walker}.

These exact \scalebox{.95}[1.0]{solutions define what came to be known as the Friedmann-Lema\^{i}tre-Robertson-Walker} (FLRW) metric, also referred to as the FRW, RW, or FL metric.
This metric starts with the assumption of spatial homogeneity and isotropy, allowing for time-dependence of the spatial component of the metric. Indeed, it is the only metric which can exist on homogeneous and isotropic spacetime.  The~assumption of homogeneity and isotropy, known as the Cosmological Principle, follows from the Copernican Principle, which states that we are not privileged observers in the universe. This is no longer true below a certain observational scale of around $100\,\mr{Mpc}$ (sometimes called the ``End of Greatness''), but~it does simplify the description of the distribution of mass in \mbox{the universe. }

The FLRW metric describes \scalebox{.95}[1.0]{a homogeneous, isotropic universe, with matter and energy uniformly} distributed as a perfect fluid. Using the definition of the metric in Equation (\ref{metric1}), it is written as:
\bq
-\dd s^2=c\dd \tau^2-R^2(t)[\dd r^2+S^0_k(r)(\dd \theta^2+\sin^2\theta\dd \phi^2)]\, ,
\eq
where $r$ is a time independent comoving distance, $\theta$ and $\phi$ are the transverse polar coordinates, and $t$ is the cosmic or physical time. $R(t)$ \scalebox{.95}[1.0]{is the scale factor of the universe. The function $S^0_k(r)$ is defined as:}
\bq S^0_k(r)=
\begin{cases}
\sin(r)&\quad (k=+1)\\
r&\quad (k=0)\\
\sinh(r)&\quad (k=-1)\end{cases} \eq
where $k$ is the geometric curvature of spacetime, the values $0$, $+1$, and $-1$ indicating flat, positively curved, and negatively curved spacetime, respectively.

Another common form of the metric defines the comoving distance as $S^0_k(r)\rightarrow r$, so that
\bq -\dd s^2=c\dd t^2-R^2(t)\left[\frac{\dd r^2 }{1-kr^2}+r^2(\dd\theta^2+\sin^2\theta\dd\phi^2)\right]\label{1}\, ,\eq
where $t$ is again the physical time, and $r$, $\theta$ and $\phi$ are the spatial comoving coordinates, which label the points of the 3-dimensional constant-time hypersurface.

The dimensionless scale factor $a(t)$ is defined as
\bq a(t)\equiv \frac{R(t)}{R_0} \, ,\eq
where $R_0$ is the present scale factor (i.e., $a=1$ at present). The scale factor is therefore a function of time, so it can be abbreviated to $a$. The metric can then be written in a dimensionless form:
\bq -\dd s^2=c^2\dd \tau^2 = c^2\dd t^2-a^2\left[\dd r^2 +S^2_k(r)(\dd\theta^2+\sin^2\theta\dd \phi^2)\right],\eq
where $S_k(r)$ can be redefined as
\bq \label{metric_form2}S_k(r)=
\begin{cases}
R_0\sin(r/R_0) & \quad (k=+1)\\
r &\quad (k=0)\\
R_0\sinh(r/R_0)&\quad (k=-1)\, .\end{cases} \eq
Equivalently, using the definition in Equation \eqref{1},
\bq
-\dd s^2=c^2\dd t^2 - a^2\left[\frac{\dd r^2}{1-k(r/R_0)^2}+ r^2(\dd\theta^2+\sin^2\theta\dd\phi^2)\right]\label{metric}\, .
\eq

The comoving distance is distance between two points measured along a path defined at the present cosmological time. It means that for objects moving with the Hubble flow, the comoving distance remains constant in time. The proper distance, on the other hand, is dynamic and changes in time. At the current age of the universe, therefore, the proper and comoving distances are numerically equal, but they differ in the past and in the future. The comoving distance from an observer to a distant object such as a galaxy can be computed by the following formula:
\bq
\chi = \int_{t_{\mr{e}}}^{t} c \frac{\dd t^{\prime}}{a(t^{\prime})} \,
\eq
where $a(t^{\prime})$ is the scale factor, $t_{\mr{e}}$ is the time of emission of photons from the distant object, and $t$ is the present time.

The comoving distance defines the comoving horizon, or particle horizon. This is the maximum distance from which particles could have travelled to the observer since the beginning of the universe. It represents the boundary between the observable and the unobservable regions of the universe.

If we take the time at the Big Bang as $t=0$, we can define a quantity called the conformal time $\eta$ at a time $t$ as:
\bq
\eta = \int_{0}^{t} \frac{\dd t^{\prime}}{a(t^{\prime})} \, .
\eq

This is useful, because the particle horizon for photons is then simply the conformal time multiplied by the speed of light $c$. The conformal time is not the same as the age of the universe. In~fact it is much larger. It is rather the amount of time it would take a photon to travel from the furthest observable regions of the universe to us. Because the universe is expanding, the conformal time is continuously increasing.

The concept of particle horizons is important. It defines causal contact. The only objects not in causal contact are those for which there is no event in the history of the universe that could have sent a beam of light to both. This is at the origin of some of the big questions about the universe associated with the Big Bang model, which gave rise to the Inflationary paradigm (see \cite[][]{Ellis1989}). We shall discuss \mbox{this later.}

\subsection{Cosmological Expansion and Evolution Histories}

The FLRW metric relates the spacetime interval $\dd s$ to the cosmic time $t$ and the comoving coordinates through the scale factor $R(t)$. The scale factor is the key quantity of any cosmological model, since it describes the evolution of the universe.
The notion of distance is fairly straightforward in Euclidean geometry. In General Relativity, however, where we work with generally curved spacetime, the meaning of distance is no longer unique. The separation between events in spacetime depends on the definition of the distance being used.

By combining the GR field equation (Equation \eqref{GR_field_eqn}) and the definition of the metric (Equation~\eqref{metric}), we obtain two independent Einstein equations, known as the Friedmann equations:%is this expression correct. YES IT IS CORRECT?
\bq
\left(\frac{\dot{a}}{a}\right)^2+\frac{kc^2}{a^2}=\frac{8\pi G}{3}\rho\label{Friedmann}
\eq
and
\bq
2\left(\frac{\ddot{a}}{a}\right) +\left(\frac{\dot{a}} {a}\right)+\frac{kc^2}{a^2}=-\frac{8\pi G}{c^2}p\label{Friedmann2}\,.
\eq

The Friedmann equations relate the total density $\rho$ of the universe, including all contributions, to~its global geometry. There exists a critical density $\rho_\mr{c}$ for which $k=0$. By rearranging the Friedmann equation and using the definition of the Hubble parameter we then obtain
\bq
\rho_\mr{c}(t)=\frac{3H^2(t)}{8\pi G}\, .
\eq

A universe whose density is above this value will have a positive curvature, that is, it will be spatially closed (k = +1); one whose density is less than or equal to this value will be spatially open ($k = 0$ or $k = -1$).

A dimensionless density parameter for any fluid component of the universe (i.e., a component for whose gravitational field is produced entirely by the mass, momentum, and stress density) can be defined by
\bq
\Omega(t)=\frac{\rho(t)}{\rho_\mr{c}(t)}=\frac{8\pi G\rho(t)}{3H^2(t)}\,.
\eq
The current value of the density parameter is denoted $\Omega_0$.

Subtracting Equation \eqref{Friedmann} from Equation \eqref{Friedmann2} yields the acceleration equation:
\bq \frac{\ddot{a}}{a}=-4\pi G\left(\frac{\rho}{3}+\frac{p}{c^2}   \right)\, .   \eq

The geodesic Equation \eqref{geodesic} allows us to compute the evolution in time of the energy and momentum of the various components particles which make up the universe.
From this evolution, we can construct the fluid equation, or continuity equation, which describes the relation between the density and pressure:
\bq
\label{fluid_eqn} \dot{\rho}+3\frac{\dot{a}}{a}\left(\rho+\frac{p}{c^2}\right)   =0\,.
\eq
This is valid for any fluid component of the universe, such as baryonic and nonbaryonic matter, or~radiation.

The foundations of the Concordance Model of cosmology depend on General Relativity. \mbox{Any modification} to the theory that changes the Einstein equations will have solutions that differ from the Friedmann equations.

The FLRW universe contains different mass-energy components which are assumed to evolve independently. This is physically valid at late cosmological times, when the components are decoupled, so the density evolutions are distinct. In Table \ref{EOStable}, we give the equation of state and the evolution of the density and scale factor for different components of the universe. The quantities in this table are explained in detail in the following sections.

\begin{table}[H]
\caption{The evolution of the various cosmological components. The quantities are the equation of state $w \equiv p /\rho c^2$,
the density $\rho$, the pressure $p$, and the scale factor $a(t)$.}
\label{EOStable}
%\small % Font size can be changed to match table content. Recommend 10 pt.
\centering\small
\begin{tabular}{lccc}
\toprule
\textbf{Component} & \textbf{\boldmath$w$} & \textbf{\boldmath$\rho=a^{3(1+w)}$} & \textbf{\boldmath$a(t)=t^{2/3(1+w)}$} \\
\midrule
Radiation (photons and relativistic neutrinos) & $1/3$ & $\sim$$a^{-4}$ & $\sim$$t^{1/2}$ \\
Dust (includes CDM, baryons and non-relativistic neutrinos) & $0$ & $\sim$$a^{-3}$ & $\sim$$t^{2/3}$ \\
Curvature &  $-1/3$ &$\sim$$a^{-2}\rightarrow a^{-4}$  & $t$\\
Cosmic strings & $-1/3$ &$\sim$$a^{-2}\rightarrow a^{-4}$  & $t$ \\
Domain walls & $-2/3$ & $ a^{-1}$ & $\sim$$t^{2}$ \\
Inflation & $\rightarrow -1$ & $\frac{1}{2} \dot{\phi}^{2}+V(\phi)$ & $\sim$$\e^{Ht}$ \\
Vacuum energy & $-1$ & constant & $\sim$$\e^{Ht}$
\\
\bottomrule
\end{tabular}
\end{table}

\subsection{Matter (Dust)}

Matter which is pressureless is referred to as ``dust''. This is a useful approximation for cosmological structures which do not interact, such as individual galaxies. Substituting $p_m=0$ in the equation of state for dust shows that the density of this component scales as:
\bq
\rho_m(a)=\frac{\rho_{m,0}}{a^3} \, ,
\eq
where $\rho_{m,0}$ is the current density. Assuming spatial flatness, the time evolution of the scale factor \mbox{is then}
\bq
a(t)=\left(\frac{t}{t_0} \right)^{2/3}\, ,
\eq
which gives us
\bq H(t)=\frac{2}{3t}\, .  \eq

This is known as the Einstein-de Sitter (EdS) solution, and it describes the evolution of $H$ in a constant-curvature homogeneneous universe with a pressureless fluid as the only component. It was first described by Einstein and Willem de Sitter in 1932 \cite{Eds1932}.

\subsection{Radiation}

In the early universe, the energy content was dominated by photons and relativistic particles (especially neutrinos). The expansion of the universe dilutes the radiation fluid, and the wavelength is increased by the expansion so that the energy decreases. From thermodynamics,
\bq E_\mr{rad}=\rho_\mr{rad}c^2=\alpha T^4\,,\eq
where $T$ is the radiation temperature and $\alpha$ is the Stefan-Boltzmann constant. The equation of state for radiation can then be derived from the fluid Equation (\ref{fluid_eqn}):
\bq \rho_\mr{rad}(a)=\frac{\rho_{\mr{rad},0}}{a^4}\quad;\quad p_\mr{rad}=\frac{\rho_{\mr{rad}}c^2}{3}\, .\eq

Combining this with the Friedmann equations, and assuming flatness ($k=0$), we obtain the time dependence of the scale factor and the Hubble parameter:
\bq a(t)=\left(\frac{t} {t_0}   \right)^{1/2}\quad;\quad H(t)=\frac{1}{2t}.
 \eq

\section{The Components and Geometry of the Universe and Cosmic Expansion }

How do we relate the expansion of the universe to its contents?
 The total density of the universe in terms of its constituent components can be written as the sum of the densities of these components at any given time or scale factor:
 \bq
\rho=\rho_m+\rho_{\mr{rad}}+\rho_\DE \, ,
\eq
where the subscript ``DE'' denotes another component of the universe, called Dark Energy.

{\begin{figure}[H]
\centering
\includegraphics[width=.8\textwidth]{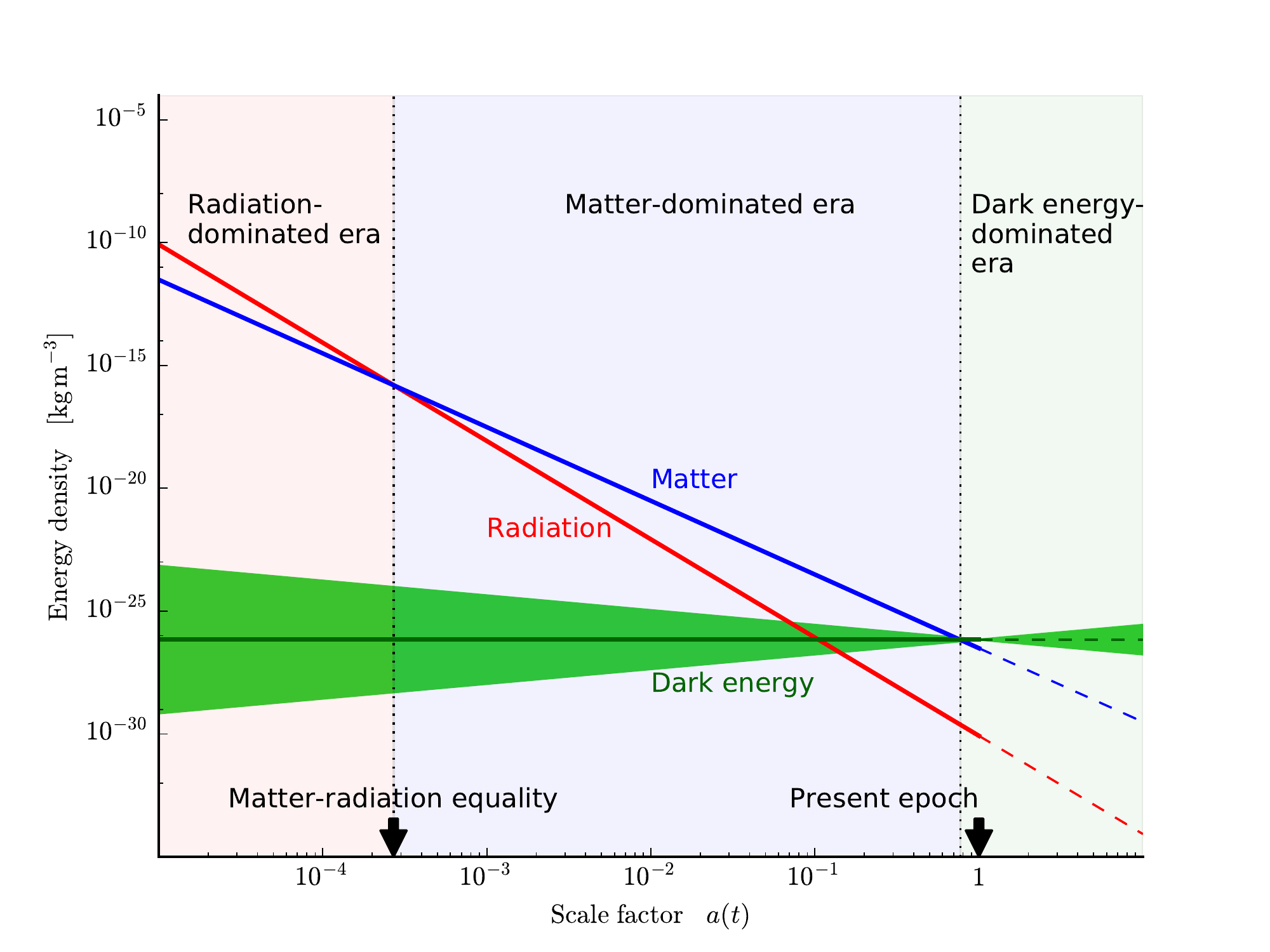}
\caption{The density evolution of the main components of the universe. The early universe was radiation-dominated, until the temperature dropped enough for matter density to being to dominate. The energy density of dark energy is constant if its equation of state parameter $w=1$. Because~the matter energy density drops as the scale factor increased, dark energy began to dominate in the recent past. At the present time ($a(t)=1$), we live in a universe dominated by dark energy. For dark energy, the green band represents an equation of state parameter $w=-1\pm 0.2$, showing how a small change in the value of this parameter can give very different evolution histories for dark energy. If the Concordance Model is correct, the universe will be completely dominated by dark energy in future epochs (shown by the dashed lines). The matter density will keep decreasing as the universe expands. Our Milky Way will merge with the Andromeda Galaxy, and eventually, the entire Local Group will coalesce into one galaxy. The luminosities of galaxies will begin to decrease as the stars run out of fuel and the supply of gas for star formation is exhausted. In the very far future, this galaxy will be in the only one in our Hubble patch, as all the other galaxies will pass behind the cosmological horizon. The night sky, save for the stars in the Local Group, will be very dark indeed. Stellar remnants will either escape galaxies or fall into the central supermassive black hole.  Eventually, baryonic matter may disappear altogether as all nucleons including protons decay, or all matter may decay into iron. In either scenario, the universe will end up being dominated by black holes, which will evaporate by Hawking radiation. The end result is a Dark Era with an almost empty universe, and the entire universe in an extremely low energy state, with a possible heat death as entropy production ceases (see, e.g., \cite[][]{Adams1997,Caldwell2003}) What happens after that is speculative.}
\label{GR_components}
\end{figure}

The total dimensionless density can then be written:
\bq
\Omega=\OM+\OR+\OD\, ,
\eq
where we  have dropped the subscript for clarity, i.e., $\Omega_{m,0}=\OM$, etc. The Friedmann Equation \eqref{Friedmann} can now be rewritten using the equations of state for the different components:
\bq
H^2(a)=\frac{8\pi G}{3} \left(\rho_m a^{-3}+\rho_{\mr{rad}} a^{-4}+\rho_\DE \e^{-3\int_a^1[1+w(a')]\dd \ln a'}\right)-\frac{k c^2}{a^2}.
\eq
This can be rearranged to give:
\bq \label{1.46}
H^2(a)=H_0^2\left[\OM a^{-3}+\OR a^{-4}+\OD \e^{-3\int_a^1[1+w(a')]\dd \ln a'}+(1-\Omega)a^{-2}\right],
\eq
or, in terms of redshift:
\bq\label{1.47}
H^2(z)= H_0^2\left[\OM{(1+z)}^{3}+\OR {(1+z)}a^{4}+\OD \e^{-3\int_0^z[1+w(z')]/(1+z')\dd \ln z'}+(1-\Omega){(1+z)}^{2}\right].
\eq

The term $1-\Omega$ is sometimes replaced by $\Omega_k$, the density due to the intrinsic geometry of spacetime. Equation \eqref{1.47} is of central importance since it relates the redshift of an object to the global density components and geometry of the universe.

The density evolution of the various components of the universe is shown in Figure \ref{GR_components}.

\section{The Hot Big Bang}

In the Standard Model, it is generally accepted that the universe arose from an initial singularity, often termed the ``Big Bang'', which occurred some $13.8$ billion years ago (as measured by \textit{Planck} \cite{Planck-Collaboration2015}). This~is not discussed here, but it should be noted that there are several proposals for the mechanism of this singularity. During this epoch, we are dealing with Planck scale physics, so most of these mechanisms involve quantum gravity. Other proposals (such as some superstring and braneworld theories) do away with the need for an initial singularity altogether.

\subsection{The Cosmic Microwave Background}

The radiation \scalebox{.95}[1.0]{density $\rho_\mr{rad}\propto a^{-4}$, so the temperature evolution of the universe from an initial $T_0$ is:}
\bq
T=\frac{T_0}{a}\,.
\eq

In other words, the universe cools down as it expands. Conversely, this means at early times, when the scale factor was close to zero, the temperature was very high (hence the term ``Hot Big Bang''). The radiation left from the early hot universe, cooled by expansion, is known as the Cosmic Microwave Background, or CMB.

The properties of atomic and nuclear processes in an expanding universe provided the first clue for the existence of a hot Big Bang. This was a remarkable achievement of the Big Bang model, because it provided an explanation for the observed abundances of chemical elements in terms of nucleosynthesis. The processes that created nuclei and atoms could only have been possible in an early universe in thermal equilibrium, with black-body spectrum which cooled down as the universe expanded. This allowed Ralph Alpher, Robert Herman, Hans Bethe and George Gamow to predict the existence and temperature of the CMB in 1948 \cite{Alpher1948,Alpher1948a,Gamow1948,Gamow1948b}. The universe therefore has a thermal as well as an expansion history. Hence the `Hot Big Bang'.

The first direct evidence for the Hot Big Bang came two decades later, with the observation of the CMB by Arno Penzias and Robert Woodrow Wilson in 1964 \cite{Penzias:1965}.

The confirmation of the thermal history of the universe, together with the discovery of charge parity violation in 1964 \cite{CPviolation1964}, provided clues about baryogenesis and the observed matter-antimatter imbalance in the universe. This inspired the first proposals for a mechanism for baryogenesis by Andrei Sakharov in1967 \cite{Sakharov1967}, followed by electroweak symmetry breaking by Vadim Kuzmin \mbox{in 1970 \cite{Kuzmin1970}.}

This is a remarkable demonstration of the success of the Concordance Model. The cosmological model fits very well with the predictions of particle physics, which in turn can be tested by cosmological observations. The Concordance Model of the structure and evolution of the universe requires a mechanism for baryogenesis as well as an explanation for Dark Matter and dark energy. \mbox{The challenge} for physical theories beyond (or within) the Standard Model is to explain the preference of matter over antimatter, and to explain the magnitude of this asymmetry. Cosmological observations can be used to address these challenges \cite{Khlopov2016}.

The CMB is an extremely isotropic source of microwave radiation, with a spectrum corresponding to a perfect blackbody at a temperature $T_0= 2.7260 \pm0.0013\,\mr{K}$ \cite{Fixsen2009}. Using the current temperature and $E_\mr{rad}=\rho_\mr{rad} c^2=\alpha T^4$, the radiation density today is given by:

\bq
\OR=2.47\times10^{-5}h^{-2}\, .
\eq

At some time in the early universe, the ambient radiation temperature corresponded to the ionisation potential of hydrogen, which is $13.6\,\mr {eV}$. During this epoch, the universe was filled with a sea of highly energetic particles and photons---a hot ionised plasma. The particles were mainly electrons and protons. Other fundamental particles (quarks) existed earlier when the ambient energy corresponded to their rest mass. At some point, as the universe expanded and cooled,  the energy of the photons was no longer sufficient to ionise the hydrogen, and within a relatively short time, all of the electrons and protons combined to form neutral hydrogen. The photons were then free to move through the universe. This process is known as decoupling and it occurred at a temperature \mbox{of $\sim$$2500\,\mr{K}$,} when the universe was approximately $380,\,000$ years old \cite{WMAP5}. It is these decoupled photons which make up the CMB. The surface on the sky from which these photons originate is known as the surface of last scattering.
\subsection{Matter-Radiation Equality}
At \scalebox{.95}[1.0]{the present epoch, neglecting dark energy, the universe is dominated by matter. This component} is characterised by the fact that the matter particles can be treated in a non-relativistic regime, whereas photons and relativistic neutrinos both behave like radiation. The total contribution to the energy density from non-relativistic components (matter) and relativistic components (radiation and relativistic neutrinos) can be written as $\Omega_{\mr{NR}}$ and $\Omega_{\mr{R}}=\OR+\Omega_{\nu}$, respectively. Using the fact that $\OM=\Omega_{\mr{m},0}a^{-3}$, the ratio of the contributions of the components is a function of the scale factor $a$:
\bq
\frac{\Omega_{\mr{R}} }{\Omega_{\mr{NR}}} = \frac{\OR+\Omega_{\nu}}{\OM}=\frac{4.15\times10^{-5}}{\Omega_{\mr{m},0}h^{2}a^{-3}} \, ,
\eq
where we explicitly use the subscript $0$ for the present-day values.

Then there must exist a scale factor for which the ratio is unity. This is given by:
\bq
a_\mr{eq}=\frac{4.15\times 10^{-5}}{\Omega_{\mr{m},0} h^2}\, ,
\eq
or, in terms of redshift,
\bq
1+z_\mr{eq}=2.4\times 10^4\Omega_{\mr{m},0} h^2\,.
\eq

The epoch at which the matter energy density equals the radiation energy density is called matter-radiation equality, and it has a special role in large-scale structure formation.

\subsection{Neutrinos}

Neutrinos have particular properties which give rise to a distinct evolution history. They are known to exist from the Standard Model of particle physics, and the Hot Big Bang model predicts the amount of neutrinos in the universe. Neutrinos can be thought of as ``dark'' matter because of their very small reaction cross-section, which implies negligible self-interaction. However, they are not \emph{cold} Dark Matter. They are simply extremely light particle that can stream out of high-density regions. They~therefore cause the suppression of perturbations on scales smaller than the free-streaming scale. Unlike~photons and baryons, cosmic neutrinos have not been observed. However, particle physics allows us to chart the history of this particle during nucleosynthesis, and to relate the neutrino temperature to the photon temperature today \cite{Hannestad:2006a, Lesgourgues:2006,Lesgourgues2014}.

The scale on which perturbations are damped by neutrinos is determined by the comoving distance that a neutrino can travel in one Hubble time at equality. For a neutrino mass $\sim$$1\,\mr{eV}$, the~average velocity, $T_\nu/m_\nu$ is of order unity at equality.  This leads to a suppression of power on all scales smaller than $k_\mr{eq}$. Note that this phenomenon depends on the individual neutrino mass, rather than the total neutrino mass. A lighter neutrino can free-stream out of larger scales, so the suppression begins at lower $k$ for the lighter neutrino species. Heavier neutrinos constitute more of the total neutrino density, and so suppress small-scale power more than lighter neutrino species, which means that we need at least two parameters to model massive neutrino phenomenology to sufficient accuracy: the neutrino mass fraction $\Onu$, or some expression of this quantity in terms of the total neutrino mass $\sum m_\nu$, and the number of massive neutrino species $N_\nu$.

Neutrinos introduce a redshift and scale dependence in the transfer function. We know that the perturbation modes of a certain wavelength $\lambda$ can grow if they are greater than the Jeans wavelength. Above the Jeans scale, perturbations grow at the same rate independently of the scale. For the baryonic and cold Dark Matter components, the time and scale dependence of the power spectrum can therefore be separated at low redshifts. This is not the case with massive neutrinos, which introduce a new length scale given by the size of the comoving Jeans length when the neutrinos become non-relativistic. In terms of the comoving wavenumber $k_\mr{nr}$, this scale is given by:
\bq
k_\mr{nr}=0.026\brn{\frac{m_\nu}{1\,\mr{eV}}}^{1/2}\OM^{1/2}\hpM
\eq
for three neutrinos of equal mass, each with mass $m_\nu$. The growth of Fourier modes with \mbox{$k>k_\mr{nr}$} is suppressed because of neutrino free-streaming. From the equation above, it is evident that the free-streaming scale varies with the cosmological epoch (since there is a dependence on $\OM$), \mbox{and therefore} the scale and time dependence of the power spectrum cannot be separated.

Neutrinos are fermions, with a Fermi-Dirac distribution with assumed  zero chemical potential.
When they decoupled from the plasma, their distribution remained Fermi-Dirac, with their temperature falling as $a^{-1}$. This decoupling occurred slightly before the annihilation of electrons and positrons, which occurred when the cosmic temperature was of the order of the electron mass ($T \approx m_e$). Neutrinos decoupled when the cosmic plasma had a temperature of around $1\, \mr{MeV}$.  \mbox{The energy} associated with this annihilation was therefore not inherited by the neutrinos, and the entropy was completely transferred to the entropy of the photon background. Thus:
\bq
(S_\mr{e}+S_\gamma)_\mr{before}=(S_\gamma)_\mr{after}\,,
\eq
where $S_\mr{e}$ and $S_\gamma$ are respectively the entropy of the electron-positron pairs and the photon background, and `before' and `after' refer to the annihilation time.

The entropy per particle species, ignoring constant factors, is $S\propto gT^3$, where $g$ is the statistical weight of the species. For bosons, $g=1$ and for fermions, $g=7/8$ per spin state.
According to the Standard Model, the neutrino has one spin degree of freedom, each neutrino has an antiparticle, and~there are three generations of neutrinos, also called ``families'' or ``species'' ($\mu$, $\tau$ and electron neutrinos). This means that the degeneracy factor of neutrinos is equal to $6$. Before annihilation, the fermions are electrons (2 spin states), positrons (2), neutrinos and antineutrinos (6 spin states). The bosons are photons (2 spin states). We therefore have $g_\mr{before}=4(7/8)+2=11/2$, while after annihilation $g=2$ because only photons remain. Applying entropy conservation and counting relativistic degrees of freedom, the ratio of neutrino and photon temperatures below $m_e$ is therefore:
\bq
\frac{T_\nu}{T_\gamma}=\brn{\frac{4}{11}}^{1/3}\,,
\eq
so that the present neutrino temperature is
\bq
T_{\nu,0}=\brn{\frac{4}{11}}^{1/3}T_\mr{CMB}=1.945 \,\mr{K} \, .
\eq
The number density of neutrinos is then
\bq\label{neutrinoeqn1}
n_\nu=\frac{6\zeta(3)}{11\pi^2}T_\mr{CMB}^3\,,
\eq
where $\zeta(3)\approx 1.202$, which gives $n_\nu\approx 112\,\mr{cm}$$^{-3}$ at the present epoch  \cite{Elgaroy:2005}.
In the early universe, neutrinos are relativistic and behave like radiation. So they contribute to the total radiation energy density $\rho_\mr{rad}$, which includes the photon energy density $\rho_{\gamma}$:
\bq
\rho_\mr{rad} = \left[ 1+\brn{\frac{7}{8}}\brn{\frac{4}{11}}^{4/3}  N_{\mr{eff}} \right]   \rho_{\gamma} \, ,
\eq
where $N_{\mr{eff}}$ is the effective number of neutrino species.
At late times, when massive neutrinos become non-relativistic, their contribution to the mass density is $m_\nu n_\nu$, giving
\bq \label{neutrinoeqn2}
\Onu = \frac {\rho_{\nu}}{\rho_{\mr{c}}} \approx\frac{\sum\limits^{N_{\nu}}_{i}m_{\nu,i}}{93.14\,\mr{eV} \, h^2} \, ,
\eq
where $m_{\nu,i}$ is the mass of individual neutrino species and $N_{\nu}$ is the number of massive neutrino species.
This expression relates the total neutrino mass $\sum{m_{\nu}}$ to the neutrino fraction $\Onu$.

It can be seen from the above that this equation can be modified through a change in the effective number of neutrino species by many factors: a non-zero initial chemical potential, or a sizeable neutrino-antineutrino asymmetry, or even a fourth, `sterile' neutrino \cite{Abazajian2001,Elgaroy:2005,Boyarsky2009}. The Standard Model predicts a value of $N_{\mr{eff}}=3.046$ for the effective number of neutrino species. This accounts for the three neutrino families together with relativistic degrees of freedom, since neutrinos are not completely decoupled at electron-positron annihilation and are subsequently
slightly heated \cite{Mangano2002}. Any significant deviation from this value could be a signature of hidden physical effects, possibly requiring a modification of General Relativity \cite{Feeney2013}.

Neutrino oscillation experiments do not, at present, determine absolute neutrino mass scales, since they only measure the difference in the squares of the masses between neutrino mass eigenstates. Cosmological observations, on the other hand, can constrain the neutrino mass fraction, and can distinguish between different mass hierarchies \cite{Elgaroy:2005}.

Observations of neutrino flavour oscillations in atmospheric and solar neutrinos, provide evidence of a difference between the masses of the different species or flavours, as well as for a non-zero mass. For three neutrino mass eigenstates $m_1$, $m_2$ and $m_3$, the squared mass differences are \cite{RevPartPhys2014}:

\vspace{-24pt}
\bq
\label{hier}\begin{split} |\Delta m_{21}^2|=m_2^2-m_1^2|& \cong 7.5\times 10^{-5}\,\mr{eV}^2\\
|\Delta m_{31}^2|=|m_3^2-m_1^2|&\cong 2.5 \times 10^{-3}\,\mr{eV}^2 \\
 \frac{|\Delta m_{21}^{2}|}{| \Delta m_{31}^{2}|}&\cong 0.03  \, .
\end{split} \eq

The ambiguity in the sign of the mass differences $\Delta m$ allows for two possible mass hierarchies: the normal hierarchy given by the scheme ${m_3\gg m_2>m_1}$, or the inverted hierarchy $m_2>m_1\gg m_3$. Given Equation \eqref{hier}, constraining the total neutrino mass to a small enough maximum value could exclude an inverted hierarchy. Conversely, a total neutrino mass ${m_\nu}$$\sim$$2$~eV is only possible with a degenerate neutrino mass scheme. Hence the interest in finding cosmological neutrino mass bounds.

The fact that cosmological constraints could be stronger than constraints from particle accelerators was noticed quite early (see \cite[][]{Primack:2001}). The `closure limit' gives us $m_\nu<90$~eV. This was first derived in the late 1960s and 1970s \cite{Gerstein:1966,Marx:1972,Cowsik:1972,Szalay1976}. Since then, cosmological neutrino bounds have improved significantly, with different methods being used e.g., luminous red galaxies \citep{Tegmark:2006}, CMB anisotropies \cite{Shiraishi:2009,
Planck-Collaboration2015}, or~weak lensing \cite{Tereno:2008,Ichiki:2008,Xia2012}.

Joint \textit{Planck} CMB and BAO observations give us $m_\nu<0.23$~eV, but various data combinations can change this figure, and strong priors on the value of the Hubble constant can provide tighter constraints \cite{Planck-Collaboration2015}.

\section{Inflation: The Second Wave of Alternative Theories}

In the late 1970s, General Relativity had been largely accepted by the scientific community. \mbox{But a series} of cosmological considerations led to renewed interest in alternative theories. These were not so much attempt to solve problems in the theory itself, but to find explanations for observations that were not explained by the theory.

General Relativity applied to the universe gave us the Hot Big Bang model: a universe expanding out of an initial highly energetic, dense state. The Hot Big Bang model was successful in explaining many interlinked phenomena which were subsequently confirmed by observation: the Hubble Law and the expansion of the universe, the thermal history of the universe, primordial nucleosynthesis, the~existence of the cosmic microwave background, the~relation between the temperature and scale factor, and finally the blackbody nature of the CMB. The remarkable fact is that these phenomena occur on extremely different scales, and are observed via different physical processes, and yet they all fit neatly within one model.

However, there are some observations which the Hot Big Bang model fails to explain. These~cosmological problems are linked to the primordial universe, the most obvious being the following (for details see \cite[][]{Olive1990, Guth:1997, Peacock:1999,Inflation_textbook, Uzan2015}, and references therein):
\begin{itemize}[leftmargin=*,labelsep=4mm]
\item The Horizon Problem
\item The Flatness Problem
\item The Monopole Problem
\end{itemize}

The horizon problem arises from the structure of spacetime. In the standard cosmological model described by the FLRW equations, different regions of the universe observed today could have not been in causal contact with because of the great distances between them which are greater than the distance that could have been traversed by light since the Big Bang. The transfer of information (i.e.,~any physical interaction) or energy can occur, at most, at the speed of light, but these regions have the same temperature and other physical characteristics. In particular, we observe causally-disconnected regions of the CMB to be in thermal equilibrium. How could this have happened?

The horizon problem was first identified in the late 1960s. This led to to early attempts to model chaotic solutions to Einstein's field equations near the initial singularity \cite{Misner1969, Belinskij1970}. In the late 1970s, Alexei Starobinsky noted that quantum corrections to General Relativity should be important in the very early universe. These corrections would lead to a modification of gravity, which induces an inflationary phase \cite{Starobinsky1979}. Starobinsky's was the first model of inflation.

The flatness problem is one the so-called ``coincidence problems'' of modern cosmology. \mbox{By the 1960s,} observations had determined that the density of matter in the universe is comparable to the critical density necessary for a flat universe. So the contribution of curvature had to be of the same order of magnitude as the contribution of matter throughout the history of the universe. This represents a fine-tuning problem. Observations of the CMB have confirmed that the universe is spatially flat to within a few percent. Why is the global geometry of the universe so flat?

The magnetic monopole problem arises from the Hot Big Bang model. Grand Unified Theories predict the production of a large number of magnetic monopoles \cite{t-Hooft1974,Preskill1984} in the early, extremely hot universe. Why have none ever been observed? If they exist at all, they are much more rare than the Big Bang theory predicts. This was noted by Zel'dovich and others in the late 1970s \cite{Zeldovich1979,Preskill1979}.

\vspace{6pt}
\noindent\emph{Hot Big Bang Plus Inflation}
\vspace{6pt}

This gave rise to the idea of a model in which the early universe undergoes a period of exponential accelerated expansion. This theory, called ``inflation'', was first formulated by Alan Guth in the 1980s~\cite{Guth1981} while he was trying to investigate why no magnetic monopoles are observed. \mbox{It was} realised that inflation solves the horizon and flatness problems, as well as explaining the absence of relic monopoles. Better still, it explains the origin of structure in the universe.

In the Standard $\LCDM$ Model, the initial perturbations from which structure evolved are assumed to have been seeded by the inflationary potential.
Reconstructing the primordial power spectrum is no easy task, and poses two main problems. Observationally, we want to extract the amplitude and scale variation from the data. Theoretically, we seek to explain the origin of the perturbations. At~present, the leading theoretical paradigm for the primordial fluctuations is inflation, which provides initial conditions for both large-scale structure and the cosmic microwave background radiation. The~theory of inflation offers a plethora of models, each of which predicts a certain power spectrum of primordial fluctuations $\mc{P}(k)$. Since the inflationary paradigm is linked to the theoretical description of the primordial power spectrum, it is necessary to briefly explain some of the main concepts here (for~the full details, see \cite[][]{Guth:1997,Inflation_textbook}).

The precise definition of inflation is any period during which the scale factor of the universe is accelerating, that is, $\ddot{a}> 0$. This expression is equivalent to other definitions of inflation:
\begin{align}
\frac{\dd}{\dd t}\frac{H^{-1}}{a}<0 \implies \epsilon \equiv -\frac{\dot{H}}{H^{2}}<1 \iff \frac{\dd^{2}a}{\dd t^{2}}>0 \iff \rho+3p<0 \,.
\end{align}

The first expression above has a remarkable physical interpretation. It means that the observable universe becomes smaller during inflation.

The basic theory of inflation states that from the initial Big Bang singularity to approximately $10^{-37}$ s, there existed a set of highly energetic scalar fields. By definition, $\Omega$ is driven towards 1 during inflation. Inflationary theories assume that gravity is described by GR, which means that the component driving inflation must satisfy ${\rho+3P<0}$.
If for example, the universe was dominated during the inflationary phase by a scalar field (or set of fields) $\phi$ with a self-interaction potential $V(\phi)$. It is the form of this potential which differentiates the various inflationary theories. Most theories assume a ``Mexican hat'' potential, with a single field, while chaotic inflation assumes a simple power law potential with a slowly varying field \cite{Linde:1989}. The action for this potential is then \cite{Lidsey:1997}
\bq
S=-\int\dd^4 x \sqrt{-g}\left[\frac{m_\mr{Pl}^2R}{16\pi}-\frac{1}{2}{(\nabla\phi)}^2+V(\phi)\right]\, ,
\eq
where $m_\mr{Pl}$ is the Planck mass. As the universe cooled, the scalar field became trapped in a false vacuum, so its energy density became constant. The potential energy, however, is nonzero, so the pressure is negative. The scale factor during inflation has the de Sitter form:
\bq
\label{scale_infl}
a(t)=\mr{e}^{({\Lambda_\mr{I}/3})^{1/2}t}\,,
\eq
where $\Lambda_\mr{I}$ represents the energy density of the inflationary field (sometimes called the inflaton).

Since the energy density of the inflaton field was very high, the associated magnitude of the negative pressure would have been very large as well. The scale factor is thought to have increased during inflation by $\sim$$\mr{e}^{65}$, and any point in the universe which found itself in a false vacuum state would have undergone inflation. The accelerated expansion lasted until the field rolled down to a minimum, when it decays into the familiar particles of the Standard Model, and the universe can then be described by an FLRW model.

The inflationary paradigm provides an explanation for the origin of structure and for the observed geometry of the universe, in addition to solving the aforementioned cosmological problems.

First, inflation solves the flatness problem. Using Equation \eqref{scale_infl}, the evolution of $\Omega$ during inflation can be written as:
\bq \label{Eqnefolds}
|\Omega(t)-1|\propto \mr{e}^{{-(4\Lambda_\mr{I}/3)}^{1/2}t}\,,
\eq
so that $|\Omega-1|$ is driven very close to 0 as $t$ increases. This explains why the universe is flat. \mbox{It also means} that this value has not deviated significantly from its initial value right after expansion. We can therefore safely assume spatial flatness throughout the history of the universe. Given~ the observational difficulties, this provides a theoretical motivation for taking the idea of a large \mbox{$\Omega_\mr{DE}$ seriously.}

Second, inflation solves the horizon problem (Figure \ref{GR_inflation}). Regions of the universe which are causally disconnected today evolved out of the same causally-connected region in the early universe. \mbox{The observed} uniformity of the CMB is no longer a problem.

Third, inflation explains why we have never observed magnetic monopoles. Due to rapid expansion of the universe during inflation, they become so rare in any given volume of space that we would be very unlikely to ever encounter one. Nor would they have sufficient density to alter the gravity and thereby the normal expansion of the universe following inflation. The problem of magnetic monopoles motivated the Guth's development of his theory in 1981 \cite{Guth1981}. The solution of the monopole problem, and problems related to other relics, was an early success of the inflationary paradigm, and inspired similar theories \cite{Sato1981,Einhorn1981}.

Fourth, the inflationary scenario provides a natural explanation for the origin of structure, providing a link between quantum mechanics and relativistic cosmological paradigm. This was realised soon after the development of the theory of inflation, and the details were worked out in the early 1980s \cite{Mukhanov1982,Mukhanov1981,Guth1982,Hawking1982,Starobinsky1982,Bardeen1983}.

An initially smooth background needs seed fluctuations around which gravitational collapse can occur. The inflationary scenario attributes their origin to quantum fluctuations in the inflaton field potential, so that the universe is not perfectly symmetric. Different points in the universe inflate from slightly different points on the potential, separated by $\delta \phi$. Inflation for these two points ends at different times, separated by $\delta t=\delta \phi/\dot{\phi}$. This induces a density fluctuation $\delta =H \delta t$ (see \cite[][]{Peacock:1999}). Since~all the points undergoing inflation are part of the same potential field, the initial fluctuations are nearly scale invariant. This means that the density amplitude on the horizon scale will also \mbox{be constant:}
\bq
\delta_\mr{H}=H\delta t=\frac{H^2}{2\pi \dot{\phi}}=\mr{constant}\,.
\eq

In summary, inflation solves the three cosmological problems listed above:
\begin{itemize}[leftmargin=2.1em,labelsep=4mm]
\item \scalebox{.93}[1.0]{The Horizon Problem. Solution: the entire universe evolved out of the same causally-connected region.}
\item The Flatness Problem. Solution: any initial curvature is diluted by the inflationary epoch and driven to zero.
\item The Monopole Problem. Solution: the rapid expansion of the universe drastically reduces the predicted density of magnetic monopoles, if they exist.
\end{itemize}

\begin{figure}[H]
\centering
\includegraphics[width=.94\textwidth]{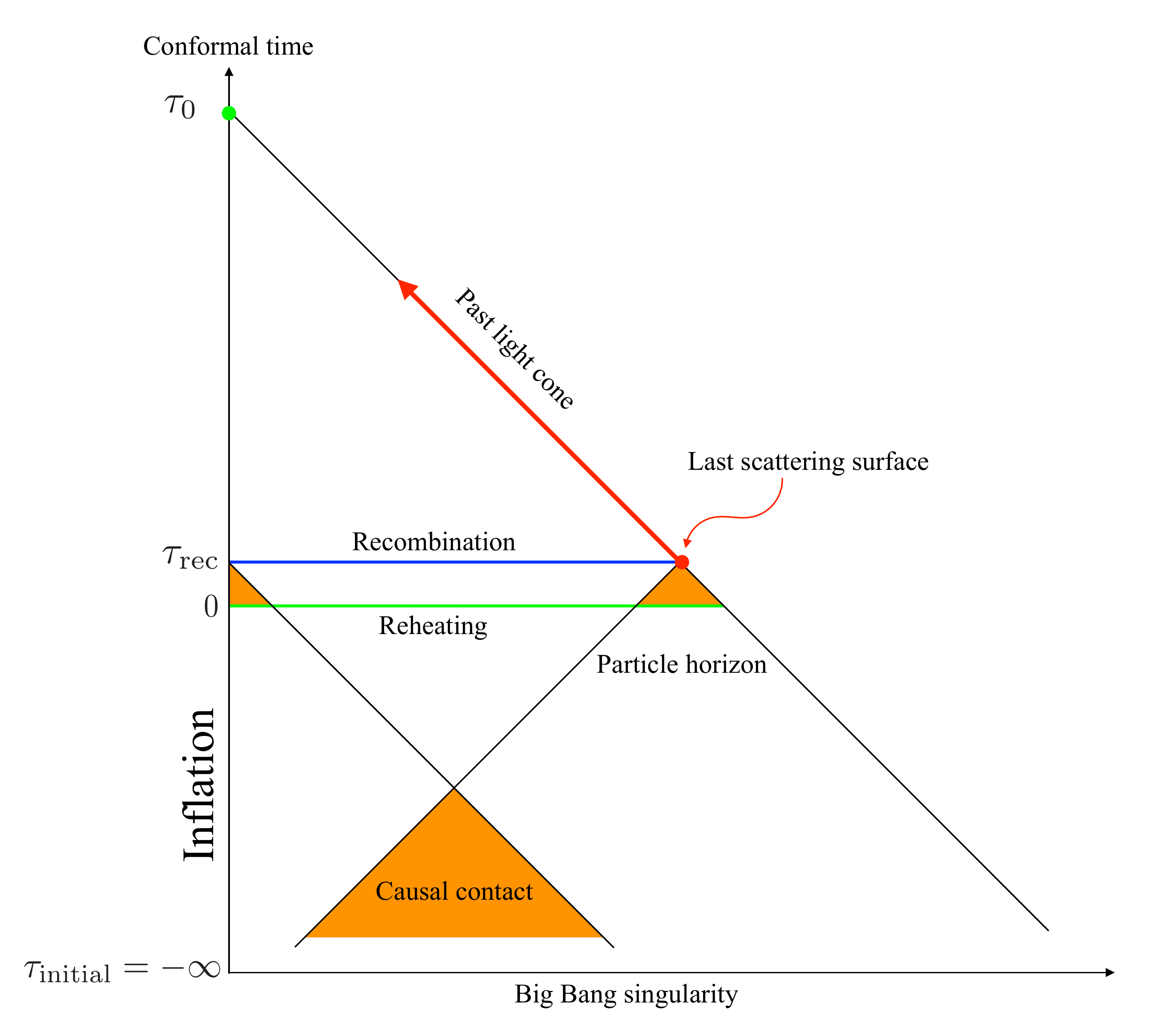}
\caption{How inflation solves the horizon problem. The light cones on the causal diagram of an inflationary FLRW model are at $\pm 45^{\circ}$. The worldlines of comoving matter are vertical on this kind of diagram. The particle horizons are horizontal lines. Here we have shown the particle horizon for the CMB. Without inflation, conformal time would only go back to $\tau_{0}$, and different regions of the CMB which we observe today along our past light cone would never have been in causal contact. Because of inflation, conformal time is extended to the Big Bang singularity, so these regions would have been in causal contact at some point in our past light cone.}
\label{GR_inflation}
\end{figure}

How long did inflation last? The answer is given by looking closely at Equation \eqref{Eqnefolds} above.
\mbox{A convenient} measure of expansion is the so-called $e$-fold number, defined as:
\bq
N \equiv \ln \left( \frac{a_{\mr{f}}}{a_{\mr{i}}}\right) = \int_{t_{\mr{i}}}^{t_{\mr{f}}} H \dd t \, .
\eq

Here, $a_{\mr{i}}$ and $a_{\mr{f}}$ are the values of the scale factor at the beginning and end of inflation,
while $t_{\mr{i}}$ and $t_{\mr{f}}$ are the corresponding proper times. The scale factor $a$ is only physically meaningful up to a normalisation constant, so the $e$-fold number is defined with respect to some chosen origin. \mbox{The reason} is that in cosmology, what is fixed is not the initial condition, but the current expansion---we cannot measure any $H$, but we measure $H_{0}$ then extrapolate backwards.

We can search for the minimum duration of inflation required to solve the horizon problem.
At~ the very least, we require that the observable universe today fits in the comoving Hubble radius at the beginning
of inflation:
\bq
(a_{0}H_{0})^{-1}< (a_{i}H_{i})^{-1} \, .
\eq
The condition is the same for the horizon and flatness problems.

If we assume that the universe was radiation-dominated since the end of inflation (giving us ${H\propto a^{-2}}$), and ignore the
relatively recent matter- and dark energy-dominated epochs, we obtain
\bq
\frac{a_{\mr{f}}}{a_{\mr{i}}} = \frac{a_0}{a_{\mr{f}}} \, .
\eq

In the general case, the condition becomes:
\bq
\frac{a_{\mr{f}}}{a_{\mr{i}}} \ge \frac{a_0}{a_{\mr{f}}} \, ,
\eq
or in terms of the number of $e$-folds,
\bq
N_{\mr{f}} - N_{\mr{i}} \ge N_{0} - N_{\mr{f}} \, .
\eq

In other words, there should be as much expansion during inflation as after inflation.

The solution to the horizon problem is the same as the solution to the flatness problem. Taking~ into account the present energy density of the universe, we need a minimum of about 50 to 60 $e$-folds. This already gives us a useful criterion for realistic inflation models. The most recent \textit{Planck} results show a preference for a higher number of $e$-folds:  ${78 < N < 157}$ \cite{Planck2015_inflation}.

Most models of \scalebox{.95}[1.0]{inflation are slow-roll models, in which the Hubble rate varies slowly \cite{Inflation_textbook, Liddle:1992, Liddle:1993}.} This model was first developed by Andrei Linde in 1982 \cite{Linde1982}. It solved a major problem in Guth's early theory. Instead of tunnelling out of a false vacuum state, inflation occurrs by a scalar field rolling down a potential energy gradient. When the field rolls very slowly compared to the expansion of the universe, inflation occurs. Hence the name ``slow-roll inflation''. However, when the gradient becomes steeper, inflation ends and reheating can occur. It is beyond the scope of this review to go into the detail of the theory, but it is necessary for us to briefly refer to the link between this theory and the spectral index of primordial fluctuations, which is an important observational parameter in the Concordance Model of cosmology.

To quantify slow roll, cosmologists typically use two parameters $\epsilon$ and $\eta$ which vanish in the limit that $\phi$ becomes constant. The first parameter is defined as:
\bq
\epsilon \equiv \frac{\dd}{\dd t}\left(\frac{1}{H} \right)=\frac{-\dot{H}}{aH^2}\,,
\eq
which is always positive, since $H$ is always decreasing. The second complementary variable which defines how slowly the field is rolling is:
\bq
\eta \equiv \frac{1}{aH\dot{\phi}^{(0)}}\left[ 3aH\dot{\phi}^{(0)}+a^2V'    \right]\,,
\eq
where $\phi^{(0)}$ is the zero-order field, and $V$ is the potential.

The scalar spectral index can be defined in terms of some function, usually a polynomial, involving the two slow-roll parameters $\epsilon$ and $\eta$. As an example we shall give two such parameterisations:
$ n=1-4\epsilon-2\eta $ \cite{Dodelson:2003} and $n=1-6\epsilon+2\eta$ \cite{Liddle:1992}. The rate of change of $n$ can also be expressed in terms of inflationary parameters: $\dd n/ \dd \ln k=16\epsilon\eta+24\epsilon^2+2\xi^2$ \cite{Kosowsky:1995}, where

 \bq
 \xi^2\equiv m_{\mr{Pl}}^4\frac{V'(\dd ^3/\dd \phi^3)}{V^2}\, ,
 \eq
 $m_\mr{Pl}$ being the reduced Planck mass ($4.342\times 10^{-6}\,\mr{g}$).

Therefore, by extracting the values of $\epsilon$ and $\eta$ from the data, using methods such as weak lensing, we can directly probe the potential of of the inflaton field. Likewise for the tilt or spectral index of the primordial power spectrum. Slow-roll inflation predicts that the spectral index of primordial fluctuations should be slightly less than $1$. The reason for this is simple. For inflation to end, the Hubble parameter $H$ has to change in time. This time-dependence changes the conditions at the time when each fluctuation mode exits the Hubble horizon and therefore gets translated into \mbox{a scale-dependence.}

Inflation accounts for the observed spatial flatness of the universe, and the absence of magnetic monopoles.
These predictions have been confirmed by various probes, most notably by precision measurements of CMB anisotropies,
starting with the Cosmic Background Explorer (COBE) \mbox{in 1992 \cite{Smoot1992,Fixsen1994,Dwek1998},}
then with the Wilkinson Microwave Anisotropy Probe (WMAP) which ran for nine years from 2001 to 2010.
WMAP data placed tight constraints on the predicted burst of growth in the very early universe,
providing compelling evidence that the large-scale fluctuations are slightly more intense than the small-scale ones,
which is a subtle prediction of many inflation models \cite{WMAP1, WMAP1_Inflation, WMAP3, WMAP5, WMAP5_Cosmo, WMAP7, WMAP9}.
Significantly, WMAP found evidence that the scalar spectral index is less than $1$ (a $2\sigma$ deviation),
implying a deviation from scale invariance for the primordial power spectrum.
\mbox{As explained} above, this is a major prediction of inflation, and this observation reinforced the evidence in favour of the theory.
Conclusive proof of a scale-dependent primordial power spectrum
(a $5\sigma$ deviation from $n_{s}=1$) was provided by the \textit{Planck} CMB anisotropy probe in
2013 \cite{Planck-Collaboration:2013aa, Planck2013_inflation} and confirmed \mbox{in 2015
\cite{Planck-Collaboration2015,Planck2015_inflation}.}

One current experimental challenge is to observe the B-modes of polarisation of the CMB
caused by primordial gravitational waves produced by inflation.
Their detection by the BICEP2 experiment was announced in early 2014.
However, more accurate modelling of the signal over the next few months, which allowed
the observation to be explained by polarised dust emission in our Galaxy,
decreased the statistical confidence of the initial result \cite{BICEP2-Collaboration2014}.
This was confirmed by \textit{Planck} data \mbox{in 2016}~\cite{Planck2016_dust}.
Upcoming large-scale structure surveys, such as the \textit{Euclid} satellite mission,
or 21-cm radiation surveys such as the Square Kilometre Array,
may measure the power spectrum with greater precision than current CMB probes,
and could provide further evidence in favour of the inflationary paradigm
\cite{Huang2012,Amendola2013,Namikawa2016}.

\section{The First Unknown Component: Dark Matter}

The first evidence for Dark Matter came from astronomy rather than cosmology. Newtonian~ physics and General Relativity both provide very precise rules for the dynamics of galaxies: the mass determines the rotation velocity. Starting in the 1920s, stronomers noticed that amount of visible matter in galaxies did not match the observed rotation curves. These curves relate the tangential velocity of the constituent stars (or gas) about the centre of the galaxy to their radial distance.  Observations of the velocities of globular clusters about galaxies showed that at large radii the velocities are approximately constant, implying that the amount of mass in the galaxies is much higher than the visible mass.

The first suggestion of the existence of hidden matter, motivated by stellar velocities, was made by Jacobus Kapteyn in 1922 \cite{Kapteyn1922}. Radio astronomy pioneer Jan Oort also hypothesized the existence of Dark Matter in 1932 \cite{Oort1932}. Oort was studying stellar motions in the local galactic neighbourhood and found that the mass in the galactic plane must be greater than what was observed. \mbox{This measurement} was later determined to be erroneous.

In 1933, Fritz Zwicky, who studied galactic clusters while working at the California Institute of Technology, made a similar inference \cite{Zwicky1933}. Zwicky applied the virial theorem to the Coma galaxy cluster and obtained evidence of unseen mass that he called \textit{dunkle Materie} in German, or ``Dark Matter''. Zwicky estimated its mass based on the motions of galaxies near its edge and compared that to an estimate based on its brightness and number of galaxies. He estimated that the cluster had about 400 times more mass than was visually observable. The gravity effect of the visible galaxies was far too small for such fast orbits, thus mass must be hidden from view. Based on these conclusions, Zwicky inferred that some unseen matter provided the mass and associated gravitation attraction to hold the cluster together.

In 1937, Zwicky made the bold assertion that galaxies would be unbound without some form of invisible matter \cite{Zwicky:1937aa}. Zwicky's estimates were off by more than an order of magnitude, mainly due to an obsolete value of the Hubble constant. The same calculation today shows a smaller fraction, using greater values for luminous mass. However, Zwicky did correctly infer that the bulk of the matter was~dark.

More evidence started to accumulate for the existence of some non-emitting component which was now being called Dark Matter \cite{Freese:2009}.
In 1959, Kahn and Woltjer \cite{Kahn1959} pointed out that the motion of Andromeda towards us implied that there must be Dark Matter in our Local Group of galaxies. Dynamical evidence for massive Dark Matter halos around individual galaxies came later, starting in the 1970s, when rotation curve data from multiple galaxies confirmed the Dark Matter halo \mbox{hypothesis \cite{Roberts1973,Einasto1974,Ostriker1974,Rubin1978}.
}
Like baryonic matter, Dark Matter is a fluid with vanishingly small pressure. Unlike baryonic matter, it has no interaction with photons, making it both dark and transparent. \mbox{It also} has a vanishingly small self-interaction beside gravity. One result of this is that the Dark Matter halos surrounding galaxies are rounder than the galaxies themselves \cite{Navarro1996} .

In the last few decades, cosmology has contributed one important piece of information: the amount of Dark Matter. The observed value of the matter density in the universe is ${\OM=0.3089\pm0.0062}$. However, the density of baryonic matter is $\OB=0.0486\pm0.0010$ \cite{Planck-Collaboration2015}. The missing mass is made up of Dark Matter.

The name ``Dark Matter'' is an indication of its nonbaryonic nature: it cannot be observed by emission of photons, so observers need to find a way around this problem. Current evidence for the existence of Dark Matter comes from a variety of sources besides galactic dynamics \cite{Freese:2009}. \mbox{The two} most important ones are CMB anisotropies and gravitational lensing. In addition, Big Bang nucleosynthesis provides evidence that some of the Dark Matter may be baryonic. The inventory of observed baryons in the local universe falls short of the total anticipated abundance from Big Bang nucleosynthesis, implying that most of the baryons in the universe are unseen \cite{Fukugita1998}.

Anisotropies in the CMB are related to anisotropies in the baryonic density field by the Sachs-Wolfe effect \cite{Sachs1967}. This means that the baryon density field variation at the time of decoupling can be linked to CMB anisotropies. If all matter were made of baryons, the amplitude of the density fluctuations should have reached $\delta$$\sim$$10^{-2}$ at the present epoch. However, we observe structures with $\delta \gg 1$ at the present epoch (e.g., galaxies and galaxy clusters). The discrepancy can only be explained by the presence of additional matter, which created potential wells for the baryons to fall into after decoupling. These potential wells would have had to be formed by a weakly interacting fluid that decoupled well before baryons and began to cluster much earlier. Such a fluid would only interact via the gravitational and possibly the weak nuclear force.  As the baryons accumulated in the potential wells, their pressure would have built up, leading to oscillations in the baryon fluid, termed ``baryon acoustic oscillations'' (BAO) \cite{Eisenstein2005,Bassett2010}. These oscillations leave an imprint on the CMB power spectrum, which has been confirmed observationally, and which constrains the mass density, leading to a further confirmation of the existence of this missing mass.

The phenomenon of gravitational lensing includes cosmic shear, weak lensing, cosmic magnification. Although the theory of cosmic shear had been worked out from the 1960s to the  early 1990s \cite{Schneider:2006}, the first detection had to await the development of instruments sensitive enough to make the required observations, and image analysis software to accurately correct for unwanted effects when measuring the shapes of galaxies. In March 2000, four groups independently announced the first discovery of cosmic shear \cite{Bacon:2000, Kaiser:2000,Van-Waerbeke:2000,Wittman:2000}. Since then, cosmic shear has established itself as an important technique in observational cosmology.

Gravitational lensing shows that the amount of lensing of galaxies around galaxy clusters is too high to be caused by the visible matter. Apart from the stars themselves, a galaxy cluster also has a gas component, but X-ray observations show that this is still not enough to account for the extra mass. The~cluster must therefore have a non-emitting halo of Dark Matter around it.

Various Dark Matter candidates have been proposed \cite{Bertone2005,Salati:2009,Garrett2011,Kappl2011,Arina2011}. However, all these candidates have one common characteristic: a very small reaction cross-section, making them extremely difficult to detect directly \cite{,Porter2011,Calore2012,Klasen2015,Mayet2016}. \scalebox{.95}[1.0]{Experiments have, however, placed limits on the mass of Weakly Interacting} Massive Particles (WIMPs), which are the current  best candidate for Dark Matter (together with axions). WIMPs are an entire new class of fundamental particle outside of the Standard Model that result from supersymmetry \cite{Jungman1996,Aartsen2013}. These results show that even the lightest Dark Matter particle should have a mass which is not below $\sim$$10$~MeV. We also know that ${\Omega_\mr{CDM}=0.2589 \pm 0.0057}$ \cite{Planck-Collaboration2015}. The conclusion is that $\Onu\ll\OM$, implying that hot Dark Matter (i.e., neutrinos) cannot account for the Dark Matter density $\Omega_\mr{CDM}$.

An alternative to Dark Matter is to explain the missing mass by means of a modification of gravity at large distances or more specifically at small accelerations. In 1983, Morderhai Milgrom proposed a phenomenological modification of Newton's law which fits galaxy rotation \mbox{curves \cite{Milgrom:1983,Milgrom:1983a,Milgrom:1983b}.} \mbox{The theory,} known as Modified Newtonian Dynamics (MOND) automatically recovers the Tully-Fischer law. The theory modifies the acceleration of a particle below a small acceleration $a_{0}$$\sim$$10^{-10}\, \mr{ms^{-2}}$, which therefore enters \scalebox{.95}[1.0]{the theory as a universal constant. The gravitational} acceleration at large distances then reads $a=\sqrt{GMa_{0}}/r$ at large distances, instead of the Newtonian law $a=\sqrt{GM/r^{2}}$.

There are two main difficulties with MOND. First, it does not explain how galaxy clusters can be bound without the presence of some hidden mass \cite{Aguirre2001,Pointecouteau2005}.
Second, attempts to derive MOND from a consistent relativistic field theory have failed.
One such attempt is the Tensor-Vector-Scalar Theory (TeVeS) \cite{Bekenstein2004} is more successful and actually relativistic but not apparently necessary since it still requires dark energy and Dark Matter.
Many models are unstable \cite{Contaldi2008}, or require actions which depend on the mass $M$ of the galaxy, thereby giving a different theory for each galaxy. Moreover, modified gravity theories have serious difficulties reproducing the CMB power spectrum and the evolution of large-scale structure \cite{Lue2004a,Dodelson2011,PhysRevD.86.083507}.

The greatest challenge to modified gravity theories, and also the clearest direct evidence of Dark Matter, comes from observations of a pair of \scalebox{.95}[1.0]{colliding galaxy clusters known as the Bullet Cluster \cite{Clowe2006}} in which the stars and Dark Matter separate from the substantial mass of ionised gas. The Dark Matter follows the less substantial stars and not the more massive gas.

The modifications of gravity proposed as alternative to the Dark Matter paradigm illustrate the need for tests of GR at large distances and low accelerations. They also illustrate the problems faced by models which favour goodness of fit over parsimony. Modified gravity theories can give an excellent phenomenological fit through an adjustment of the values of the extra parameters, but there is no universal principle to determine these values. This requirement for simplicity and predictivity is met by General Relativity.

\section{The Second Unknown: Dark Energy and the New Wave Alternative Theories}
\label{section_DE}

The current motivation for alternative theories seems to be the search for an explanation of the observed accelerating expansion of the universe.
Let us consider the justification for the dark energy paradigm within the inflationary $\Lambda\mr{CDM}$ model, and the process which led to its acceptance by the scientific community (see \cite{Peebles2003}).

Round about the time that GR was developed, the universe was thought to be static.
There was no compelling reason to think otherwise. Einstein realised that his equations implied a non-static universe, so in 1917 he revised his field equations of GR to read \cite{Einstein:1917}:

\bq G_{\mu\nu}  - \Lambda g_{\mu\nu}  = G_{\mu\nu} -8\pi G \rho_\Lambda g_{\mu\nu} =8\pi G T_{\mu\nu}
\eq
where $\rho_\Lambda = \Lambda/8 \pi G$ is proportional to the cosmological constant $\Lambda$.

It can be seen from this equation that Einstein did not consider the cosmological constant to be part of the stress-energy term.
One could, of course, put $\rho_\Lambda g_{\mu\nu}$ on the right-hand side of the equation and count it as part of the source term of the stress-energy tensor and simply consider  $\rho_\Lambda$ to be the vacuum energy.
This is not just a semantic distinction.
When $\rho_\Lambda$ takes part in the dynamics of the universe,
then the field equation is properly written with $\rho_\Lambda$, or its generalisation, as part of the stress-energy tensor:
\bq G_{\mu\nu}=8\pi G(T_{\mu\nu}+\rho_\Lambda g_{\mu\nu})\,.\eq

The equation describing gravity is then unchanged from its original form---there is no new physical theory. Instead, there is a new component in the content of the universe.

This component must satisfy Special Relativity (that is, an observer can choose coordinates so that the metric tensor has Minkowskian form). An observer moving in spacetime in such a way that the universe is observed to be homogeneous and isotropic would measure the stress-energy tensor \mbox{to be }
\bq   T_{\mu\nu}=
   \begin{pmatrix} % or pmatrix or bmatrix or Bmatrix or ...
      \rho & 0 & 0& 0 \\
      0 & p &0 &0 \\
      0&0&p&0\\
      0&0&0&p
   \end{pmatrix}.
   \eq

This means that the new component in the stress-energy tensor looks like an ideal fluid with negative pressure:
\bq p_\Lambda=-\rho_\Lambda .\eq

In modern concordance cosmology, this component is usually termed ``dark energy''. If the equation of state parameter of dark energy is constant, i.e., $w(z)=-1$, then its energy density will be constant regardless of the expansion of the universe.

Einstein inserted the cosmological constant because he felt that the non-static universe predicted by the formalism of GR was incorrect, given the data available in 1917 \cite{Einstein:1917}. At the time, observations of the universe were limited primarily to stars in our own galaxy, with observed low velocities, so there was solid observational evidence justifying the assumption that the universe was static. Einstein's goal was to obtain a universe that satisfied Mach's principle of the relativity of inertia. However, observational evidence started to accumulate for another paradigm. In 1917, Vesto Slipher \cite{Slipher:1917} published his measurements of the spectra of spiral nebulae, which showed that most were shifted towards the red. The breakthrough came when the linear redshift-distance relation was formulated by Hubble~\cite{Hubble29}, who showed that the universe was expanding. Einstein then dropped his support for the cosmological constant.

In the FLRW cosmological model, the expansion history of the universe is determined by the mass density of the different components, whose sum is normalised to unity:
\bq \label{unity}   \Omega_{m,0} + \Omega_{\mr{rad,0}} + \Omega_{X,0} +\Omega_{k,0} = 1\,,
\eq
where the 0 subscript indicates the present epoch. We use the term $\Omega_X$ to show that this equation does not assume anything about the nature of the additional energy component (dark energy). In fact we could have used $\OL$ or $\OD$ in the current Concordance Model.

Big bang nucleosynthesis and observations of large scale structure provide a good determination of the mass content of the universe, allowing $\OM$ and $\OR$ to be fixed. However, observations in \mbox{the 1980s} and 1990s started to show inconsistencies with the cosmological model at the time, which was a matter-dominated, expanding universe with a present-epoch Hubble constant of $H_0\simeq~ 0.7\, \mr {kms}^{-1}~\mr{Mpc}$ and $\OL=0$ \cite{Paal1992,Krauss1998}. This was the so-called ``age problem'', where the predicted age of the universe seemed to be younger than the age of the oldest stars.  Angular-diameter distances to the last scattering surface at $z=1100$ measured from the CMB are in fact $1.7$ times smaller than those predicted by an isotropic and homogeneous universe containing only pressureless matter  (see~\cite{Rasanen:2009}).  Since the inflationary scenario, which by then was well established, predicts a flat $\Omega_\mr{total}=1$ universe, there was a problem with the cosmological model.

It was realised that one of the three assumptions of the cosmological model had to be wrong. Either the universe contains exotic matter with a negative pressure, or standard General Relativity is wrong, or the universe is not homogeneous and isotropic. The solution could also lie in some combination of the three. Most of the research since the late 1990s has followed the first approach, and the term ``Concordance Model'' refers to an FLRW universe, following General Relativistic cosmology, containing dark energy.

Within the FLRW framework, two main proposals were put forward: one was $\Lambda\mr{CDM}$, in which there is a contribution to the energy density from a term similar to the cosmological constant (or the cosmological constant itself), and the other was $\nu+\mr{CDM}$, where the missing mass came from massive neutrinos ($m_\nu\simeq 7\mr{eV}$) (e.g., \cite[][]{Blasone2004,Capolupo2007}).

The first strong evidence of dark energy came in 1998 and 1999, when observations of the luminosities of type Ia supernovae indicated that the \scalebox{.95}[1.0]{expansion of the universe is accelerating \cite{Riess:1998aa, Perlmutter:1999}.} Concurrently, other observations constrained the neutrino mass to $m_\nu\ll 7\mr{eV}$, thus discounting the $\nu+\mr{CDM}$ model and confirming $\Lambda\mr{CDM}$ as the Concordance Model (e.g., \cite[][]{Sahni2000,Peebles2003}). It is not clear when the term ``dark energy''  was first used, but it seems to have been around 1998. The term is analogous to ``Dark Matter'', which had been in use for some time \cite{Turner:1999}.

Since then, numerous observations have confirmed cosmic acceleration, including supernovae, the cosmic microwave background, large-scale structure and baryon acoustic oscillations \mbox{(see, e.g., \cite[][]{Eisenstein2005a,Frieman2008,Astier2012}).} The values of the present epoch matter and radiation components are \mbox{well established:}
\bq
 \Omega_{m,0}\equiv \frac{8\pi G \rho_{m,0}}{3H_0^2} \sim 0.3 \quad, \quad\Omega_{\mr{rad},0}\equiv\frac{8\pi G\rho_{\mr{rad},0}}{3H_0^2}\sim 1\times 10^{-4}\,,
 \eq
   where $H_0$ is the present value of the Hubble parameter $H(a=1)$.

The data also indicate that the universe is currently nearly spatially flat:
\bq
|\Omega_K|\ll 1\, .
\eq
This is normally taken to imply that the spatial curvature $K=0$, since
\bq
\Omega_{k,0}=0\equiv\frac{-K}{a_0^2H_0^2}\sim 0\, .
\eq

Thus it also justifies the inflationary paradigm. However, inflation only tells us that ${\Omega_K\rightarrow 0}$, so~that the curvature may have had a nonzero value in the past. In the present universe, however,
the~distinction is negligible. In any case, Equation (\ref{unity}) implies that there has to be a nonzero $\Lambda$ (a~constant term added to the Einstein equation) such that
\bq
 \Omega_{\Lambda,0}\equiv \frac{\Lambda}{3H_0^2}\sim 0.7\, .
 \eq

  Inserting these values into the Friedmann equation leads to the dramatic conclusion that the expansion of the universe is accelerating:
   \bq
 \ddot{a}_0=H_0^2\left(\OL -\frac{1}{2}\OM-\OR    \right)>0\, ,
 \eq
  where $a_0$ is the present value of the scale factor $a(t)$.

Note that this conclusion only holds if the universe is homogeneous and isotropic (i.e., \mbox{a Friedmann-Lema\^{i}tre model).} In such a universe, the distance to a given redshift $z$ and the time elapsed since that redshift are tightly related via the only free function, $a(t)$. If the universe is isotropic around us, but not homogeneous, that is, a non-Copernican Tolman-Bondi-Lema\^itre model \cite{Enqvist2008}, then this relation would be lost and present data might not imply acceleration. A Copernican model where this relation again breaks down is the inhomogeneous universe, where the acceleration can be produced via nonlinear averaging---the backreaction of inhomogeneities.

Dark energy is a fluid component whose equation of state is:
\bq
p_\mr{DE}=wc^2\rho_\mr{DE}\,.
\eq

This is the equation of state in its most general  form, since $w$ can be any function of redshift, scale factor or cosmic time, with the constraint that $w\le 0$ (i.e., the fluid has a negative pressure). Assuming that $w=w(a)$, we have the following density-scale relation:
\bq
\rho_\DE(a)=\rho_{\DE,0}\e^{-3\int_a^1[1+w(a')]\dd (\ln a')}\,.
\eq

It can be seen that in the special case of a constant $w=-1$, the fluid equation implies that the density is constant.

If dark energy is a cosmological constant, it still leaves the question of its physical nature. \mbox{Its observed} value of $\Lambda \approx 3 \times 10^{-122}c^{3}/\hbar G$ \scalebox{.95}[1.0]{is so small that it is hard to interpret as the vacuum energy. }

One possibility is that the final value of the cosmological constant is zero, and that cosmic acceleration is due to \scalebox{.95}[1.0]{the potential energy of a scalar field, with some sort of mechanism to dynamically} relax it to a small value. This notion leads to models of dark energy which
invoke a slowly-rolling cosmological scalar field to source accelerated expansion, smilar to cosmological inflation.

How can a scalar field drive cosmic acceleration? The action of a scalar field minimally coupled to Einstein gravity is
\bq
S = \int\dd^4x\sqrt{-g}\left(\frac{m_{\mr{Pl}}^2R}{2}-\frac{1}{2}(\partial\phi)^2-V(\phi)\right)~.
\eq

The stress-energy tensor for the scalar field is given by
\bq
T_{\mu\nu}^\phi = \partial_\mu\phi\partial_\nu\phi-g_{\mu\nu}\left(\frac{1}{2}(\partial\phi)^2+V(\phi)\right)~.
\label{scalarTmunu}
\eq

For a homogeneous field such that $\phi = \phi(t)$, a cosmological scalar acts like a perfect fluid with equation of state $w = P/\rho$ given by
\bq
w_\phi = \frac{\frac{1}{2}\dot\phi^2-V(\phi)}{\frac{1}{2}\dot\phi^2+V(\phi)}\, .
\eq

The observed expansion leads us to a value of $w_\phi \simeq -1$, which requires a very slowly-rolling field: $\dot\phi^2\ll V(\phi)$. There has also been a lot of interest in constructing quintessence models which can produce an equation of state of the ``phantom'' type ($w_\phi <-1$).

The equation of state of dark energy that has the potential to distinguish between dark energy candidates. The most important distinction that can be made between different dark energy models  is whether the energy density of this component is constant, filling space homogeneously, or whether it is some form of quintessence field whose energy density can vary  in time and space. There is a multitude of alternative models, such as $f(R)$ \cite{Sotiriou2010,Fr2010,Capozziello2011}, or Chameleon Models  \cite{Khoury2004,Durrer:2008a}. It is therefore useful to consider the redshift evolution of $w$, so that $w$ is an arbitrary function of redshift $z$. There are a number of different parameterisations of $w(z)$ (\cite{Linder:2003}, and references therein).

The most common parameterisation is sometimes termed the Chevallier-Linder-Polarski (or CPL) parameterisation \cite{ChevPol2001,Linder:2003}:
\bq w(z)=w_0- \frac{\dd w}{\dd a}(1-a)\,,\eq
where the scale factor $a=1/(1+z)$.
If we define
\bq w_a=-\frac{\dd w}{a \dd \ln a}\,,
\eq
then the equation becomes
\bq\label{Linder_DE} w(a)=w_0+w_a(1-a)\,,
\eq
which is the most commonly used form.

It has been shown that this parameterisation is stable and robust over large redshift ranges \cite{Linden2008}.  A  wide range of functional forms of $w(a)$ can be parameterised by the $w_0-w_a$ combination. However, there are some dark energy models which it cannot reproduce (see \cite[][]{Wang:2006,Johri:2007,Avelino:2006,Scherrer2015}).

The problem with the dark energy paradigm, stated simply, is that the parameters are not constrained well enough to rule out certain models. We have fairly good bounds on the dark energy density, but the dark energy equation of state is still poorly constrained. Even for a constant $w$ model, corresponding to $\LCDM$, the bounds are such that a time-varying $w(a)$ could mimic a constant $w$, thereby disguising underlying physics (see \cite[][]{Chongchitnan2010,Astashenok2013,Debono2014}).

\section{The Evolution of Large-Scale Structure}

After the epoch of matter-radiation equality, and before the onset of dark energy domination, the mass-energy content of the universe became dominated by matter. From an initially smooth background (as evidenced by CMB observations), structures have evolved to a scale of more than $100\,\mr{Mpc}$, with the term ``large scale structure'' being used to refer to objects modelled on this scale. \mbox{At this scale,} the mass-energy inhomogeneities can be modelled as perturbations on a homogeneous and isotropic unperturbed background spacetime. Below this scale we observe galaxy clusters, individual galaxies, and stars. The model of structure formation must be accurate enough to provide an good description of the universe on a wide range of scales.

The standard model for the formation of structure assumes that at some early time there existed small fluctuations, which grew by gravitational instability. The origins of these fluctuations are unclear, but they are thought to arise from quantum fluctuations of the primordial universe, uncorrelated and with Gaussian amplitudes, which were then amplified during a later inflationary phase \cite{Inflation_textbook}.
The~assumption that the amplitudes of the relative density contrasts is much smaller than unity means that we can think of the primordial fluctuations as small perturbations on a homogeneous and isotropic background density. This ensures that we can describe them using \mbox{linear theory.}

Heuristically, the mechanism of structure formation can be understood in terms of gravitational self-collapse. Matter collapses gravitationally around initial mass overdensities. This increases the relative density of that region, causing further collapse of more matter, and amplifying the effect. The~linear theory of structure formation needs to be relativistic, because the perturbations on any length scale are comparable or larger than the horizon size at sufficiently early times. The horizon size is defined as the distance $ct$ which light can travel in time $t$ since the Big Bang.  Dissipative effects and pressure also affect structure formation, as explained below (for details of the theory, see \cite[][]{Bardeen:1980, Lifshitz:1946}).

The relative density is the density $\rho$ at a particular point in space $\vect{x}$ relative to the mean $\ol{\rho}$ at some time parameterised by the scale factor $a$, and can be expressed as a dimensionless density contrast:
\bq \delta(\bs{\mr{x}},a)=\frac{\rho(\bs{\mr{x}},a)-\ol{\rho}(a)}{\ol{\rho}(a)} \,.
\eq

This quantity can be understood as the dimensionless density perturbation of some background matter distribution.

There are two types of density perturbations that can occur within a matter-radiation fluid. If the fluid could be compressed adiabatically in space,  the perturbations have a constant matter-to-radiation ratio everywhere. Since the energy density of radiation is proportional to $T^4$, and the number density is proportional to $T^3$, the energy densities of radiation and matter are \mbox{related by:}
\bq
\delta_\mr{rad}=\frac{4}{3}\delta_\mr{m}\,.
\eq

Isocurvature perturbations occur when the entropy density is perturbed, but not the energy density. Since the total energy density remains constant, \scalebox{.95}[1.0]{there is no change in the spatial curvature and}
\bq
\rho_\mr{rad}\delta_\mr{rad}=\rho_m\delta_\mr{m}\,.
\eq

Perturbations can occur at different scales, or `modes'. The latter term is used when the amount of perturbation on a particular scale is expressed using Fourier analysis. The Fourier transform pair of $\delta(\vect{x})$ is:
\begin{align}
\hat{\delta}(\vect{k} )&=\int \dd^3 x\delta(\vect{x})\e^{i\vect{k}.\vect{x}}\quad;\notag\\
\delta(\vect{x} )&=\int \frac{\dd^3 k}{(2\pi)^3}\hat{\delta}(\vect{k})\e^{-i\vect{k}.\vect{x}},
\end{align}
with each mode assumed to evolve independently. In the Einstein-de Sitter regime, linear adiabatic perturbations scale with time as follows:
\bq
\delta\propto
\begin{cases}
a(t)^2&\text{(radiation domination)}\\
a(t)&\text{(matter domination)}
\end{cases}
\eq
while isocurvature perturbations are initially constant and then decline:
\bq
\delta\propto
\begin{cases}
\text{constant}&\text{(radiation domination)}\\
a(t)^{-1}&\text{(matter domination)}\,.
\end{cases}
\eq

In both cases, the overall shape of the spectrum of the perturbations over all modes is preserved, while the amplitude changes with time.  The evolution described above is affected on small scales by a number of processes.

\subsection{Evolution on Small Scales}
During the \scalebox{.95}[1.0]{radiation-dominated epoch the growth of certain modes is suppressed. This behaviour} can be modelled in terms of the horizon scale $\lambda_H(a)$, which is the distance $ct$ that light could have travelled since the initial singularity (a comoving horizon size). A mode $k$ is said to enter the horizon when  $\lambda=\lambda_H(a)$, where $\lambda=(2\pi)/k$. If $\lambda<\lambda_H(a_\mr{eq})$ then a mode enters the horizon during the radiation-dominated epoch. The time scale for collapse of matter during this epoch is larger than the typical expansion time scale ($t\sim 1/H(a)$) due to the relatively rapid expansion $\rho_\mr{rad} \propto a^{-4}$. \mbox{The growth} of these modes is therefore suppressed. After the epoch of matter-radiation equality ($a=a_\mr{eq}$), these perturbations can then start to collapse gravitationally. We can define the suppression factor for a particular mode as the factor by which the amplitude is reduced had it not entered \mbox{the horizon:}
\bq\label{power_supression}
f_\mr{sup}=\left(\frac{a_\mr{enter}}{a_\mr{eq}}   \right)^2=\left(\frac{k_0}{k}     \right)^2
\eq
where the mode evolves as $\propto a^2$ until it enters the horizon at $a_\mr{enter}$ and is suppressed until $a_\mr{eq}$, when its evolution resumes as $\propto a$. The second equality in the above equation comes from applying an Einstein-de Sitter approximation where $k_0=1/\lambda_H(a_\mr{eq})$ (see \cite[][]{Bartelmann:2001}).

Pressure opposes gravitational collapse for modes with a wavelength less than the Jeans \mbox{length \cite{Jeans1},} sometimes called the free-streaming scale, defined as
\bq
\lambda_J=c_s\sqrt{\frac{\pi}{G\rho}}\,.
\eq

During the radiation-dominated epoch, the sound speed $c_s=c/\sqrt{3}$ and the Jeans length is always close to the horizon size. The Jeans length then reaches a maximum at $a=a_\mr{eq}$ and then begins to decrease as the sound speed declines. This means that on scales larger than the comoving horizon size, perturbations are only affected by gravity, and the spectrum starts to turn over at this point (where the effects of pressure begin to dominate). The comoving horizon size at $z_\mr{eq}$ is given by:
\bq R_0r_H(z_\mr{eq})\approx\frac{16.0}{\OM h^2}\mr{Mpc}\,.\eq

Another important scale occurs where photon diffusion erases perturbations in the matter-radiation fluid. This process is termed Silk damping \cite{Silk1968}. The scale at which it occurs is characterised by the distance travelled by the photon in a random walk by the time of last scattering:
\bq
\lambda_S\approx 16.3(1+z)^{-5/4}( \OB^2 \OM h^6)^{-1/4}\mr{Gpc}\,.
\eq

All of the effects mentioned above are particularly important where the behaviour of massive neutrinos is concerned. Heuristically we can understand the complexity of their behaviour by considering them as a component whose equation of state changes as the universe evolves. From a component which behaves like photons (since the particles have a very small mass and relativistic speeds), massive neutrinos lose energy and start behaving like baryonic matter \cite{Bond:1980,Silk2012}.

\subsection{Growth oF Perturbations in the Presence of Dark Energy}

All of the above effects were described in an Einstein-de Sitter universe. In a universe with a smooth non-clustering dark energy component below the horizon scale, the matter perturbation fields evolves according to:
\begin{align}
\ddot{\delta}+2H\dot{\delta}-(3/2)H^2\OM\delta & =0 \notag\\
\delta''+(2-q)a^{-1}\delta^{\prime}-(3/2)\OM a^{-2}\delta&=0,
\end{align}
where a dot denotes a time derivative and a dash denotes a derivative with respect to $a$. The term $q$ is the deceleration parameter. This can be interpreted in the following way: the perturbations grow according to a source term which involves the amount of matter ($\OM$) but the growth is suppressed by the friction term due to the expansion of the universe. The latter is also known as the Hubble drag.

If we define the growth as the ratio of the amplitude of a perturbation at a time $a$ to some initial amplitude, i.e.,
\bq
D(a)=\frac{\delta(a)}{\delta(a_\mr{initial})}\,,
\eq
the equation becomes, for a general dark energy scenario where $w=w(a)$  (see \cite[][]{LinderJenkins:2003})
\bq
D''+\frac{3}{2}\left( 1-\frac{w(a)}{1+X(a)}    \right)\frac{D'(a)}{a}-\frac{3}{2}\left(\frac{X(a)}{1+X(a)}\right)\frac{D}{a^2}=0\,,
\eq
where
\bq
X(a)=\frac{\OM}{\OD}\e^{-3\int_a^1\dd \ln a'w(a')}
\eq
is the ratio of the matter density to the dark energy density. For large $X$ (i.e., ${\OM\sim 1}$ where \linebreak$\OD\sim 1-\OM$) we recover the matter-dominated behaviour ($D\sim a$). To parameterise deviations from this behaviour we define the ``normalised growth'' as ${G=D/a}$. The evolution equation is then:
\bq
G''+\left[\frac{7}{2}-\frac{3}{2}\left(\frac{w(a)}{1+X(a)}\right)              \right]\frac{G''}{a}+\frac{3}{2}\left(\frac{1-w(a)}{1+X(a)}\right)\frac{G}{a^2}=0\,.
\eq

This equation allows us to physically interpret the effects of dark energy. In the presence of dark energy, the Hubble drag term is increased, so that growth is suppressed in a universe with an accelerating expansion. This is similar to the suppression due to radiation dominance.
\subsection{The Power Spectrum of Matter}
\label{matter_power_spectrum}

In an FLRW universe, the homogeneity and isotropy assumption means that any statistical properties must also be homogeneous and isotropic. The implication for the matter perturbation field is that its Fourier modes must be uncorrelated (due to homogeneity). Usually, we assume that the mode amplitudes are Gaussian. This assumption is well motivated since the theory for the seed fluctuations assumes that they have a quantum origin. Due to the central limit theorem, the sum of a sufficiently large number of mode amplitudes will tend towards a Gaussian distribution \mbox{(see, e.g., \cite[][]{Bernardeau2002,Percival2005}).}

Such a field, with uncorrelated modes, and a Gaussian distribution of mode amplitudes is called a Gaussian random field, and can be entirely described by its two-point correlation function:
\bq\label{Gaussian_random_field}
\expec{\delta(\vect{x})\delta^*(\vect{y})   }=C_{\delta\delta}(|\vect{x}-\vect{y}|)\,.
\eq

The angled brackets denote an ensemble average (an average over a multitude of realisations). The value of $\delta$ at a given point in the universe will have a different value in each realisation, with a variance $\expec{\delta^2}$.
Since we can only observe one realisation of our universe (in other words, at most only a finite region in this one universe), we apply the ergodic principle: The average over a sufficiently large volume is equal to the ensemble average.

In Fourier space, the correlation function can be written as:
\bq \expec{\hat{\delta}(\vect{k})\hat{\delta}^\ast(\vect{k}^{\prime})} =\int {\dd}^3 x\mr{e}^{i\vect{k}.\vect{x}}\int {\dd}^3 x^{\prime}\mr{e}^{-i\vect{k}^{\prime}.\vect{x}^{\prime}}\expec{\delta (\vect{x})\delta^\ast(\vect{x}^{\prime})} .  \eq

Replacing $\vect{x}^\prime=\vect{x}+\vect{y}$, and substituting Equation \eqref{Gaussian_random_field}, this can be written as:
\begin{align}
\expec{\hat{\delta}(\vect{k})\hat{\delta}^\ast(\vect{k}^{\prime})} &=\int {\dd}^3 x\mr{e}^{i\vect{k}.\vect{x}}\int \dd^3 y\mr{e}^{-i\vect{k}^\prime.(\vect{x}+\vect{y})}C_{\delta\delta}(|\vect{y}|)\\
&=(2\pi)^3\delta_D(\vect{k}-\vect{k}^{\prime}) \int \dd^3 y \mr{e}^{-i\vect{k}.(\vect{y}) C_{\delta\delta}(|\vect{y}|)}\\
&=(2\pi)^3\delta_D(\vect{k}-\vect{k}^{\prime})P_\delta(|\vect{k}|) \, .
\end{align}

The power spectrum has been defined as the Fourier transform of the correlation function:
\bq  P_\delta(|\vect{k}| )= \int \dd^3 y\mr{e}^{i\vect{k}.(\vect{y})}C_{\delta\delta}(| \vect{y} |)\, .  \eq

The standard convention in cosmology is to abbreviate $P_\delta(|\vect{k}|)$ to $P(k)$, where $k=|\vect{k}|$. The power spectrum can be expressed in dimensionless form as the variance per $\ln k$, so that:

\bq \Delta^2(k)=\frac{{k^3}P(k)}{2\pi^2}\, .\eq

\subsubsection{Nonlinear Evolution}

The power spectrum gives us the evolution of the initial matter density fluctuations. However, the~linear evolution breaks down at small scales, when complex structures begin to form, and~overdensities can no longer treated as perturbations on a smooth background. This is the nonlinear regime of gravitational evolution. The scale above which nonlinearities cannot be ignored is approximately set by ${\Delta(k_\mr{NL})\simeq 1}$, which corresponds to ${k_\mr{NL}\simeq 0.2\,\hpM}$ in most cosmological models.
The standard way to model nonlinear evolution is by using phenomenological fits based on $N$-body simulations.

One strategy is to build a models based on a stable clustering hypothesis \cite{HKLM91,JMW95,Peacock:1996ys}, which assumes that the nonlinear collapsed objects form isolated, virialised systems that are decoupled from the expansion of the universe.

A different approach is used in the halo model \cite{Smith:2003aa}. Here, the density field is decomposed into individual clumps of matter with some density profile and varying mass. By using this model to calculate the number of clumps within a given volume, the galaxy halo profile can be calculated. This is the equivalent of the power spectrum for these matter halos. A functional relation between the linear power spectrum and this halo profile is then derived and calibrated using large $N$-body simulations. This relation is then used to calculate the nonlinear power spectrum, using a fitting formula, for instance (see, e.g., \cite{Seljak:2000,PeacockSmith2000}).

Whichever approach is used, it must account for a range of small-scale physical processes, such as baryonic physics, stellar formation, galactic magnetic fields, and Dark Matter and neutrino properties, which have become more significant with the ability of future astrophysical experiments to probe the nonlinear regime with ever increasing precision
(see, e.g., \cite{Huterer:2005,van-Daalen2011,Semboloni2011,Bird2012,Hearin2012,Takahashi2012,Villaescusa-Navarro2014}).

The need for realistic models of the universe has come to the fore in recent years, due the massive improvement in the quality and volume of cosmological measurements. Most of the $N$-body simulations rely on a perturbative approach. They use an FLRW cosmological background (perfectly homogeneous and isotropic)), and assume that any sub-horizon inhomogeneous structure of the universe will contribute to an average expansion on horizon-sized volumes driven by the horizon-averaged density. With the next generation of cosmological probes, these may not be accurate enough to model the universe realistically. There are several ongoing efforts to build fully relativistic, nonlinear, inhomogeneous and asymmetric models using numerical methods \cite{Bentivegna2016,Mertens2016,Giblin2016}. This is a significant contribution to the study of the backreaction effect and the question of the expansion rate of the universe.

\subsubsection{The Primordial Perturbations}

In the very early universe, the tiny initial perturbations are thought to have formed a Gaussian random field whose covariance function is diagonal and nearly scale-invariant. This form, known as the Harrison-Peebles-Zel'dovich spectrum, was assumed in most cases within the Standard Model, as~it corresponds very closely to the observed power spectrum in the universe.

This type of spectrum was first proposed in the 1970s by Edward Robert Harrison \cite{Harrison:1970}, \mbox{Yakov Zel'dovich \cite{Zeldovich:1972},} and Phillip James Edwin Peebles \cite{Peebles:1970}, who were working independently, as~the spectrum for initial density fluctuations. This hypothesis was subsequently closely borne out by observations.
The defining characteristic of a Harrison-Peebles-Zel'dovich spectrum  spectrum is that it describes a fractal metric, where the degree of perturbation is the same on all scales (hence the term ``scale-invariant''), so that $P(k)\propto k$. If we assume scale invariance for the power spectrum on large scales, and combine this with Equation \eqref{power_supression}, this implies the following general shape for the matter power spectrum in the Einstein-de Sitter scenario:

\bq
P(k)\propto\begin{cases}
k& \text{for}\quad k\ll k_0\\
k^{-3}& \text{for} \quad k\gg k_0\, .
\end{cases}
\eq

The actual form of the spectrum depends in non-trivial ways on the parameters in the cosmological model, including the `slope' of the initial power spectrum $n_s$, where ${P(k)\propto k^{n_s}}$. \mbox{In a scale-invariant} spectrum in the linear regime, the fiducial value of $n_s$ is taken to be 1.

Cosmological observations were consistent with scale invariance up until the early data releases from the \textit{Wilkinson Microwave Anisotropy Probe} (WMAP) \cite{WMAP1, Mukherjee2003}, but showed some tension. Subsequent observations cast further doubts on scale invariance \cite{WMAP3}. In 2013, the \textit{Planck} probe, mapping the anisotropies in the CMB, led to an important and conclusive result: it ruled out scale invariance at over $5\sigma$. The primordial power spectrum was found to be scale-dependent, with $n_{s}=0.9603\pm0.0073$ \cite{Planck2013_inflation}. This was confirmed by the 2015 data release, which found that $n_{s}=0.968\pm0.006$ \cite{Planck2015_inflation}. This significant result is a powerful demonstration of the importance of multiple sources of data, joint observations from different probes, and the increasing reliance on complex statistical techniques for cosmological model selection \cite{Trotta2007,Trotta2007a,Bridges2009,Vazquez2012}.

Thus it can be seen that the current Concordance Model, consisting of $\LCDM$ with an inflationary epoch in the early universe, was built in stages over the last 100 years. It is the result of a process of accumulation of evidence and testing of competing models and theories. At each step, theory provides the basis for adjustments to the model, and observations from different probes provided the evidence (Figure \ref{GR_chart}). The current model should in no way be seen as ``true''. It is merely the best model that fits all the data available so far. Future data may very well require an adjustment to the Concordance~Model.

\begin{figure}[H]
\centering
\includegraphics[width=.52\textwidth]{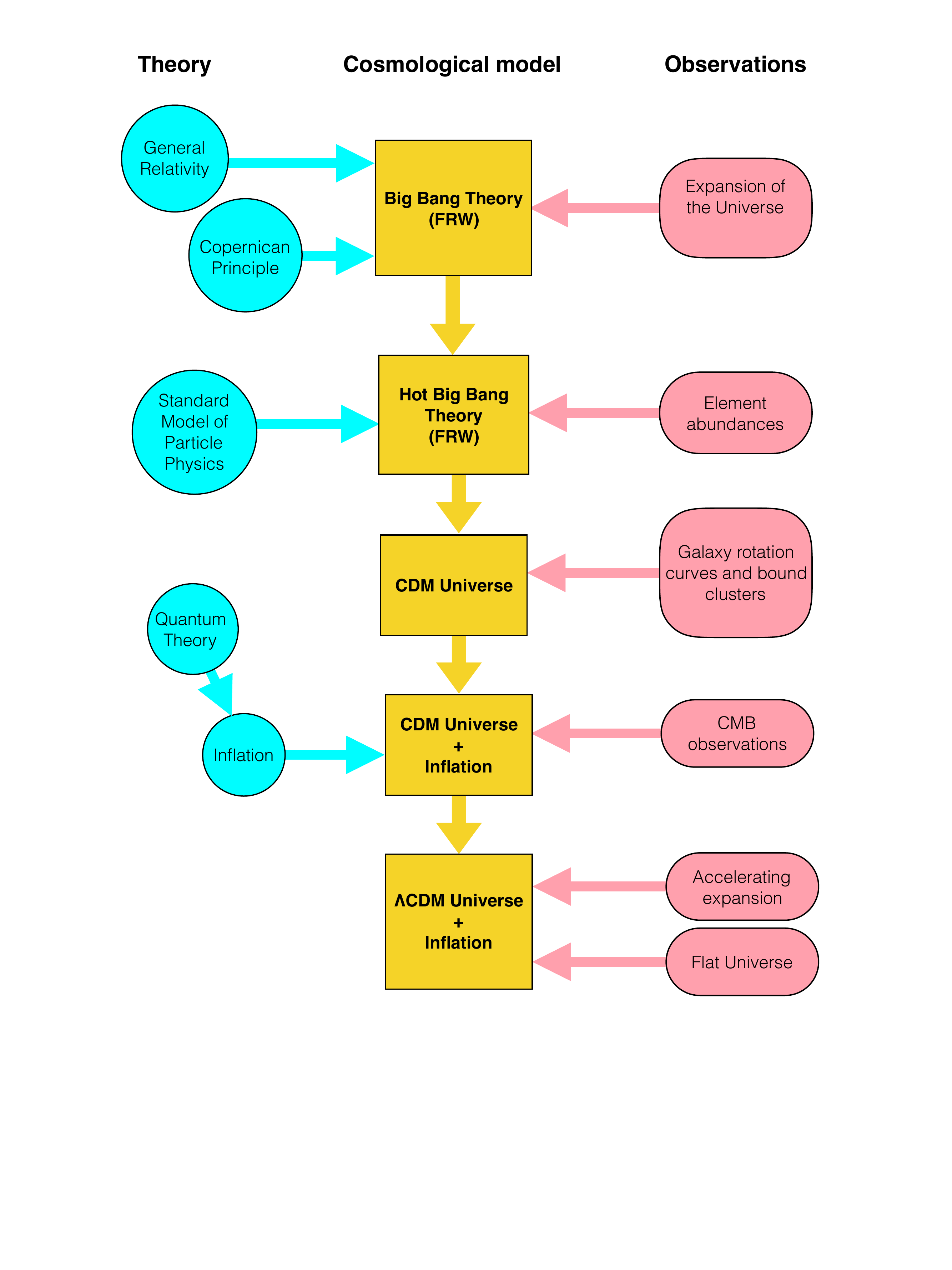}
\caption{How the Concordance Model of Cosmology was developed. Theories and observations motivated the development of cosmological models, which were adjusted as new observations challenged the older models.}
\label{GR_chart}
\end{figure}

\section{How Do We Test General Relativity?}
The assumption of metric coupling given by Equation (\ref{metriccoupling}) has been tested accurately many times over the last 100 years, at scales from $10^{-4}$ m in laboratories \cite{Hoyle2004,Kapner2007}, up to $10^{14}$ m in the Solar \mbox{System \cite{Baker2015}.} These experiment test the implications of the metric coupling: the spacetime independence of non-gravitational constants, the isotropy of the coupling of all matter field to a unique metric tensor, the universality of free-fall, and gravitational redshift.

The standard way to test constraints on $S_{\mr{gravity}}$ and therefore the dynamics of General Relativity is by using the parameterised post-Newtonian (PPN) formalism \cite{Will93}. This assumes that gravity is described by a metric over all scales. The idea is to write the most general form that $g_{\mu \nu}$ can take in the presence of matter, when considering correction of order $1/c^{2}$ with respect to the Newtonian limit. This method was first used in 1923 by Arthur Eddington \cite{Eddington1923}. In its simplest form, it provides us with two phenomenological parameters $\beta^{\mr{PPN}}$
and $\gamma^{\mr{PPN}}$ entering the Schwarzchild metric in \mbox{isotropic coordinates:}
\bq
g_{00}=-1+\frac{2Gm}{rc^{2}}+2\beta^{\mr{PPN}}\left(\frac{2Gm}{rc^{2}}  \right)^{2} \quad \mr{and} \quad g_{ij}=\left(1+2 \gamma^{\mr{PPN}}\frac{2Gm}{rc^{2}}\right)\delta_{ij} \, .
\eq

According to General Relativity, $\beta^{\mr{PPN}} =\gamma^{\mr{PPN}} =1$. The experimental constraints on these parameters are summarised in Table \ref{ppntable}.

\begin{table}[H]
\caption{Some constraints on parameterised post-Newtonian (PPN)
parameters from recent tests.}
\label{ppntable}
\small % Font size can be changed to match table content. Recommend 10 pt.
\centering
\begin{tabular}{lll}
\toprule
\textbf{Method}& \textbf{Constraint} & \textbf{Experiment} \\
\midrule
Shift of perihelion of Mercury &  $|2 \gamma^{\mr{PPN}} - \beta^{\mr{PPN}} -1|<3 \times 10^{-3}$ &Data to 1990 \cite{Shapiro1990}\\
\midrule
Lunar laser ranging & $|4 \beta^{\mr{PPN}} - \gamma^{\mr{PPN}} -3|=(4.4\pm 4.5) \times 10^{-4}$ & Data to 2004 \cite{LLR2004,LLR2009} \\
\midrule
Very long baseline interferometry& $ |\gamma^{\mr{PPN}}-1|=4\times 10^{-4}$ & Data from 1979 to 1999  \cite{Shapiro2004} \\
\midrule
Time-delay variation&  $ \gamma^{\mr{PPN}}-1=(1.2\pm2.3)\times 10^{-5}$  & Cassini spacecraft \cite{Bertotti2003}\\
\midrule
\multirow{2}{*}{Planetary perihelion precessions} & $\beta^{\mr{PPN}} -1  = (-2\pm3)\times 10^{-5}$ &Solar System \\

&  $\gamma^{\mr{PPN}} -1= (4\pm 6)\times 10^{-5}$ &ephemerides to 2013  \cite{Pitjeva2013} \\
\bottomrule
\end{tabular}
\end{table}

It should be noted that the PPN formalism assumes that there is no characteristic length scale for the gravitational interaction, and therefore it does not allow testing of finite-range effects. \mbox{These too have been} constrained to be very close their General Relativistic value of zero. \mbox{The deviation} of amplitude $\alpha$ from a Newton potential on a characteristic scale $\lambda$ is typically $\alpha < 10^{-2}$ on scales ranging from a few millimetres to Solar System size \cite{Hoyle2004}. This implies no deviation from GR over more than 15 orders of magnitude in length scale.

The Solar System tests constrain Eddington's two PPN parameters to a tiny region very close to~$1$. The formalism has been generalized to include eight \scalebox{.95}[1.0]{additional phenomenological parameters \cite{Will93}} to describe any possible deviation from GR at the first post-Newtonian order. They have all been constrained to be very close to their General Relativistic values. The latest data use Solar System ephemerides, which include perihelion measurements for Mercury and the other planets, as well as lunar ephemerides. Independent teams of astronomers have estimated corrections to the standard first-order post-Newtonian General Relativistic formalism to be all statistically compatible with \mbox{zero \cite{Fienga2011,Pitjeva2013}.}  This leads us to conclude that General Relativity is the only theory consistent with Solar System experiments at the post-Newtonian order.

It should be noted that, so far, it is only the first-order post-Newtonian, static,
Schwarzschild-like part of the spacetime metric that has been modelled and tested using
Solar System dynamics. The~first-order post-Newtonian gravitomagnetic or Lense-Thirring part of the spacetime metric has neither been modelled nor tested yet in the Solar System. However, it has recently been pointed out that this could be possible in the next few years  by focussing on particular models (Sun and planets, planets and spacecraft, or planets
and planets) \cite{Iorio2011a}. This would open the field to the possibility of
constraining the PPN parameters using a more complete model of General Relativistic effects.

What about larger or smaller scales? The size of the universe was around $10^{-35}$ m in the beginning, and the present size is around $10^{26}$ m. GR remains untested at these extreme scales. \mbox{But one fundamental} assumption of GR is that it describes gravity at all scales (i.e., the theory assumes the analytical continuity of solutions).

The evolution of the universe itself therefore provides a useful test of General Relativity. The~challenge lies in testing a wide range of potentials (weak and strong field regimes) over cosmological scales. The strong field regime is tested using compact objects. However, over cosmological scales, the kind of objects that would produce gravitational potentials approaching the strong field regime simply do not exist. One solution is to observe the evolution of the universe and check whether the evolution of large-scale structure corresponds to the predictions of General~Relativity.

\section{Cosmological Tests}

General Relativity has been submitted to 100 years of Solar System tests, which it has passed with flying colours. The discoveries of the last two decades: cosmic acceleration, the scale dependence of the primordial power spectrum, and also the open question of Dark Matter, have made it clear that GR must also be tested at astrophysical and cosmological scales.

This presents a serious challenge. We derive our knowledge of the universe from measurements of distances and times, and statistical properties like the distribution of matter. The main difficulty in extending the Solar System tests to cosmological scales is that these measurements depends strongly on the construction of cosmological models.

These models depend on four hypotheses:
\begin{enumerate}[leftmargin=2.3em,labelsep=4mm]
\item[(1)] A theory of gravitational interactions.

\item[(2)] A description of the matter in the universe and the non-gravitational interactions such as electromagnetic emissions.

\item[(3)]A hypothesis on the symmetry.

\item[(4)]A hypothesis on the topology, or the global structure of the universe.
\end{enumerate}

Some of the hypotheses are hard to verify, and some have unverifiable implications. Assuming the symmetry of our solutions means that we also assume the laws of physics are the same throughout the universe, including its unobservable part outside the cosmological horizon (delimited by a radius of around $15.7\,\mr{Gpc}$). This is a very strong assumption, but one that is unverifiable.

Any cosmological model needs all four hypotheses. The first two are the physical theories, but their equations cannot be solved without some kind of assumption on the symmetry of the solutions, given by the third hypothesis. The fourth hypothesis is then an assumption on the global properties of these solutions.

The simplest $\LCDM$ Concordance Model assumes that gravity is described by General Relativity (Hypothesis 1), and that the universe contains the particles and fields of the Standard Model of particle physics \cite{RevPartPhys2014}, together with Dark Matter and a cosmological constant (Hypothesis 2). \mbox{The Einstein} equations require an effective stress-energy tensor averaged out on large scales, so the model requires an extra assumption on the averaging procedure. This averaging and the validity or otherwise of this assumption is the subject of research on the backreaction effect. The model assumes the Copernican Principle (Hypothesis 3), and the continuity of the solutions of Einstein equations across all spatial sections (Hypothesis 4).

The number of alternative cosmological models proposed, especially since the 1990s, is too numerous to list here. Many of them are impossible to rule out simply by fitting them to the data, because their predictions are so close to those of $\LCDM$. It is easier to test the four hypotheses rather than trying to test the observables predicted by the models. As with any null test, a significant violation would indicate the need to modify one or more hypotheses.

As far as tests of GR (the theory of gravity in $\LCDM$) are concerned, many experiments test more than one assumption. We list the main experiment types, and the assumptions they test, in Table \ref{Tableoftests}.

\begin{table}[H]
\caption{What we are actually testing. Experimental tests of General Relativity (GR) often probe more than one assumption, and isolating the effects is a challenge in itself.}
\label{Tableoftests}
\small % Font size can be changed to match table content. Recommend 10 pt.
\centering
\begin{tabular}{ll}
\toprule
\textbf{Experiment}& \textbf{Assumption Tested } \\
\midrule
Solar System tests & Metric coupling \\
Quadrupolar shift of nuclear energy levels & Isotropy \cite{Hughes1960,Drever1961,Allmendinger2014}\\
Lunar laser ranging and orbiting gyroscopes & Universality of freefall \cite{LLR2012} and structure of metric \\
Space-borne clocks & Gravitational redshift \cite{Delva2015}\\
Shift in perihelion of planets & Structure of metric \cite{Iorio2015} \\
Time invariance of physical constants & Metric coupling \cite{Uzan2011} \\
Detection of gravitational waves & Lorenz gauge condition and inhomogeneous wave equation\\
\bottomrule
\end{tabular}
\end{table}

\subsection{Testing the Description of Matter and Non-Gravitational Interactions}

General Relativity can be tested by measuring the fundamental constants of the theory \cite{Flambaum2007,Lea2007,Uzan2011,Rich2015}. Any variation would require a modification of GR \cite{Barrow2002,Barrow2013}. A local measurement of a fundamental constant, such as a determination of the fine-structure constant from the Oklo phenomenon \cite{Fujii2000,Uzan2003,Lamoreaux2004}, is actually a cosmological-scale measurement along the time dimension. Astrophysical probes such as 21-cm radiation \cite{Khatri2007} or the cosmic microwave background \cite{Nakashima2010}, can be used to test the constancy of the fine-structure constant $\alpha$, or to constrain simultaneous variations of $\alpha$ and Newton's gravitational constant $G$ \cite{Martins2010}. A non-constant $G$ would have serious implications both for Newtonian physics and for General Relativity. This motivates ongoing efforts to devise new methods to measure this quantity, and to push the limits of experimental accuracy in order to test the constancy of $G$ \cite{Anderson2015,Pitkin2015,Anderson2015a,Iorio2016a,Feldman2016}.

\subsection{Testing the Assumption of Symmetry}

The assumption of symmetry can be tested by checking for any deviations from isotropy. \mbox{This requires} statistical ensembles of data, so the best observables are the CMB, and, on a smaller scale, large-scale structure \cite{Lahav2002,Hansen2004,Schwarz2015}.

\subsection{Testing the Gravitational Interactions}

The gravitational interactions have been tested many times at laboratory scales and all the way up to Solar System scales. The challenge today is to test them at cosmological scales.

Tests of General Relativity may be placed in three broad categories: laboratory, astrophysical, and cosmological. Experiments cannot span all length scales, and so they cannot test all theories. \mbox{We are} forced to design experiments which can test alternative theories, or the effects of GR, at particular length scales (Table \ref{TableDEtests}).

Laboratory tests probe effects from sub-millimetre scales to a few hundred metres. Galileo's experiments on weights dropped from the Leaning Tower of Pisa \cite{SEGRE1989435} are one example of a laboratory-scale test. The torsion balance experiments of Cavendish and E\"{o}tv\"{o}s \cite{Eotvos1890}, and their modern versions \cite{Gundlach1997,Schlamminger2008,Adelberger2009102,Wagner2012}, are another.

Astrophysical tests probe gravity from Solar System scales all the way up to galactic cluster scales. They include experiments such as laser ranging off the Moon (lunar laser ranging) and several of the Earth's artificial satellites \cite{Dickey1994,Murphy2013,Pearlman2002,Appleby2016}, and proposals to extend laser ranging to other planets in the Solar System  (planetary laser ranging) \cite{Degnan2008,Iorio2011,Dirkx2015,Smith2006,Chen2013}, radar astrometry on near-Earth objects \cite{Margot2009}, observations of the precession of the perihelion as well as higher-order effects for Mercury and other planets~\cite{Fienga2011,Fienga2016,Iorio2016}, observations of spacecraft in the Solar System \cite{Pitjeva2013}, actual or proposed measurements of the acceleration of the Pioneer \cite{Anderson2002,Nieto2005,Turyshev2010} and New Horizons probes \cite{Nieto2008,Iorio2013,Iorio2016d}, observations of the orbits of compact objects such as neutron stars \cite{Damour1992,Lyne2004,Kramer2009},  and measurements of galaxy rotation curves \cite{McGaugh2000,Famaey2012}. In the last decade, observations of extrasolar planetary systems have been proposed and used in order to test GR \cite{Adams2006,Adams2006a,Adams2006b,Iorio2006,Jordan2008,Pal2008,Iorio2011b,Iorio2011c,Iorio2016b,Iorio2016c}. All of these systems are characterised by their spherical symmetry. Many can be approximated by a test object orbiting at a distance $r$ around a central mass $M$. The gravitational field of the more massive central object is then probed by observing the orbit of the test~body.

\begin{table}[H]
\caption{Bridging the length scales to test the cosmological model. Experimental tests of gravity and dark sector couplings, at their typical length scales. Massive gravity (MG) screening mechanisms would show up at short ranges, while smooth dark energy manifests itself at cosmological scales. We give the experimental accuracy from current and future experiments planned over the next decade. The comparison between growth and expansion history comes from combined BAO, supernova, weak lensing, redshift distortion,  cosmic microwave background (CMB) lensing, and cluster data. Lensing effects and dynamical mass comparisons can be carried out over a range of scales: inside galaxies, using strong lensing and stellar velocities, and on cosmological scales, using cross-correlations. This is a route to testing screening effects in alternative theories. Laboratory and Solar System tests can also probe dark sector couplings besides short-range effects, but many of the constraints obtained depend on the cosmological model.}
\label{TableDEtests}
\small % Font size can be changed to match table content. Recommend 10 pt.
\centering
\scalebox{0.85}[0.85]{
\begin{tabular}{llll}
\toprule
\textbf{Test}& \textbf{Length Scale}	& \textbf{Theories Probed} & \textbf{Current Status (and Future)} \\
\midrule
Growth vs. expansion history & $100~\,\mr{Mpc} - 1~\,\mr{Gpc}$ & GR with smooth dark energy & $10\%$ accuracy ($2\%$--$4\%$)\\
Lensing vs. Dynamical mass & $0.01$\,--\,$100\,\mr{Mpc}$ & Test of GR  & $20\%$ accuracy ($5\%$)\\
Astrophysical tests & $0.01~\mr{AU}$\,--\,$1~\,\mr{Mpc}$ & MG screening mechanisms & $\sim$$10\%$ (Up to 10 times improvement)\\
\multirow{2}{*}{Laboratory and Solar System tests} & \multirow{2}{*}{$1~\mr{mm}$\,--\,$1~\mr{AU}$} & \multirow{2}{*}{PPN parameters in MG} & Constraints are model dependent \\
& & & (Up to a tenfold improvement)\\
\bottomrule
\end{tabular}}
\end{table}

For such systems, the deviation of the metric from the Minkowski form is characterised by the magnitude of the Newtonian gravitational potential
\bq
\epsilon = \frac{GM}{rc^{2}} \, .
\eq

The strongest gravitational fields accessible to an observer occur in the limit $\epsilon \rightarrow \mathcal{O}(1)$. In this limit, the central object is a black hole and the test object is close to the event horizon.

We know that the Riemann curvature tensor is an essential quantity in General Relativity. The~approximate magnitude of this tensor is expressed by the Kretschmann scalar (the fully contracted Ricci scalar) for the Schwarzchild metric:
\bq
\xi \propto \frac{GM}{r^{3}c^{2}}\, .
\eq

Note that this expression is more complicated for rotating objects \cite{Henry2000}. For the purposes of our description, however, we may use this simple expression, which we call the curvature.

The third broad category is the cosmological tests. First, the measurements are made on a statistical ensemble. The masses must be treated as power spectra. Second, the gravitational field assigned to each wavenumber $k$ is very weak. The description of the system often depends on the cosmological model. At this scale, the description must take into account the background perturbations and the expansion of the universe.

The position of the various systems on a potential-curvature parameter space is shown in \mbox{Figure \ref{GR_curv_pot}.} The potential accessible to observers is bounded by the line $\epsilon \simeq 1$. There is no limit to the maximum curvature that can be observed except for the Planck limit, where $r\simeq 1.6\times 10^{-33}\, \mr{cm}$. \mbox{This lies} many orders of magnitude above the boundaries of the figure. General Relativity is a complete theory, and does not fail below the Planck scale. To test or even probe GR below the Planck scale is a different matter, and a goal that has not yet been achieved. Many alternative theories of gravity that attempt to explain cosmic acceleration do so through modifications to GR at the Planck scale, so it is important to explore ways in which GR can be tested at these scales.

\begin{figure}[H]
\centering
\includegraphics[width=\textwidth]{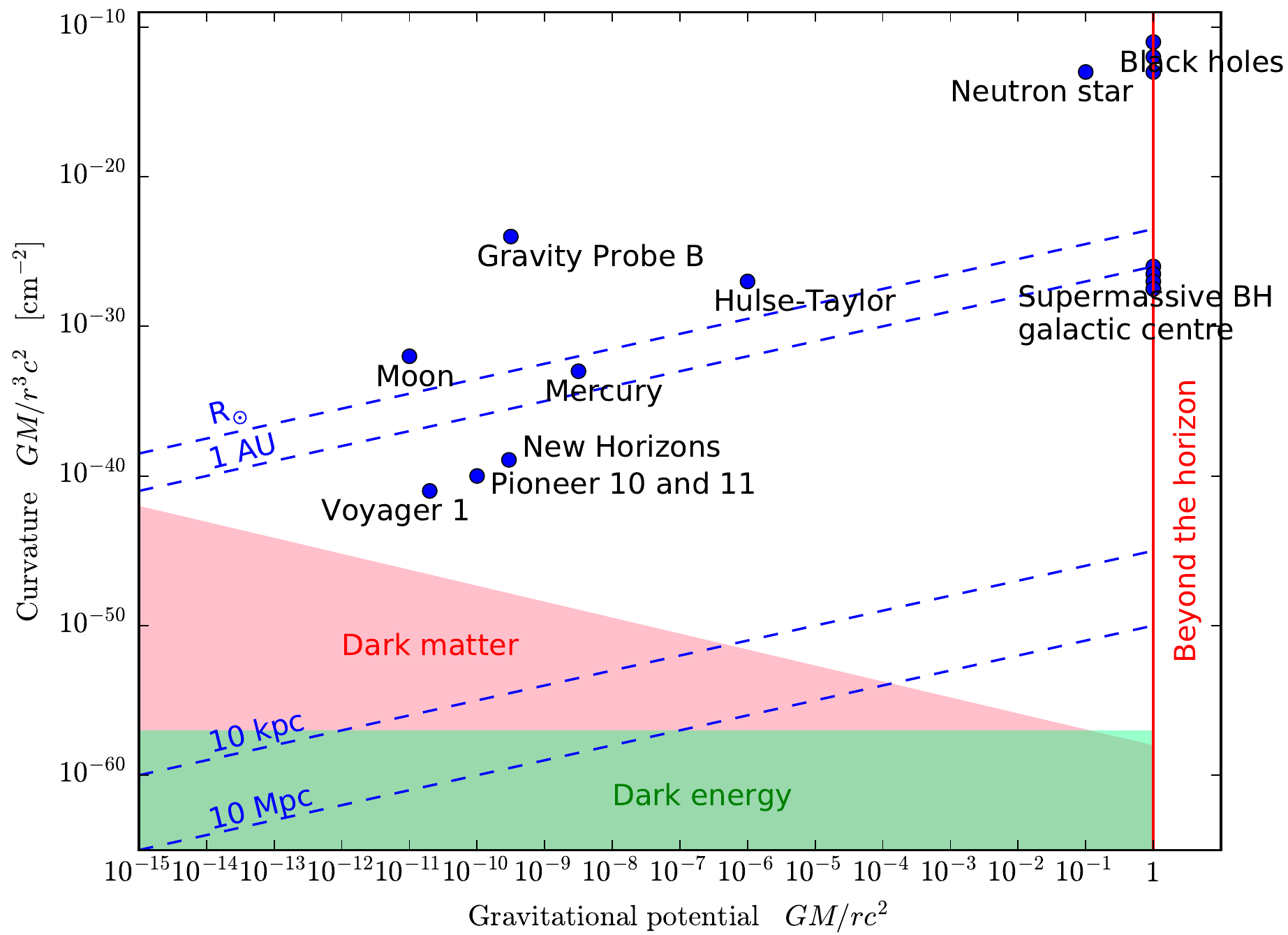}
\caption{The parameter space for quantifying the strength of a gravitational field. The horizontal axis measures the potential. The vertical axis measures the spacetime curvature of the gravitational field at a radius $r$ away from a central object of mass $M$. Regions of this parameter space with potential greater than $1$ represent distances from a gravitating object that are smaller than the event horizon radius and are therefore inaccessible to observers. The red vertical line on the right-hand side of the plot marks the horizon limit. This is a schematic plot, and in no way do we show an exhaustive list of objects and systems that have been used or could be used to test GR. The region of Solar System-scale tests is broadly bounded by the Moon, Gravity probe B \cite{GravityProbeB,GravityProbeB2}, Mercury, and the Pioneer and New Horizons spacecraft. We have included the Voyager spacecraft in the diagram, even though, unlike the Pioneer probe, it was never suitable for tests of GR. The famous Hulse-Taylor binary pulsar \cite{Hulse1975}, although a neutron star binary, is also roughly in the region of Solar System tests. Black holes and neutron stars are in the strong field regime, and the former are at at the limit of the event horizon boundary. Adapted from \cite{Psaltis2008}.}
\label{GR_curv_pot}
\end{figure}

What is the minimum curvature? The unperturbed FLRW metric is isotropic, and the unperturbed Kretschmann scalar is a function of time only. The curvature for the homogeneous universe drops as the universe expands. The present universe has a curvature which is just above the boundary of the region marked `Dark energy', since dark energy is not yet completely dominant over pressureless matter. However, the curvature will approach this limit asymptotically. In this paradigm, \mbox{this represents} a fundamental minimum curvature scale. It is shown in the figure by the region labelled ``Dark energy''.

Galaxies are astrophysical probes of GR. Their innermost regions can be modelled as test particles orbiting a central supermassive black hole. Galactic rotation curves, which can only be explained by introducing Dark Matter, describe galaxy velocities up to the outermost regions. Systems below a constant acceleration scale of approximately $1.2 \times 10^{-10}\, \mr{ms}^{-2}$ cannot be modelled without adding a contribution to the gravitational filed in the form of Dark Matter. This constant acceleration is the diagonal boundary of the region labelled ``Dark matter'' in the figure.

Note that the regions of parameter space occupied by Dark Matter and dark energy overlap---there is a degeneracy between these two unknown components of the cosmological model. However, it is not impossible to distinguish between their effects, since their properties are different. In particular, Dark Matter forms clumps just like baryonic matter, while dark energy does not.

GR has not been tested in the region between \scalebox{.95}[1.0]{curvatures of $\sim$$10^{-40}$ and $\sim$$10^{-50}$. This corresponds} to the region between Solar System scales and cosmological scales probed by galaxy surveys and the CMB. The challenge lies in finding systems which span these scales. In theory they do exist, \mbox{in the form} of galaxies, clusters and superclusters. Their rotation curves transition from Schwarzchild orbits in their innermost regions, to outer regions dominated by Dark Matter. However, our observations are hampered by the fact that we are limited by the resolution of our telescopes, so we can only observe the outer regions. In addition, the untested region is situated between a region where GR is extremely well-constrained (by Solar System tests), and a region where Dark Matter and dark energy have to be invoked, and where we must take the cosmological model into account, because the effect of large-scale structure on background dynamics becomes non-negligible. It has been suggested that this backreaction effect may be at the origin of the observed cosmic acceleration.

Cosmological tests of General Relativity may be broadly classified as follows:

\begin{itemize}[leftmargin=*,labelsep=4mm]

\item{Tests of the consistency between the expansion history and the growth of structure. A discrepancy
in the equation of state parameter of dark energy $w$, inferred from the two approaches can indicate a
breakdown of the GR-based smooth dark energy cosmological paradigm.}

\item{Detailed measurements of the linear growth factor across different scales and redshifts.}

\item{Comparison of the cosmological mass distribution inferred from different probes, especially redshift
space distortions and lensing. }
\end{itemize}

\section{Possible Modifications of GR and Cosmological Implications}
\label{ModificationsSection}

It is useful to identify the regimes in which modifications to GR may appear. This enables us to get a clearer picture of the capabilities and limitations of current and future experiments to tests these alternative theories.

\subsection{Weak and Strong-Field Regimes}

In order to test gravity in the strong-field regime, we need to observe compact objects with a very high density \cite{Psaltis2009,Johannsen2012,Kramer2015}. Black holes are good candidates. Their compactness and mass takes $GM/rc^{2}$ close to the maximum limit of unity. However, they have a serious drawback. Because of the ``no hair theorem'', they are not characterised by any coupling to a scalar field and therefore cannot be used to test for this effect, and to discriminate between scalar-tensor theories and GR.

Neutron stars, on the other hand, are still very compact bodies, but they can be strongly coupled to a hypothetical scalar field. This property has been used to test relativistic parameters by observing the  Hulse-Taylor binary pulsar PSR B1913+16 \cite{Weisberg2003}, and the neutron star--white dwarf binary {PSR~J1141-6545} \cite{Bailes2003}. Gravitational time delay has been \scalebox{.95}[1.0]{tested using other binary pulsars \cite{Stairs2002,Kramer2005,Kramer2006,Desai2016,Huang2016},} and pulsars have also been suggested as ideal probes of the Lense-Thirring effect \cite{Iorio2009,Kehl2016}. General Relativity passes these tests with flying colours.

\subsection{Small and Large Distances}
Distance-dependent modifications can be induced by a massive degree of freedom, which will cause a Yukawa-like coupling. General Relativity is very well constrained on the size of the Solar System, and there are several tests constraining Yukawa \scalebox{.95}[1.0]{interactions at Solar System scales \cite{Iorio2012,Deng2013,Li2014},} but there are no constraints on scales larger than $10 \hpM$, at least without assuming some cosmological model. Some theories put forward to explain cosmic acceleration, such as Chameleon mechanisms~\cite{Khoury2004a}, are essentially modifications of GR at cosmological distance scales.

\subsection{Low and High Accelerations}
Galaxy \scalebox{.95}[1.0]{rotation curves and galaxy dynamics motivated the Dark Matter paradigm. The Tully-Fischer} law \cite{Tully1977} tells us that Dark Matter cannot be explained by a modification of General Relativity at a fixed distance. MOND instead explains it by modifying gravity at low accelerations below the typical acceleration $a_{0}$$\sim$$10^{-8}\,\mr{cm\cdot s}^{-2}$.

\subsection{Low and High Curvature}

Curvature $R$ is important in distinguishing possible extensions of the Einstein-Hilbert action. A curvature-dependent may become important even if the potential $\Phi$ remains small. In the Solar System, the curvature $R_{\odot} \sim 4\times 10^{-28}\,\mr{cm}^{-2}$. The curvature of the homogeneous universe according to the Friedmann equation is
\bq
R_{\mr{FLRW}}(z) =3H_{0}^{2}\left( \OM(1+z)^{3} +4\OL\right) \, ,
\eq
from which we can see that the curvature of the universe evolved with time, from $\sim$$10^{-33}\,\mr{cm}^{-2}$ at the time of nucleosynthesis, to $\sim$$10^{-56}\, \mr{cm}^{-2}$ at $z=1$. The curvature scale associated with the cosmological constant is $R_{\Lambda}=(1/6)\Lambda$, so the phenomenology of the cosmological constant occurs in low curvature~regime
\bq
R<R_{\Lambda}\sim 1.2\times 10^{-30}R_{\odot}\,.
\eq

This is of particular interest to paradigms which seek to explain cosmic acceleration through the backreaction effect. For cosmological-scale perturbations, we are always in the weak field regime. However, the curvature perturbation associated with large-scale structure becomes of the order of the background curvature at redshift $z$$\sim$$0$, even if we are still in the weak field limit. This means that the effect of large-scale structure on the background dynamics may be non-negligible \cite{Dunsby2010}.

In summary, in order to explain the dark energy or Dark Matter problem by modifying General Relativity, the modifications have to be either at large scales (typically Hubble scales), low accelerations (typically below $a_{0}$, or small curvatures (typically $R_{\Lambda}$). The regions corresponding to dark energy and Dark Matter curvature and potentials are shown in Figure \ref{GR_curv_pot}.

\subsection{Cosmological Probes}

The idea of testing General \scalebox{.95}[1.0]{Relativity using large-scale structure was first proposed in 2001 \cite{Uzan2001}.} This relies on the ingenious idea that if gravity is well-described by General Relativity, and the universe well-described by $\LCDM$, then on sub-Hubble scales, and considering only scalar perturbations, the~spacetime metric can be written as
\bq
\dd s^{2} = -(1+2\Phi)\dd t^{2}+ (1-2\Psi)a^{2}(t)\gamma_{ij}\dd x^{i} \dd x^{j}
\eq
where $\Phi$ and $\Psi$ are the two potentials, and $\gamma_{ij}$ is the metric of the spatial section
\cite{Peebles:1980,Rich2001,Uzan2006}. The Einstein equations reduce to the Poisson equation
\bq
\Laplace \Psi = 4\pi G \rho_{\mr{matter}} a^{2}\delta_{\mr{matter}}
\eq
and
\bq
\Psi- \Phi =0 \, ,
\eq
since the matter anisotropic stress is negligible. The spectrum of the two gravitational potentials has to be proportional to the matter power spectrum. The scale dependence of the gravitational potential $P_{\Phi}$ and of the matter distribution $P(k)$ are related by:
\bq
k^{4}P_{\Phi}(k,a)=\frac{9}{4}\OM H^{2}a^{-2}P(k,a)\, .
\eq

If the Poisson equation is modified by some modification of gravity, the matter power spectrum changes shape according to the cosmological model assumed, so the above relation is not model-independent. However, the fact that the two spectra differ is independent of the cosmological model. It therefore provides a test of the underlying gravitational theory. Such a test can be carried out by comparing weak lensing data to galaxy surveys.

Various similar approaches have been proposed \cite{Uzan2010}. This approach allows us to test various classes of alternative theories by means of large-scale structure. Some of these theories include Dvali-Gabadadze-Porrati models \cite{DGP,Lue2004,Song2008}, quintessence \cite{Benabed2001,Koivisto2007,Tsujikawa2013,Baldi2011,Koivisto2015}, and scalar-tensor \mbox{theories \cite{Schimd2005,Rodriguez-Meza2010,Goenner2012,Takushima2014}.} The difficulty lies in finding a parameterisation of the perturbation equations which is consistent with the one used for the background evolution, since both assume the same theory for gravitational interactions.

The cosmic microwave background is another potential testing ground for General Relativity. The amplitude and position of the peaks of the cosmic microwave background allows us to probe the potential wells present during the recombination era by extracting information on the primordial power spectrum created during inflation. \textit{Planck} data, alone or in combination with weak lensing has been used to test GR \cite{Dossett2015,Di-Valentino2016} and modified gravity \cite{Hu2013,Pettorino2013,Planck-Collaboration2015a}.

Despite the fact that GR is extremely well tested on laboratory and Solar System scales, cosmology provides plenty of scope for alternatives to General Relativity. Let us consider cosmic acceleration, which is now a confirmed observation. It requires an explanation. The cosmological constant paradigm rests on three assumptions: that the observations are correct, that GR is the correct theory of gravity, and that our FLRW model of the universe is correct.

More than two decades of precision measurements have removed any doubt that the observed acceleration may be due to incorrect modelling of experimental errors. So in order to do away with the cosmological constant, we must seek possible answers in the other two assumptions.

Backreaction is an ``alternative'' to the alternative theories, or to the Dark Matter paradigm. It is firmly within the General Relativity paradigm, and seeks to explain cosmic acceleration by modelling the universe and the structures within it, and therefore its expansion history, in more detail than is the case with the FLRW model. It keeps the first and last assumption, but does away with the assumption of an FLRW universe. However, is it enough to explain the observed acceleration?

If GR is assumed to be the correct theory together with the other two assumptions, cosmological measurements are usually interpreted as providing evidence for Dark Matter and a nonzero cosmological constant or dark energy. This poses conceptual problems. Why is the observed value of the cosmological constant so small in Planck units? It also poses a coincidence problem. Why is the energy
density of the cosmological constant so close to the present matter density?
No dynamical solution of the cosmological constant problem is possible within GR---the cosmological constant is not the attractor of some dynamical function.

This opens the field to possible modifications of GR. Should GR be modified at low and high energies? This is a serious challenge for theorists. Einstein's theory is the
unique interacting theory of a Lorentz-invariant massless spin-2 particle. New physics in the gravitational sector must introduce additional degrees of
freedom. These additional degrees of freedom must modify the theory at low or high energies, or both, while being consistent with GR in the intermediate-energy regime, that is, at~length scales $1\upmu \leqslant \ell \leqslant10^{11} \,\mr{m}$, where the theory is extremely well tested.

Figure \ref{GR_length_pot} illustrates some of the difficulties in testing the completeness of General Relativity. There~is an obvious ``scale desert'' between Solar System scales, and cosmological scales. At the other end of the scale, there is a gap which is often overlooked: between sub-millimetres scales, which is the current limit where GR has been tested, and the Planck scale. A modification of gravity at very small scales would be apparent in this regime. Even ignoring quantum effects, it is a serious challenge to test gravity in a regime in which vastly stronger forces come into play.

\begin{figure}[H]
\centering
\includegraphics[width=.9\textwidth]{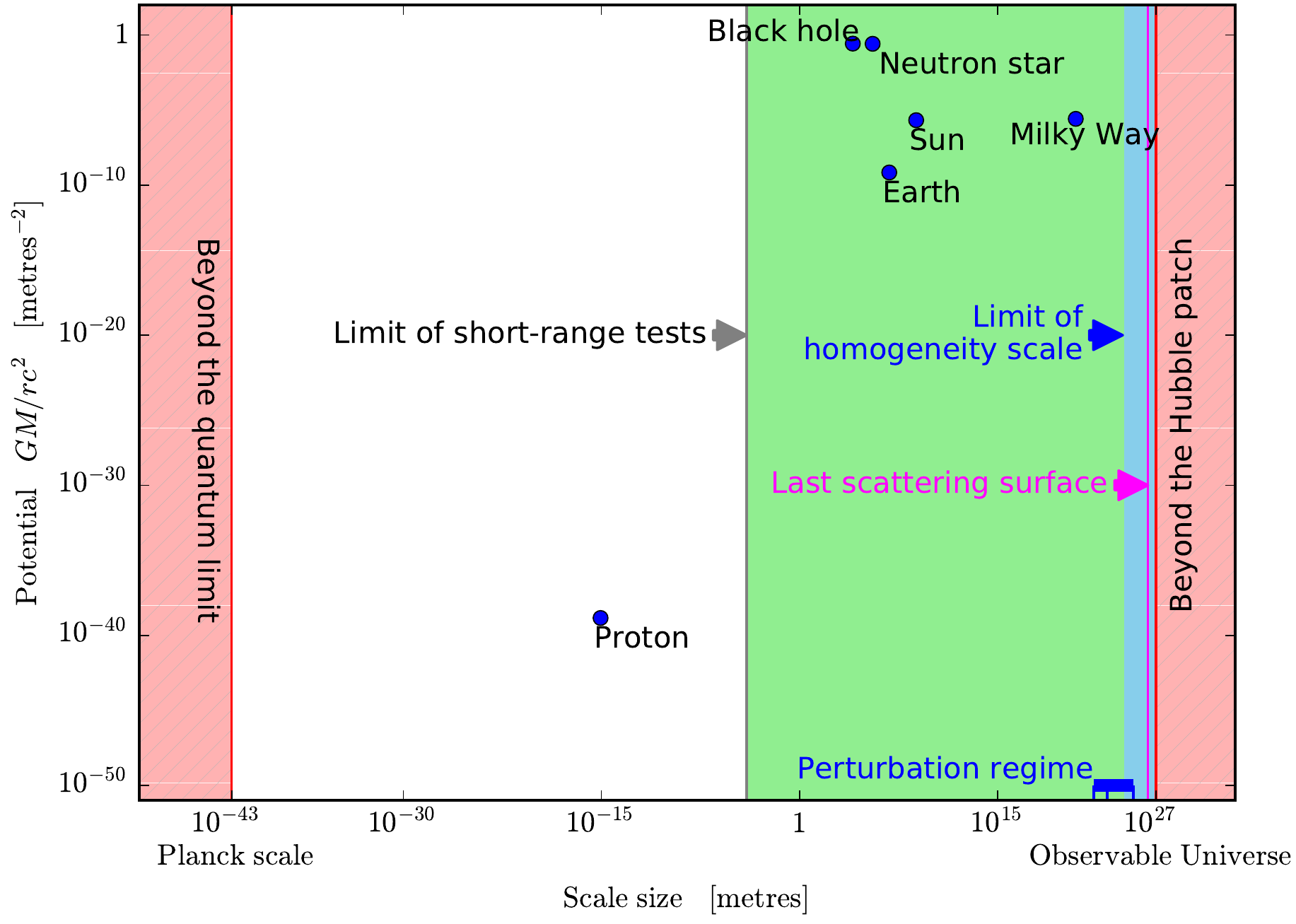}
\caption{The parameter space for experiments. The horizontal axis is the typical length scale of the object in question. The vertical axis measures the gravitational potential. The red vertical line on the left-hand side marks the Planck scale. The vertical line on right-hand side of the plot marks radius of the observable universe, or Hubble radius. Experimental verification of GR is impossible beyond these limits. The radius of the surface of last scattering is only slightly smaller than our Hubble radius. Assuming GR implies assumptions far beyond the range that has been experimentally tests. For instance, if we define the Planck mass as $m_{\mr{Planck}}=\sqrt\hbar c/G$, we are assuming that the gravitational constant remains constant down to the Planck length. This extrapolates the inverse square law over a scale of more than $10^{30}$ from what ha s been tested. The green region is where Solar System tests have been carried out. Beyond $\sim$$100\,\mr{Mpc}$, assumption of a homogeneous and isotropic metric becomes accurate enough to use in physical models. The blue region shows length scales at which the FLRW metric is valid. By way of comparison to this parameter space, the nonlinear regime for perturbation theory, which gives use the matter power spectrum, covers $\sim$$10^{22}$ to $\sim$$10^{23}$ m, the linear, quasi-static regime covers  $\sim$$10^{23}$ to $\sim$$10^{25}$ m, and beyond that is the superhorizon regime. These length scales ranging from $1\, \mr{Mpc}$ to above $1\, \mr{Gpc}$ fit on a tiny part of the horizontal axis above, shown by the thick blue horizontal line.}
\label{GR_length_pot}
\end{figure}

\section{The Nature of Dark Energy and the Implications for General Relativity}

The Concordance Model of cosmology assumes that General Relativity is correct, an assumption which is justified by the tests which GR has undergone. Within this model, $\sim$$95\%$ of the content of the universe is unaccounted for. Dark matter, which makes up around $25\%$ of the mass-energy of the universe is a matter-like component which is cold (sub-relativistic) and weakly interacting.  \mbox{The discrepancy} between the observed acceleration of the expansion of the universe and the predictions of GR leads to the conclusion that there must be a cosmological component with a negative equation of state parameter making up around $70\%$ of the mass-energy content of the universe: dark energy.

But the dark energy paradigm does not fix the nature of this component. There exist many theories which attempt to explain its nature. In particular, one can ask whether the basic assumptions of the Concordance Model---homogeneity and isotropy---are correct.

If the universe is not homogeneous and isotropic, the FLRW equations are no longer valid. Over~the last decade, one line of research has attempted to explain the accelerated expansion by exploring the implications of an inhomogeneous universe on a General Relativistic cosmology. This effect is the backreaction. This approach does away with dark energy.

There are various ongoing investigations on the effect of the backreaction due to an inhomogeneous universe
\cite{Rasanen2004,Rasanen:2009, Wiltshire:2009,Dunsby2010,Rasanen2011, Buchert2015, Rasanen2015,Rasanen2016,Boehm2013},
with different lines of research offering different interpretations of the Buchert equations, where the Friedmann equations are supplemented by an additional backreaction term \cite{Buchert:2000}. Whether one can explain all of the observed expansion history of the universe as a consequence of the growth of inhomogeneities without invoking some additional fluid component is the subject of ongoing debate \cite{Buchert:2008,Buchert2015}. The backreaction, even if it turns out to be incapable of replacing the dark energy paradigm, is still a subject worth investigating, if anything as a correction to the homogeneity assumption, which obviously breaks down at small scales.

If we assume that the Copernican Principle holds, then the universe is well described by a Friedmann-Robertson-Walker spacetime. The dynamics of the background expansion are determined by the content of the universe: the list of fluids (perfect fluids, due to the assumption of the Copernican Principle), with their equations of state. Within the dark energy paradigm, we can distinguish two main strategies for formulating hypotheses:
\begin{enumerate}[leftmargin=2.3em,labelsep=4mm]
\item[(1)] there is some new kind of component in the universe, or
\item[(2)] there is some new property of gravity.
\end{enumerate}

Let us first recall that General Relativity rests on two assumptions: the gravitational interaction is described by a massless spin-2 field, and matter is minimally coupled to the metric, which implies the weak equivalence principle. The Einstein-Hilbert action described by Equation (\ref{EHaction}) implies that
\bq
S_{\mr{gravity}}=\frac{c^{3}}{16\pi G}\int R\sqrt{-g} \dd^{4}x + S_{\mr{matter}}[\mr{matter}; g_{\mu\nu}]\, , \label{EHaction2}
\eq
where $R$ is the Ricci scalar of the metric tensor $g_{\mu\nu}$, and $S_{\mr{matter}}[\mr{matter}; g_{\mu\nu}]$ is the action of the matter fields. If we only consider field theories, this gives us a useful classification scheme for the different theories that seek to explain the nature of dark energy.

The first strategy listed above assumes that gravitation is described by General Relativity, and introduces new forms of gravitating components beyond the Standard Model of particle physics. \mbox{This adds} a new term $S_{\mr{DE}}[\psi;g_{\mu\nu}]$ to the action in Equation \eqref{EHaction2} while keeping the Einstein-Hilbert action of all the standard and Dark Matter unchanged.

The second strategy modifies gravity, and therefore extends the action, either by modifying the Einstein-Hilbert action of the coupling of matter. These theories also involve new forms of matter.

A cosmological constant $\Lambda$ is the simplest modification which can be made to gravity, and it is equivalent to dark energy with a constant equation of state. To explain the observed acceleration, \mbox{the new} form of matter must have an equation of state $w<-1/3$. Dark energy can also be attributed to the energy of the vacuum, although the energy predicted by the Standard Model of particle physics is either $0$ (using super-symmetry), or $10^{120}$ orders of magnitude larger than the observed cosmological value \cite{Banks2014}. There are ongoing attempts to solve this `fine-tuning problem' using string \scalebox{.95}[1.0]{theory \cite{Arkani-Hamed2005,Susskind:2007,Bauer2010},} causal sets \cite{Bombelli1987,Ahmed2004}, or by using anthropic arguments \cite{Weinberg1987,Garriga2004} .

The other approach is to attribute dark energy to a scalar field whose potential has evolved in some way that it currently exerts a negative pressure. Such fields, in theories within the framework of GR, are termed Quintessence. Their distinguishing feature is that they allow the equation of state of dark energy to evolve. Alternatives to Quintessence within the same approach include K-essence, Phantom Fields, or the Chaplygin Gas \cite{Armendariz-Picon2001,Kamenshchik2001,de-Putter2007,Durrer:2008}.

Another strategy is to depart from General Relativity and modify the laws of gravity and posit dark energy as the manifestation of an effect arising from extra dimensions, or higher-order corrections. Within this category, the more successful theories have been of two types. Dvali-Gabadadze-Porrati (DGP) dark energy considers the universe as a 4D brane within a 5D Minkowskian bulk \cite{DGP}. The~weakness of gravity relative to the other forces is explained  by gravity ``leaking'' into the higher dimensions as it acts through the bulk \cite{Deffayet2002}, whereas the other forces act within the brane. The other class of theories is $f(R)$ gravity, where the Ricci scalar $R$ in the Lagrangian is replaced by some function $f(R)$. Such theories correspond to scalar-tensor gravity with vanishing Brans-Dicke parameter \cite{Amendola:2007}.

We can therefore identify four main classes of theories, as shown in Figure \ref{GR_classes}. Classes 1 and 2 assume GR and introduce new forms of gravitating matter, the difference being that in class 2, the distance-duality relationship may be violated due to mechanisms such as photon decay. Classes 3 and 4 modify gravity in some way. In class 3, a new field introduces a long-range force so that gravity is no longer described by a massless spin-2 graviton. In this class, there may be a variation of the fundamental constants. Class 4 includes theories in which there may exist an infinite number of new degrees of freedom, such as brane models or multidimensional models. These models predict a violation of the Poisson equation on large scales. The constraints which may be placed on these theories, and the experiments to test them, correspond to the regimes in which gravity is modified, as summarised in Section \ref{ModificationsSection}.

An exhaustive list of dark energy or cosmic acceleration theories is beyond the scope of this review. We simply note that most of them give different predictions for the equation of state of dark energy, its evolution, or the expansion history of the universe . To distinguish between these proposals we need to track the evolution of these parameters throughout the history of the universe.

\begin{figure}[H]
\centering
\includegraphics[width=.95\textwidth]{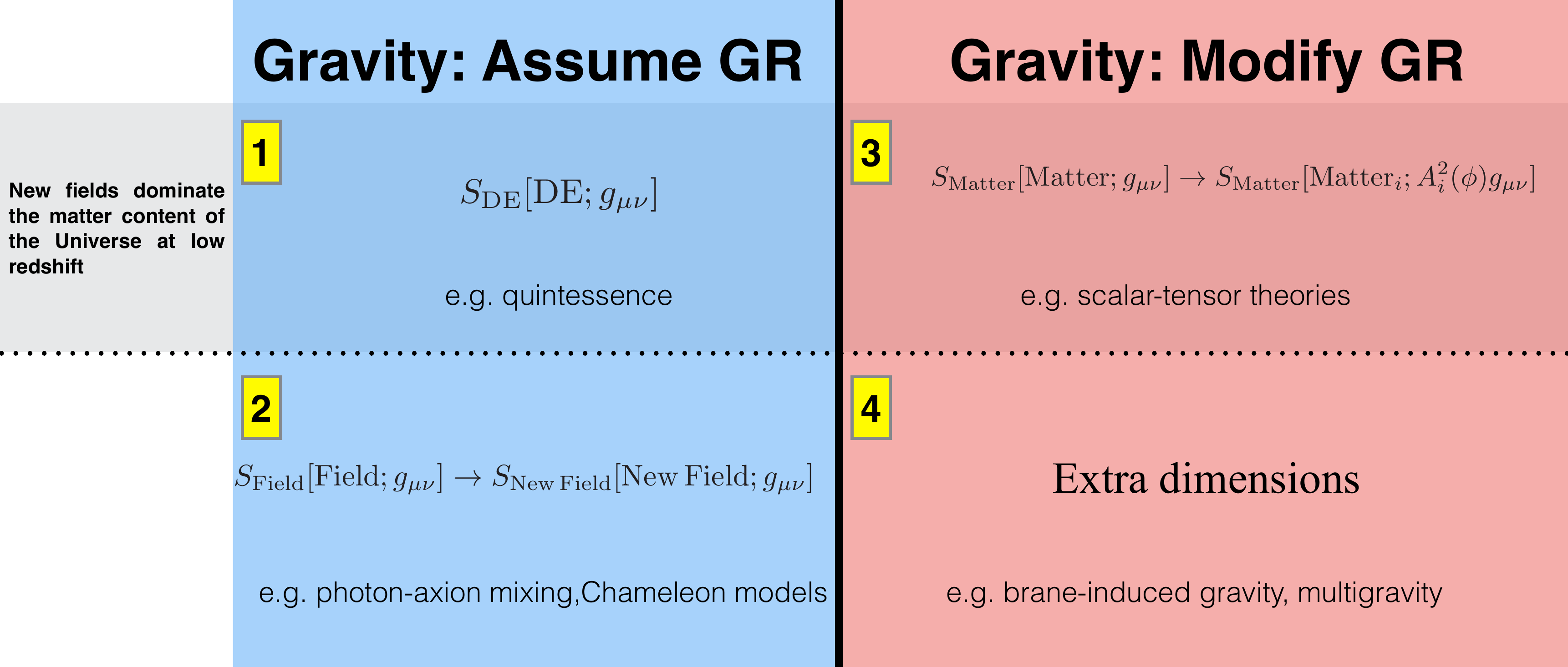}
\caption{The four main classes of dark energy theories, within the two broad strategies, classified as modifications of the General Relativistic action. Classes 1 and 2 assume the gravitational metric coupling of GR, whereas classes 3 and 4 modify this metric coupling, and are therefore modifications of gravity. In the upper line of classes (1 and 3), new fields dominate the matter content of the recent universe. Adapted from \cite{Uzan2006}.}
\label{GR_classes}
\end{figure}

\section{The Current Status of General Relativity}

General Relativity has been subjected to a multitude of tests in its 100 years of existence.
\mbox{As of 2016,} the main predictions of GR have been tested and confirmed. Whereas it is sufficient for most purposes at ordinary accelerations and energy scales to use Newtonian calculations, General Relativity has found its way into daily life, in the Global Positioning System and geodesy. \mbox{The postulates} of General Relativity have been confirmed with ever-increasing accuracy. Deviations from the Einstein Equivalence Principle are now constrained to below $\sim$$10^{-14}$, deviations from Local Lorentz Invariance down to $\sim$$10^{-20}$, and deviations from Local Position Invariance down to $\sim$$10^{-6}$.

General Relativity has been probed down to scales of $\sim 10^{-6}$~metres in laboratories, and up to $1000$~AU in space-based experiments and observations. Astrophysical and cosmological observations have probed GR at scales from $1\,\mr{Mpc}$ up to gigaparsec scales. In the Solar System, the dynamics of GR have been tested with radar and laser-ranging. We can track the ephemerides of the planets in the Solar System, right up to the minor outlying planets (such as Sedna, with an aphelion just short of $1000$~AU). The farthest objects whose trajectory has been followed from the moment they were ``thrown'' are now outside the Solar System. They are the Pioneer 10 and 11 spacecraft, and are now around 70  AU distant, and Voyager 1, which is at a distance of 135 AU, making it the only object to have reached interstellar space. Communication with the Pioneer spacecraft ceased in 2003. \mbox{The Pioneer} spacecraft exhibited an anomalous constant acceleration towards the Sun which could not be explained using GR. This prompted a reexamination of all the recorded data. It is now generally accepted that the anomaly is caused by thermal radiation, and that once this is accounted for, there is no remaining anomalous acceleration \cite{Turyshev2012}.

Gravitational waves, among the last untested predictions of GR, were first detected in late 2015, ushering a whole new era of observational astrophysics, in which the strong field regime can be probed and where the predictions of GR can be tested.

The evidence for General Relativity is extremely strong. The theory has passed all tests in the weak-field limit at Solar System scales, including tests of the assumptions (the Equivalence Principle) and the predictions specific to GR (frame-dragging, gravitational time dilation), and in the strong field with the observation of a binary black hole merger and the resulting gravitational waves.\mbox{ We give} a summary of the experimental milestones in Table \ref{GRtests}. However, there are still issues that allow room \mbox{for speculation. }

The first is the question of the completeness of GR. Is it valid at all scales? There is a scale gap in our tests of GR between laboratory, Solar System and galactic scales, and \mbox{cosmological ones}. \mbox{This gives} rise to a multitude of domains of validity for different alternative theories, whereas ideally we should seek a universal theory that can explain phenomena at all scales.

The second issue is the accuracy of our approximations. General Relativity may be conceptually simple in that it is based on a minimum number of postulates. {But the resulting field equations}, when applied to real physical systems, can be very hard to solve. We~get around this by making approximations such as spherical symmetry, or homogeneous and isotropic {perfect~fluids}, which~allows us to obtain analytical solutions. However, the accuracy of these approximations may not always be good enough. This is evident in the case of cosmological perturbations and large-scale~structure. {At~which} scales is it valid to use a~Friedmann-Lema\^itre-Robertson-Walker~metric? {Can~the} backreaction explain some or all of the observed cosmic acceleration? For smaller systems such as aspherical collapsing bodies, we still need accurate models in order to match theory and observation. Such questions have spurred the development of numerical methods in General Relativity. Physicists now have the necessary computing power to go far beyond simple first-order approximations.

The remaining open questions are of a cosmological nature: Dark Matter and dark energy, and~inflation. Dark matter and dark energy account for around $25\%$ and $70\%$ of the mass-energy content of the universe, respectively. They are not a problem for the theory of \mbox{gravity itself,} \mbox{but it} does mean that we do not know the nature of $95\%$ of the content of the universe. Inflation solves \mbox{a number} of cosmological problems, but whatever theory we choose to explain inflation, we still need to introduce new physics beyond GR.

% start a new page
\newpage
% change it to landscape
\paperwidth=\pdfpageheight
\paperheight=\pdfpagewidth
\pdfpageheight=\paperheight
\pdfpagewidth=\paperwidth
\newgeometry{layoutwidth=297mm,layoutheight=210 mm, left=2.7cm,right=2.7cm,top=1.8cm,bottom=1.5cm, includehead,includefoot}
\fancyheadoffset[LO,RE]{0cm}
\fancyheadoffset[RO,LE]{0cm}
%%%%%%%%%%%%%%%%%%%%%%%%%%%%%%%%%%%%%%%%%%%%%%%

\begin{table}[H]
\caption{The major predictions of GR, and their experimental status.}
\label{GRtests}
\small % Font size can be changed to match table content. Recommend 10 pt.
\centering
\begin{tabular}{llll}
\toprule
\textbf{Effect} &  \textbf{Milestones}  &   \textbf{Current, Future and Proposed Experiments}  &   \textbf{Status}  \\
\midrule

Mass equivalence  &  Galileo \cite{Adler1978,SEGRE1989435} &  E\"ot-Wash Group \cite{Wagner2012} &  Confirmed\\
 & E\"otv\"os \cite{Eotvos1890}  & Lunar Laser Ranging  \cite{LLR2012}\\
 & & MICROSCOPE\cite{Berge2015} \\
 &  & STEP \cite{Overduin2012}\\
 & & Galileo Galilei satellite \cite{Nobili2009}\\
 \midrule

 Gravitational time & Eddington solar eclipse \cite{Dyson1920}&  Quantum interference of atoms \cite{Muller2010}&  Confirmed\\ dilation \cite{Bardeen1975}
 &Pound-Rebka experiment \cite{PhysRevLett.4.337}& ACES \cite{Turyshev2016} &\\
 & Space-borne hydrogen masers \cite{Vessot1980} &  Galileo 5 and 6 satellites\cite{Delva2015} & \\
 && Einstein Gravity Explorer \cite{Schiller2009}&
\\  \midrule

Precession of orbits  &  Orbit of Mercury (Einstein \cite{Ein15})   &  Binary pulsar observations  \cite{Antoniadis2013,Iorio2014,Fienga2015}  &  Confirmed\\
 & &Solar System and extrasolar planets \cite{Iorio2011c,Pitjeva2013}&
 \\  \midrule

De Sitter precession  &  Gravity Probe B \cite{Everitt2011} & Binary pulsars \cite{Kramer2016} & Confirmed \\
&Lunar laser ranging \cite{Bertotti1987,Shapiro1988,Williams1996}&Improved lunar laser ranging \cite{Merkowitz2010,Martini2013}&\\
&Binary pulsars\cite{Kramer1998,Breton2008}& &\\
\midrule

Lense-Thirring precession &  Gravity Probe B \cite{Everitt2011}  & LARASE \cite{Lucchesi2015}   &  Confirmed\\
& LARES \cite{Iorio2009,Kehl2016,Ingram2016}&Laboratory tests \cite{Bosi2011}&\\
&&Solar System bodies \cite{Iorio2011a}&\\
&&Binary pulsars \cite{Iorio2009,Kehl2016}&\\
&&Black holes \cite{Bardeen1975,Stone2012,Franchini2016,Ingram2016} &\\
\midrule
Gravitational waves  &  LIGO \cite{GW2015}  &  Advanced LIGO \cite{Martynov2016}&  Recently confirmed \\
 &      &      eLISA \myurl{https://www.elisascience.org}  & in two events \cite{GW2015,Abbott2016a}\\

\midrule
Strong field effects &  PSR J0348 + 0432 \cite{Antoniadis2013}  & Black holes \cite{Johannsen2012,Johannsen2016} & Recently confirmed\\
&& Binary pulsars \cite{Berti2015} &  \\
\midrule
Orbital precession due to &&Low-orbit satellites \cite{Soffel1987,Heimberger1989,Soffelbook,Iorio2015a} & Not yet observed \\
oblateness of central body&& Stars orbiting black holes \cite{Iorio2015a} &\\
&&Juno spacecraft around Jupiter \cite{Iorio2013a}&\\
%&&&\\
\bottomrule
\end{tabular}
\end{table}

%%%%%%%%%%%%%%%%%%%%%%%%%%%%%%%%%%%%%%%%%%%%%%%
% change everything back
\newpage
\restoregeometry
\paperwidth=\pdfpageheight
\paperheight=\pdfpagewidth
\pdfpageheight=\paperheight
\pdfpagewidth=\paperwidth
\headwidth=\textwidth

\section{Future Developments}

The current Concordance Model of cosmology was built in successive (and sometimes concurrent) steps. General Relativity applied to a spacetime under the Copernican Principle, \mbox{filled with} pressureless matter, produced the Einstein-de Sitter model. Motivated by the observed Hubble expansion, it resulted in the Big Bang model. Clues from element~abundances, baryon~assymmetry, and knowledge of nucleosynthesis from the Standard Model of particle physics meant that the universe had to have a thermal history, which resulted in the Hot Big Bang~Model. When~evidence for missing mass became incontrovertible, cold Dark Matter had to be added to the inventory of cosmic components. This model worked well, but not well~enough. It~could not explain the observed homogeneity of the universe across regions which were causally disconnected, nor~\mbox{its flatness}. So~Inflation was introduced. Observations of an accelerated cosmic expansion motivated a search for explanations within the then current paradigm, which~resulted in various hypotheses: \mbox{a curved} geometry, supermassive neutrinos, or perhaps a particular cosmological~topology. \mbox{In the end}, the paradigm had to be shifted yet again with the introduction of dark energy.

The Concordance Model can explain the observations with just six parameters: the physical baryon density parameter $\OB h^{2}$, where $h$ is the Hubble parameter, the physical Dark Matter density parameter $\OC h^{2}$, the age of the universe $t_{0}$, the scalar spectral index $n_{s}$, the curvature fluctuation amplitude $\Delta_{\mr{R}}^{2}$, and the reionisation optical depth $\tau$. That such a degree of fit is offered by such \mbox{a simple} model is remarkable.

The success of the Concordance Model has been its ability to include physical effects at extremely different scales, from primordial nucleosynthesis to large-scale structure evolution, in~one~coherent~theory. However,~this does not allow us to state that the
$\LCDM$ model is~correct. It merely implies that deviations from $\LCDM$ are too small compared to the current observational uncertainties to be inferred from cosmological data alone. This leaves room for some very fundamental open questions, which we have described in this review.

The science of cosmology finds itself at a critical point where it has to make sense of the vast quantity of data that has
become available. Different probes have allowed us to piece together interlocking information which, so far, confirms the Concordance Model. The cosmic microwave background has provided conclusive evidence of a flat geometry, super-horizon features, the correct harmonic peaks, adiabatic fluctuations, Gaussian random fields, and most recently, a departure from scale invariance. We have not yet observed primordial inflationary gravitational waves. Large-scale structure observations, which probe the recent universe, provide firm evidence in favour of the Concordance Model's explanation of  the evolution of density perturbations and the growth of structure, and provide a bridge between the effects of long-range gravitational interactions and shorter-range~forces.

The recent detection of gravitational waves in two events (possibly three), one hundred years after they were predicted by Einstein, directly validates General Relativity in several ways. It shows that GR is correct in the strong-field regime, that black holes really exist, and black hole binaries too, and it proves that gravitational waves are a real physical phenomenon, and not just a mathematical artefact of GR.

We are fast approaching the point where cosmological observations will be limited only by cosmic variance, i.e., no more data will be available from our Hubble patch \cite{Starkman2010}. What does that imply for the development of new theories? How do we test the predictivity of these theories without new data?

We will have to look for new effects in old data, and for new correlations in future data. Large-scale structure is extremely useful in testing General Relativity and \mbox{cosmological models}, \mbox{but the} future may bring other observational windows. In particular, the next fifty to one hundred years may see the development of a gravitational wave astronomy. This is likely to be an even more significant development than even CMB astronomy. The observation of primordial gravitational waves would provide vital information on the inflationary epoch \cite{Domcke2016,Ito2016}. B-mode~polarisation of the CMB offers an indirect pathway to the observation of this gravitational wave   \linebreak background~\cite{Fidler2014,Namikawa2015}. The second major development could be the observation of a cosmic neutrino background~\cite{Abazajian2015}, which is the result of neutrino decoupling in the lepton era. This would push the observations along our past light cone even further back in redshift, providing information on the universe before recombination and the CMB.

\subsection{Plausible Conclusions from Incomplete Information}

The statistical questions facing cosmologists pose some particular problems. We observe a finite region of our universe,
which is itself a single realisation of the cosmological theory. We can only observe whatever is on or inside our past light cone, as shown in Figure \ref{GR_cone}. Not only are we limited by cosmic variance, we also have just a single data point for the cosmological model.

There exist several alternatives to GR and to $\LCDM$ that have not been ruled out by experiment. Constructing viable physical models is not just a question of fitting the model to the data. It is a~question of model selection, which requires robust statistical techniques that allow us to make sensible decisions using our incomplete information. Bayesian inference provides a quantitative framework for plausible conclusions \cite{Trotta:2008aa,Hobson:2010aa}. We can identify three levels of Bayesian inference.

\begin{enumerate}[leftmargin=2.2em,labelsep=4mm]
\item[(1)] Parameter inference (estimation). We assume that a model $M$ is true, and we select a prior for the parameters $\boldsymbol{\theta}$, or the $\mr{Prob}({\boldsymbol{\theta}} | M)$.
\item[(2)] Model comparison. There are several possible models $M_i$. We find the relative plausibility of each in the light of the data $D$, that is we calculate the ratio $\mr{Prob}(D|M_{i})/\mr{Prob}(D|M_{0})$.
\item[(3)] Model averaging. There is no clear evidence for a best model. We find the inference on the parameters which accounts for the model uncertainty.
\end{enumerate}

At the first level of Bayesian inference, we can estimate the allowed parameter values of \mbox{the theory}. If~we assume General Relativity as our theory, we still need to fix the values of the various~constants. This is the rationale behind ongoing efforts to measure quantities like Newton's constant $G$ ever more accurately, despite the fact that the parameter has been around for three~centuries. What are the energy densities of the various components of a $\LCDM$ universe? If we include a dark energy equation of state parameter $w$, what value does it take?

Next, we can ask which parameters we should include in the theory.  Should we~include a cosmological constant in General Relativity? Or should we include dynamical dark energy~parameter? Although current data are consistent with the six-parameter $\LCDM$ model based on GR, there are more than twenty candidate parameters which might be required by future data (see \cite{Liddle:2004aa}). We cannot simply include all possible parameters to fit the data, since each one will give rise to degeneracies that weaken constraints on other parameters, including the $\LCDM$ parameter set (e.g., \cite{Metcalf1998,Debono:2009,Howlett2012}). The~landscape of alternative cosmological models is even larger if we relax our assumptions on the theory of gravity \cite{Joyce2015}.

The goal in data analysis is usually to decide which parameters need to be included in order to explain the data. For physicists, those extra parameters must be physically motivated. \mbox{That is}, \mbox{we need} to know the physical effects to which our data are sensitive, so that we can relate these effects to physics. At the current state of knowledge, we have to acknowledge the possibility of more than one model. We therefore require a consistent method to discard or include parameters. This is the second level of Bayesian inference---model selection.

Bayesian model selection penalises models which introduce wasted parameter space. We can always construct a theory that fits the data perfectly, even better than GR, but we would need to introduce extra free parameters (e.g., extra fields or couplings between matter and the metric). This is the mathematical equivalent of Occam's razor. We seek a balance between goodness of fit (the degree of complexity) and predictive power (consistency with prior knowledge). General Relativity fits all the data with the minimum number of parameters. In cosmology, $\LCDM$ is the best model because it only involves one new parameter and no new fields.

The problem of model selection in relation to GR is as old as the theory itself. \mbox{But only} recently have cosmologists have started to use Bayesian methods for cosmological model \linebreak selection (e.g., \cite{Jaffe:1996, Mukherjee:2006, Kilbinger:2009a, Wraith:2009}), when the astrophysical data began to have the necessary statistical power to enable model testing. Bayesian techniques are starting to be applied to General Relativity itself~\cite{Del-Pozzo2011}. With the next generation of astrophysical probes in the pipeline, model selection is likely to grow in importance~\cite{Trotta:2008aa}.

Bayesian model selection cannot be completely free of assumptions. In cosmology, there is some model structure which depends on a number of unverifiable hypotheses about the nature of the universe. The Copernican Principle is one such hypothesis \cite{Ellis:1975aa,Ellis:2009,Ellis:2006aa}.

The third level of Bayesian inference is model averaging. In the current scenario, there is firm evidence for General Relativity as the best model for gravitational interactions. However, it is still useful to quantify our degree of certainty (or doubt), for one simple reason: we do not have the final list of alternative theories. In other words, when we choose GR against any number of alternative theories, we have no knowledge about other alternatives outside that list (such as theories yet to be developed, for example). At best, we know which alternatives are ruled out by the data. This level of Bayesian inference is the application of a principle that has been called `Cromwell's Rule' \cite{Wiley2009}: even if all the data show our theory to be correct, we should allow a non-zero probability, even if tiny, that the theory is false.

The utility of alternative theories becomes evident when we apply Bayesian model selection. For~science to advance by falsification, it is not enough to claim that the present theory is~false. We~need to know which alternative theory is favoured instead. Newtonian gravity would have likely have survived the 1919 eclipse if General Relativity had not been formulated.

\subsection{Experimental Progress}

There has been rapid progress in constraining cosmological parameters and models over the last two decades, with a multitude of experiments observing the CMB, large-scale structure, galaxies~and~supernovae. We will just provide a summary of the most recent data sets.

The first are anisotropies in the CMB, where the main statistic is the angular power spectrum of fluctuations $C_{\ell}$, and polarisation of the CMB.
The most recent and current are:
WMAP ({\myurl{http://map.gsfc.nasa.gov}}),
\textit{Planck} ({\myurl{http://www.cosmos.esa.int/web/planck}}),
Atacama Cosmology Telescope (ACT) ({\myurl{http://act.princeton.edu}}),
South Pole Telescope (SPT) ({\myurl{https://pole.uchicago.edu}}),
Atacama Cosmology Telescope polarisation-sensitive receiver (ACTPol)~\cite{Niemack2010},
SPTPol, Spider \linebreak({\myurl{http://spider.princeton.edu}}),
Polarbear ({\myurl{http://bolo.berkeley.edu/polarbear}}),
Background Imaging of Cosmic Extragalactic Polarization (BICEP2) ({\myurl{http://bicepkeck.org},\myurl{https://www.cfa.harvard.edu/CMB/bicep2}}),
Keck Array ({\myurl{http://bicepkeck.org},\myurl{https://www.cfa.harvard.edu/CMB/keckarray}}).

The second source of data are surveys cataloguing the angular positions and redshifts of individual galaxies, leading to the power spectrum of fluctuations $P(k,z)$, or the two-point correlation function $\xi(r)$.
The recent and current experiments are:
Baryon Oscillation Spectroscopic Survey (BOSS) using Sloan Digital Sky Survey (SDSS) data ({\myurl{http://www.sdss3.org/surveys/boss.php}}),
Dark Energy Survey (DES) ({\myurl{https://www.darkenergysurvey.org}}),
Weave ({\myurl{http://www.ing.iac.es/weave/science.html}}),
Hobby-Eberly Telescope Dark Energy Experiment (HETDEX) ({\myurl{http://www.hetdex.org}}),
Extended Baryon Oscillation Spectroscopic Survey (eBOSS) ({\myurl{https://www.sdss3.org/future/eboss.php}}),
Mid-Scale Dark Energy Spectroscopic Instrument (MS-DESI) ({\myurl{https://www.skatelescope.org}}),
Canadian Hydrogen Intensity Mapping Experiment (CHIME) ({\myurl{http://chime.phas.ubc.ca}}), Baobab,
MeerKAT ({\myurl{http://www.ska.ac.za/science-engineering/meerkat}}), and
ASKAP ({\myurl{http://www.atnf.csiro.au/projects/askap/index.html}}).

The third source of data are weak lensing, which use the fact that images of distant galaxies are distorted and correlated
by intervening gravitational potential wells to produce statistics such as the
convergence power spectrum $C^{\kappa}_{\ell}$ \cite{Beynon2010,Harnois-Deraps2015}.
Some current experiments are:
Dark~Energy Survey (DES) ({\myurl{https://www.darkenergysurvey.org}}),
Red Cluster Sequence Lensing Survey (RCSLens) ({\myurl{http://www.rcslens.org}}),
Canada-France Hawaii Telescope Lensing Survey (CFHTLenS) ({\myurl{http://www.cfhtlens.org}}),
New Instrument of Kids Arrays (NIKA2) ({\myurl{http://ipag.osug.fr/nika2}}), and
Hyper Suprime-Cam (HSC) ({\myurl{http://hsc.mtk.nao.ac.jp/ssp}}).

The final source of data are catalogues of peculiar velocities. By measuring redshifts and radial distances of galaxies and
clusters it is possible to reconstruct a radially projected map of large-scale~motions. Progress in this field will come from all three data sets above.

The main science goal of the next generation of cosmological probes is to test the Concordance Model of cosmology.
Some major experiments will be operational in the next decade.

\textit{Planck}, decommissioned in 2013, marked a major milestone in CMB experiments. Its proposed successor is the ground-based programme CMB-S4 \cite{Carlstrom2016}, which should reach sensitivities below $10^{-3} {\upmu}$~K whose main aim is to achieve higher resolutions, probe larger scales, and measure new observables such as polarisation.

There are various future experiments to map the mass-energy content of the universe (including~baryons and Dark Matter), either in the planning phase, or close to completion. The~Large Synoptic Space Telescope (LSST) ({\myurl{https://www.lsst.org}}), which should achieve first light in 2019, \mbox{is a ground-based} telescope which will map the entire sky.

The \textit{Euclid} space telescope ({\myurl{http://www.euclid-ec.org}}), due for launch in 2020, will map galaxies and large-scale structure over the whole sky at visible and near-infrared wavelengths, providing a~catalogue of 12 billion sources at 50 million . \textit{Euclid} will probe the recent universe, when galaxies have formed and dark energy starts to dominate.  Its main scientific objective is to understand the origin of the accelerated expansion of the universe by probing the nature of dark energy using weak-lensing observables (which include cosmic shear, higher-order distortions, and~cosmic~magnification), and galaxy-clustering observations \cite{Amendola2013}.

The Square Kilometre Array (SKA)  ({\myurl{https://www.skatelescope.org}}), which will begin operations in 2020, is a multi-wavelength radio telescope, built across multiple sites to achieve the largest collecting area ever \cite{Santos2015,Bull2016}. It should provide the highest resolution images of the radio sky, thus providing maps of large-scale structure, and observing pulsars which should provide direct tests of General Relativity. The SKA will observe the epoch between the emission of the CMB and the formation of the first galaxies. Neutral hydrogen surveys (or 21 cm intensity mapping) \cite{Hall2013}, offer~yet another promising probe, as do galaxy redshift surveys \cite{Lahav2004}.

Joint observables will be key in the next generation of experiments. These include the Sachs-Wolfe effect, and the Sunyaev-Zel'dovich effect \cite{Dupe2011}. We will also need to extract more observables from the CMB. We need information on in order to constrain inflationary models.

As experiments probe larger scales at better resolutions (low $\ell$ and high $\ell$), the data analysis and the cosmological tests (dark energy, Dark Matter, the properties of cosmic neutrinos) will require accurate calculations on the growth of large-scale structure, which can only be achieved using $N$-body simulations
\cite{Stabenau2006,Adamek2013,He2015,Winther2015,Eingorn2016,Hahn2016,Kamdar2016,Kamdar2016a}.

At the other end of the length scale, there is particle physics. The Very Large Hadron~Collider, the successor to the Large Hadron~Collider ({\myurl{https://home.cern/topics/large-hadron-collider}}) (\mbox{which can achieve} energies around $14$~TeV), is still in its conceptual phase. It could, if built, probe~energies around $100$~TeV, allowing it to test physics beyond the Standard Model, possibly including Supersymmetry and Grand Unified Theories, and thus provide clues on the nature of Dark Matter and dark energy (see, e.g., \cite{Smoot2014}).

Tremendous progress is also being made in Milky Way astrophysics. With the launch on 2013 of the Gaia mission ({\myurl{http://www.cosmos.esa.int/web/gaia}}), we will soon have improved data on Solar System ephemerides, and on the orbits and tidal streams in the Milky Way. This will allow precise tests of GR at Galactic scales. In particular, tidal streams provide an opportunity to probe GR at late cosmological times and to close the gap between astronomical and cosmological scales.

Now that gravitational waves have been detected, gravitational wave physics is set to become an~established branch of astrophysics. The Laser Interferometer Gravitational-Wave Observatory (LIGO) ({\myurl{https://www.ligo.caltech.edu}}),  Advanced LIGO (aLigo) ({\myurl{https://www.advancedligo.mit.edu}}) and  Virgo ({\myurl{http://www.virgo-gw.eu}}) will subject GR to a battery of test in the strong regime. Also~targeting the regime of strong curvatures and potentials, the planned Event Horizon Telescope \linebreak({\myurl{http://www.eventhorizontelescope.org}}) is a network of millimetre and sub-millimetres telescopes being used for very-long baseline interferometry to directly image supermassive black holes in galactic~centres.

\subsection{Theoretical and Computational Progress}

On the theoretical front, there are four main lines of development in theories \mbox{of gravity}. \mbox{They are} all motivated by current open questions in physics. The ongoing attempt to find a~Grand Unified Theory continues to motivate the development of string theory \cite{Rickles2014} as a framework \mbox{for gravity}. \mbox{Aside from} the theoretical difficulties of a mathematically complex theory, the~challenge for string theory is to produce physical predictions which can be experimentally~tested. The~second approach is brane theories or supergravity, in which spacetime has more than \mbox{four dimensions}. The~resulting field theory combines Supersymmetry (from particle physics) and General Relativity~\cite{Brane2010}. The third approach is quantum gravity~\cite{Woodard2009}, in which spacetime,\mbox{ as a dynamical field}, is a~\mbox{quantum object}. This~implies a violation of Lorentz Invariance near the Planck scale, which in turn means that some particle decays forbidden by Special Relativity are allowed, and possibly charge-parity-time violations too. This motivates the search for signatures of quantum gravity in particle \mbox{physics experiments.}

The final approach, which is closer to the theoretical framework of the Concordance Model of cosmology, is a phenomenological use of General Relativity. This includes the various alternative theories that seek to explain cosmic acceleration and the missing mass, and also the various theories which provide dark energy and Dark Matter candidates.

The early alternatives to General Relativity were motivated by\mbox{ theoretical considerations}. \mbox{The current} alternatives are mainly motivated by the open questions in cosmology. Cosmology is now in the age of Big Data. In the last decade, the data finally caught up with the theory, and we are now in a position to test many of the current alternative theories using statistical techniques.

The principles that underpin statistical techniques are simple enough. Under minimal assumptions about signal and noise, it is simply a question of maximising the Gaussian likelihoods. However, in practice they are extremely complex. The foreground contamination, the signal, the possibly non-Gaussian noise and the systematics all have to be modelled. This requires \mbox{data simulation}, \mbox{so the} number of maps that is generated is far larger than the number that is actually observed by the instrument (see, e.g., \cite{Norris2011,Planck-Collaboration2015b,Planck-Collaboration2015c,Juric2015,Desai2015,Dodson2016}).

A low signal-to-noise ratio, and higher-resolution observations of fainter signals over a larger frequency range, result in massively big data sets. For statistical probes such as the CMB and large-scale structure, these data sets have to be analysed as a whole, in order to correlate data. In addition to the volume of data collected, this generates a huge amount of computational data, and requires the appropriate computing power to carry out multiple complex calculations. We are witnessing a Moore's Law in cosmology, where the data volume of experiments increases by a factor of around $1000$ every ten years, as shown in Figure \ref{GR_date_size}.

In addition to the data volume challenge, we have an algorithmic challenge. The number of computations for cosmological data analysis depends on four main factors: the number of observations $N_{\mr{t}}$, the number of pixels $N_{\mr{p}}$, the number of multipoles (a function of the resolution and frequency range) $N_{\ell}$, and the number of iterations $N_{\mr{i}}$. The first three quantities determine the data volume. The data simulation scales at least as $\mathcal{O}(N_{\mr{t}})$. The map-making scales as
$\mathcal{O}(N_{\mr{i}} N_{\mr{t}} \log N_{\mr{t}})$. The~maximum-likelihood power spectrum estimation scales as $\mathcal{O}(N_{\mr{i}} N_{\mr{l}} N_{\mr{p}}^{3})$.

\begin{figure}[H]
\centering
\includegraphics[width=\textwidth, trim=0 1cm 0 1.1cm,clip]{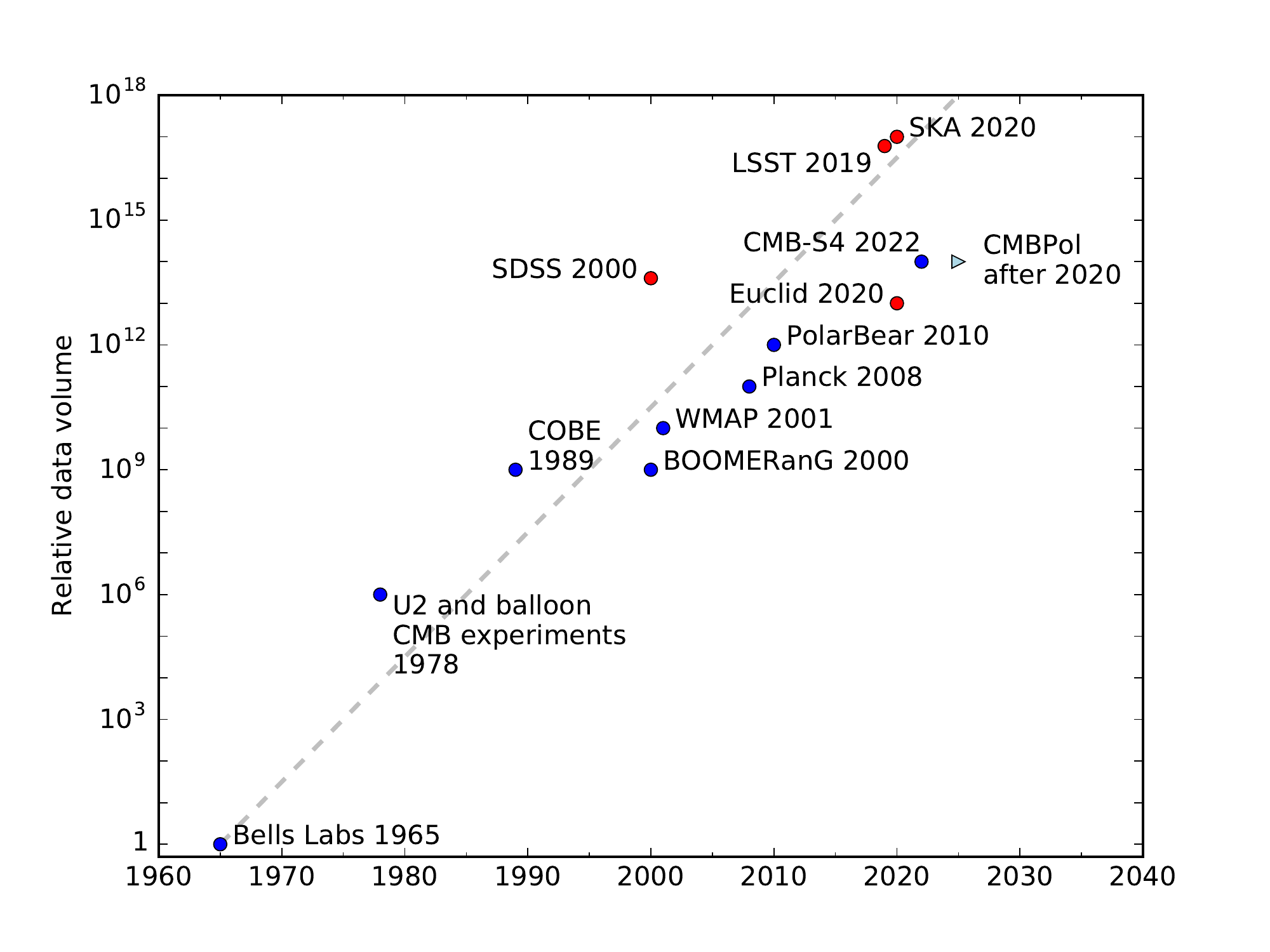}
\caption{The growing data volume of experiments. The data volume of each experiment is shown as an order-of-magnitude  multiple of the data volume of the 1965 Bells Labs experiment which detected the CMB. The labels show the year of `first light' for each experiment. CMB surveys are marked by blue dots, while red dots show large-scale structure surveys. A first light date for CMBPol (l\mbox{ight blue triangle}) has not yet been fixed.  Note that the vertical axis is logarithmic: the date volume increases about a thousandfold every ten years (grey dashed line). Note too that we plot CMB, large-scale structure, space and ground-based probes on the same graph. Ground-based probes will always tend to have a larger data volume than space-borne probes, due to the bandwidth limit on data transmission from spacecraft. Longer-running experiments will also have a larger data volume.}
\label{GR_date_size}
\end{figure}

\textit{Planck} marked a milestone in data science. It was the first CMB experiment in which the whole data treatment process was parallelised, and where Monte Carlo methods were used in order to cut down on the number of data realisation iterations that were carried out.

There are two main considerations in data analysis: the amount of data, and the complexity of the theory. Future experiments will require sophisticated techniques, and considerable computing power to process the vast amounts of data. Gaia is already collecting data \cite{Arviset2016}, \mbox{while Euclid}, \mbox{the SKA}, Enhanced LIGO, and CMB-S4 will soon be operational. Farther ahead, the Evolved Laser Interferometer Space Antenna (eLISA) ({\myurl{https://www.elisascience.org}}), due for launch \mbox{in 2034}, \mbox{will also} generate huge volumes of data, and will require the necessary computing power to test GR directly, by comparing the data to simulated black hole collapse, inspiralling binaries.

What direction will experimental cosmology  take over the coming decade? Extracting fainter signals, such as CMB polarisation, or going to very high resolutions requires larger data volumes to provide a higher signal-to-noise ratio, and it requires more complex models to control fainter systematic effects. Even the Solar System tests of General Relativity will depend upon vastly greater data volumes and computational complexity to get the full relativistic~ephemerides, given~the ever-increasing number of objects being tracked in the Solar System and the significantly greater precision of the data for each~object.

This enables us to make a sensible prediction on future developments. The science we are able to extract from present and future data sets will be determined by the limits on our computational capability, and our ability and willingness to exploit it.

\subsection{Conclusions}

GR may well survive for another 100 years. After all, Newtonian gravity was around for \mbox{200 years}. \mbox{GR has} just reached its peak, when data and computing power have caught up with t\mbox{he theory}. \mbox{We are} at a pivotal moment in the history of GR. We are on the point of confirming beyond reasonable doubt all its predictions throughout its entire domain of validity.

We have seen how modern cosmology is faced with big questions which touch the very foundations of physics. What is this form of matter which interacts only with gravity and apparently with nothing else? Why is the expansion of the universe accelerating? What caused the universe to undergo a period of rapid expansion soon after the Big Bang? These questions, motivated by cosmological observations, lead to questions about fundamental physics. Are there forces and interactions besides the four we know of, that is, gravity, electromagnetism, and the strong and weak nuclear forces? Are there particles beyond the Standard Model? What determines the value of the fundamental constants of nature? What is the real structure of spacetime? Are there \mbox{extra dimensions}?

Science needs data, so each of these questions must be addressed through careful experiment. The challenge of modern experimental physics is to probe nature at extreme distances and energies, well outside the capabilities of the instruments that were available to Einstein. It has certainly come a long way, as shown by the detection of gravitational waves in 2015, a feat which was thought to be impossible by many of Einstein's contemporaries.

General Relativity is not the final theory of gravity, for there is no such thing. As General Relativity turns 100, we would do well to celebrate it with a healthy does of scientific scepticism. Long live General Relativity, and a big welcome to its eventual replacement, whether in our lifetime or not.

%%%%%%%%%%%%%%%%%%%%%%%%%%%%%%%%%%%%%%%%%%%
\vspace{6pt}
%
%%%%%%%%%%%%%%%%%%%%%%%%%%%%%%%%%%%%%%%%%%%
%%% optional
%\supplementary{The following are available online at www.mdpi.com/link, Figure S1: title, Table S1: title, Video S1: title.}
%
%%%%%%%%%%%%%%%%%%%%%%%%%%%%%%%%%%%%%%%%%%%
\acknowledgments{George F. Smoot acknowledges support through his Chaire d'Excellence Universit\'e Sorbonne Paris Cit\'e and the financial support of the UnivEarthS Labex programme at Universit\'e Sorbonne Paris Cit\'e (ANR-10-LABX-0023 and ANR-11-IDEX-0005-02).
I.D. acknowledges that the research work disclosed in this publication is partially funded by the
REACH HIGH Scholars Programme---Post-Doctoral Grants. The grant is part-financed by the European Union, Operational Programme II---Cohesion Policy 2014--2020.}
%
%%%%%%%%%%%%%%%%%%%%%%%%%%%%%%%%%%%%%%%%%%%
\authorcontributions{All the authors conceived the idea and contributed equally.}
%Pls add it.
%For research articles with several authors, a short paragraph specifying their individual contributions must be provided. The following statements should be used ``X.X. and Y.Y. conceived and designed the experiments; X.X. performed the experiments; X.X. and Y.Y. analyzed the data; W.W. contributed reagents/materials/analysis tools; Y.Y. wrote the paper.'' Authorship must be limited to those who have contributed substantially to the work reported.}
%
%%%%%%%%%%%%%%%%%%%%%%%%%%%%%%%%%%%%%%%%%%%
\conflictofinterests{The authors declare no conflict of interest.}
%
%%%%%%%%%%%%%%%%%%%%%%%%%%%%%%%%%%%%%%%%%%%
%%% optional

\abbreviations{The following abbreviations are used in this work:\\
\begin{description}[font=\normalfont, labelindent=0pt, labelsep=60pt, leftmargin=60pt, style=multiline, parsep=6pt]
\vspace{-6pt}
\item[AU] Astronomical Unit
\item[CDM] Cold Dark Matter
\item[CMB] Cosmic Microwave Background
\item[EEP] Einstein Equivalence Principle
\item[ECKS] Einstein-Cartan-Kibble-Sciama
\item[eV] electronvolt
\item[FLRW] Friedmann-Lema\^{i}tre-Robertson-Walker
\item[GR] General Relativity
\item[Gy] Gigayear ($10^{9}$ years)
\item[$\Lambda$CDM] $\Lambda$ Cold Dark Matter
\item[LLI] Local Lorentz invariance
\item[LLR] Lunar laser ranging
\item[LPI] Local position invariance
\item[Mpc] Megaparsec
\item[PPN] Parameterised post-Newtonian
\item[SR]  Special Relativity
\item[TeV] teraelectronvolt ($10^{12}$ electronvolts)
\end{description}

%GR: General Relativity\\
%SR: Special Relativity\\
%CDM: Cold Dark Matter\\
%EEP: Einstein Equivalence Principle\\
%AU: Astronomical Unit\\
%Mpc: Megaparsec\\
%Gy: Gigayear ($1\times 10^{9}$ years)\\
}
%
%%%%%%%%%%%%%%%%%%%%%%%%%%%%%%%%%%%%%%%%%%%

%%%%%%%%%%%%%%%%%%%%%%%%%%%%%%%%%%%%%%%%%%
\bibliographystyle{mdpi}
\renewcommand\bibname{References}
%=====================================
% References, variant A: internal bibliography
%=====================================
%\input{SmootDebonoGR_SUBMISSION_MODIFIED.bbl}

%=====================================
% References, variant B: external bibliography
%=====================================
%\bibliography{GR_bibliography}

%%%%%%%%%%%%%%%%%%%%%%%%%%%%%%%%%%%%%%%%%%
%% optional
%\sampleavailability{Samples of the compounds ...... are available from the authors.}

%%%%%%%%%%%%%%%%%%%%%%%%%%%%%%%%%%%%%%%%%%
\end{document}